\documentclass[iop,revtex4,numberedappendix,appendixfloats]{emulateapj}

\usepackage{amsmath}
\usepackage{mathrsfs}
\usepackage{lscape}
\usepackage{wasysym}

\usepackage{grffile}
\usepackage{subfigure}

\usepackage{hyperref}

\makeatletter
\newcommand\software{\@testopt\@software{[}}%
\def\@software[#1]#2{{\vskip6pt{\large\it Software:} #2}}%
\makeatother

\newcommand{\vect}[1]{\boldsymbol{#1}}
\newcommand*\diff{\mathop{}\!\mathrm{d}}
\newcommand*\Diff[1]{\mathop{}\!\mathrm{d^#1}}
\newcommand{\pdf}{\ensuremath{pdf}}
\newcommand{\RM}{{\sl RoadMapping}}

\usepackage[usenames,dvipsnames]{xcolor}
\definecolor{darkorange}{HTML}{FF7F00}
\definecolor{brightorange}{HTML}{FDBF6F}
\definecolor{darkgreen}{HTML}{33A02C}
\definecolor{brightgreen}{HTML}{B2DF8A}
\usepackage{tikz}
\newcommand{\tikzcircle}[2][black,fill=black]{\tikz[baseline=-0.5ex]\draw[#1] (0,0) circle (#2);}%
\newcommand{\tikzsquare}[2][black,fill=black]{\tikz[baseline=-0.5ex]\draw[#1] ([xshift=-2pt,yshift=-2pt]10,0) rectangle ++(#2,#2);}%

\shorttitle{Action-based Dynamical Modeling: The Influence of Spiral Arms}
\shortauthors{Trick et al.}

\begin{document}

\title{Action-based Dynamical Modeling for the Milky Way Disk:\\The Influence of Spiral Arms\\}

\author{Wilma H. Trick\altaffilmark{1,2}, Jo Bovy\altaffilmark{3,5}, Elena D'Onghia\altaffilmark{4,5}, and Hans-Walter Rix\altaffilmark{1}}

\altaffiltext{1}{Max-Planck-Institut f\"ur Astronomie, K\"onigstuhl 17, D-69117 Heidelberg, Germany}
\altaffiltext{2}{Correspondence should be addressed to trick@mpia.de.}
\altaffiltext{3}{Department of Astronomy and Astrophysics, University of Toronto, 50 St. George Street, Toronto, ON, M5S 3H4, Canada}
\altaffiltext{4}{Department of Astronomy, University of Wisconsin, 2535 Sterling Hall, 475 N. Charter Street, Madison, WI 53076, USA}
\altaffiltext{5}{Alfred P. Sloan Fellow}


\begin{abstract}
\RM{} is a dynamical modeling machinery developed to constrain the Milky Way's (MW) gravitational potential by simultaneously fitting an axisymmetric parametrized potential and an action-based orbit distribution function (DF) to discrete 6D phase-space measurements of stars in the Galactic disk. In this work we demonstrate \RM{}'s robustness in the presence of spiral arms by modeling data drawn from an $N$-body simulation snapshot of a disk-dominated galaxy of MW mass with strong spiral arms (but no bar), exploring survey volumes with radii $500~\text{pc} \leq r_\text{max} \leq 5~\text{kpc}$. The potential constraints are very robust, even though we use a simple action-based DF, the quasi-isothermal DF (qDF). The best-fit \RM{} model always recovers the correct gravitational forces where most of the stars that entered the analysis are located, even for small volumes. For data from large survey volumes, \RM{} finds axisymmetric models that average well over the spiral arms. Unsurprisingly, the models are slightly biased by the excess of stars in the spiral arms. Gravitational potential models derived from survey volumes with at least $r_\text{max}= 3~\text{kpc}$ can be reliably extrapolated to larger volumes. However, a large radial survey extent, $r_\text{max}\sim 5~\text{kpc}$, is needed to correctly recover the halo scale length. In general, the recovery and extrapolability of potentials inferred from data sets which were drawn from inter-arm regions appear to be better than those of data sets drawn from spiral arms. Our analysis implies that building axisymmetric models for the Galaxy with upcoming Gaia data will lead to sensible and robust approximations of the MW's potential.
\end{abstract}


\section{Introduction} \label{sec:intro}

An important basis for learning more about the Milky Way's (MW) overall gravitational potential and orbit distribution function (DF) is to find the ``best possible'' axisymmetric model for the Galaxy. Given such a model the identification and characterization of non-axisymmetries like spiral arms or stellar streams in stellar phase-space (and chemical abundance) data would then become more straightforward.

Several approaches to constrain an axisymmetric potential and/or orbit DF have recently been put forward: \citet{2013ApJ...779..115B} and \citet{2014MNRAS.445.3133P} fitted potential and DF simultaneously to stellar kinematics in the disk and got precise constraints on the overall potential; \citet{2015MNRAS.449.3479S} and \citet{2016MNRAS.460.1725D} investigated extended DFs for the disk and halo respectively (given a fiducial potential), that included the metallicity of each star, in addition to the distribution in orbit space.

In this work we will continue our investigation of the \RM{} approach (\emph{``Recovery of the Orbit Action Distribution of Mono-Abundance Populations and Potential INference for our Galaxy''}). The first application of \RM{} was performed by \citet{2013ApJ...779..115B}. \citet{2016ApJ...830...97T}, hereafter Paper I, subsequently performed a detailed analysis of the strengths and limitations of the approach. \RM{} presumes that simple stellar populations in the MW disk---be it mono-abundance populations (MAPs), i.e., stars with the same $[\mathrm{Fe}/\mathrm{H}]$ and $[\alpha/\mathrm{Fe}]$ \citep{2012ApJ...751..131B,2012ApJ...753..148B,2012ApJ...755..115B,2016ApJ...823...30B}, or maybe also mono-age populations \citep{2013ApJ...773...43B,2014MNRAS.442.2474M,2016MNRAS.456.3655M,2014A&A...572A..92M,2016ApJ...823..114N,2017ApJ...834...27M}---follow simple orbit DFs, like, e.g., the quasi-isothermal DF (qDF) by \citet{2011MNRAS.413.1889B}. That MAPs in the MW are well described by the qDF was first shown by \citet{2013MNRAS.434..652T}. The qDF is expressed in terms of the orbital actions $\vect{J}=(J_R,J_\phi=L_z,J_z)$, which are integrals of motion, quantify the amount of the orbit's oscillation in each of the coordinate directions $(R,\phi,z)$, and are therefore excellent orbit labels. Given an assumed gravitational potential one can calculate the orbital actions from the stars' current phase-space positions $(\vect{x},\vect{v})$ (\citealt{2012MNRAS.426.1324B,2016MNRAS.457.2107S}; see also \citet{2014ApJ...795...95B} for a general method to compute $(\vect{x},\vect{v})\longrightarrow(\vect{J},\vect{\theta})$). Only if this assumed gravitational potential is close to the true potential, the action distribution of the stellar MAP in question will follow an orbit DF of qDF-shape. This is the idea on which \RM{} builds, and which allows us to simultaneously fit potential and orbit DF to observations.

\citet{2013ApJ...779..115B} employed this approach to measure the Milky Way's surface density profile within $1.1~\text{kpc}$ using 43 MAPs in the Galactic disk from the SDSS/SEGUE survey \citep{2009AJ....137.4377Y}. To avoid spiral arm effects, they did not use in-plane motions. Their potential model had only two free parameters (disk scale length and relative halo-to-disk contribution to the radial force at the solar radius). To account for missing model flexibility they constrained the surface density for each MAP only at one best radius. The profile they derived in this fashion had a scale length of $R_\text{s}=2.5~\text{kpc}$ and was---in the regime $R>6.6~\text{kpc}$---later confirmed by \citet{2014MNRAS.445.3133P} using a different action-based procedure.

Given the success of this first application and in anticipation of the upcoming data releases from Gaia in 2016-2022 \citep{2013CEAB...37..115E}, Paper I improved the \RM{} machinery and studied its strengths and breakdowns in detail, by investigating a large suite of mock data sets. Under the prerequisite of axisymmetric data and model, we found that \RM{}'s modeling success is stable against minor misjudgments of DF or selection function, and that---if the true potential is not contained in the proposed family of model potentials---one can still find a good fit that returns the correct forces, given the limitations of the model. Paper I also found that measurement uncertainties of the order of those by the final Gaia data release should be good enough (within $3~\text{kpc}$ from the Sun) to allow for precise and unbiased modeling results. 

The MW is, however, not axisymmetric. The bulge contains a strong bar \citep{1980ApJ...236..779L,1991ApJ...379..631B,1991MNRAS.252..210B,1997MNRAS.288..365B,2000MNRAS.317L..45H,2013MNRAS.435.1874W} and the disk itself is threaded by spiral arms \citep{1958MNRAS.118..379O,1976A&A....49...57G,2009PASP..121..213C,2014ApJ...783..130R}, and (ring-like) overdensities \citep{2002ApJ...569..245N,2008ApJ...673..864J,2015ApJ...801..105X}, which induce non-circular motions and asymmetries in stellar number counts. There is also kinematic evidence in the disk for moving groups \citep{1998AJ....115.2384D,2005A&A...430..165F,2009ApJ...700.1794B,2010ApJ...717..617B} and streaming motions (in 21 cm or velocities) \citep{2015ApJ...800...83B,2013MNRAS.436..101W,2012MNRAS.425.2335S}, both of which are likely caused by non-axisymmetric perturbations to the gravitational potential.

As \RM{} and related approaches can only build axisymmetric models, this is an important breakdown of modeling assumptions which was not investigated in Paper I. In this paper we want to understand in which respects \RM{} will still give reliable constraints on the MW's gravitational potential in the presence of spiral arms.

Our investigation makes use of an $N$-body simulation snapshot of a spiral galaxy with strong spiral arms presented in \citet{2013ApJ...766...34D} (see their Figure 8, top left panel). From this snapshot we draw mock data in regions with different spiral arm strengths. We then apply the \RM{} machinery to these data sets and test how well we recover the local and overall gravitational potential.

In Paper I we confirmed and tested separately the robustness of \RM{} in the case that the data came from a different model family---for either the potential or DF---than assumed in the dynamical modeling. What would happen if both potential and DF model families were slightly wrong at the same time? The set-up of this study will automatically cover this important test case. The potential and orbit DF model that we are using were picked as a pragmatic compromise between (i) being a reasonable choice given the initial axisymmetric set-up of the galaxy simulation, and (ii) because of their simplicity, computational advantages, and---in case of the qDF---because that's what we are planning to use in the MW. Given that the simulation has evolved away from its axisymmetric beginnings, we expect our chosen model to be reasonable, but not particularly well-suited to model this galaxy.

Spiral arms introduce another---but possibly minor---breakdown of the modeling assumptions: In a non-axisymmetric gravitational potential the three actions will not be strict integrals of motions anymore \citep{2008gady.book.....B,2011A&A...527A.147M,2012A&A...548A.127M,2012MNRAS.422.1363S,2012MNRAS.421.1529G,2016ApJ...824...39V}. It will be interesting to see if the action-based DF in \RM{} modeling is still informative, even if it only uses the approximate actions estimated in an axisymmetric potential.

In the MW we expect the central bar to introduce additional non-axisymmetries in the Galactic disk---but presumably not stronger ones than the spiral arms. As the galaxy simulation in this work does not have a central bar, we do not investigate specific bar effects here.

Though non-axisymmetry could be a severe problem for \RM{}, we show in this paper that \RM{} potential estimates are still surprisingly accurate, which makes us optimistic that they will also be so for the MW. 

This paper is organized as follows. Section \ref{sec:simulation} describes the $N$-body simulation snapshot of a spiral galaxy that we are going to model in this study, explains how we extract 6D stellar phase-space data from it, and how we quantify the spiral arm strength. There we also review similarities and differences between the simulation in this work and what we know about the MW. Section \ref{sec:RoadMapping} summarizes the \RM{} dynamical modeling framework, and introduces the DF and potential model that we will fit to the data. Section \ref{sec:results} is dedicated to presenting the results: In Section \ref{sec:results_part1} we discuss in detail the \RM{} modeling results derived from one data set within a survey volume with radius $r_\text{max}=4~\text{kpc}$ around the Sun. Section \ref{sec:results_part2} then investigates a whole suite of \RM{} analyses, corresponding to survey volumes of different sizes and different positions within the galaxy and with respect to the spiral arms. In Section \ref{sec:discussion} we discuss the results and give an outlook to the application of \RM{} to Gaia data. We conclude in Section \ref{sec:conclusion}.

\section{Data from a galaxy simulation} \label{sec:simulation}

\RM{} requires 6D phase-space coordinates $(\vect{x}_i,\vect{v}_i)$ for a large set of stars that move independently in a collisionless galactic potential. If we want to test \RM{} on a simulated galaxy, it is most convenient to apply it to an $N$-body simulation with a huge number of low-mass ``star'' particles. In that way, we can directly take the positions and the motions of individual particles as independent tracers of the potential, just as with the stars in the MW, without having to use an error-prone prescription to turn a single particle into many stars. The high-resolution simulations with its millions of particles by \citet{2013ApJ...766...34D} satisfy this requirement.

\begin{figure}[!htbp]
	\subfigure[Surface mass density of particles. \label{fig:simulation_xy}]{\includegraphics[width=\columnwidth]{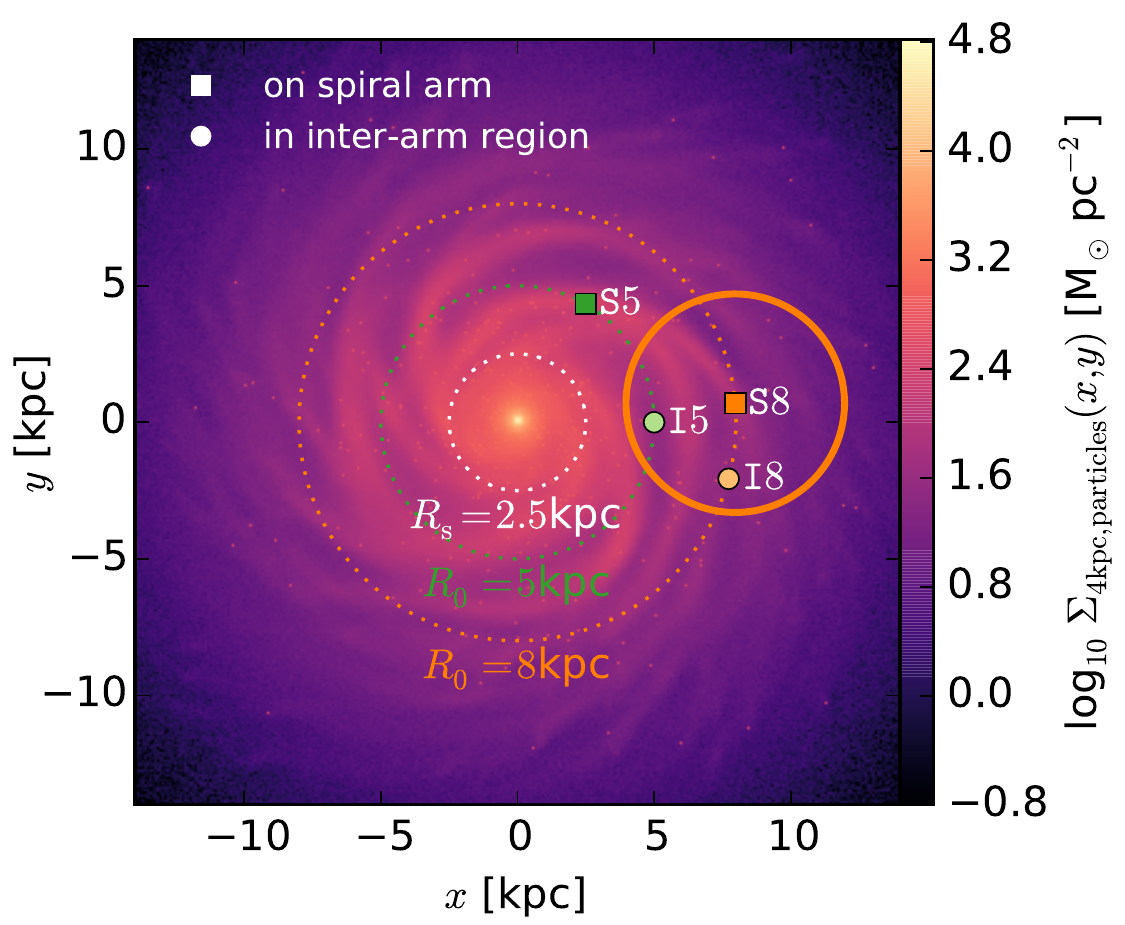}} %
    \subfigure[Mass density of particles. \label{fig:simulation_Rz}]{\includegraphics[width=\columnwidth]{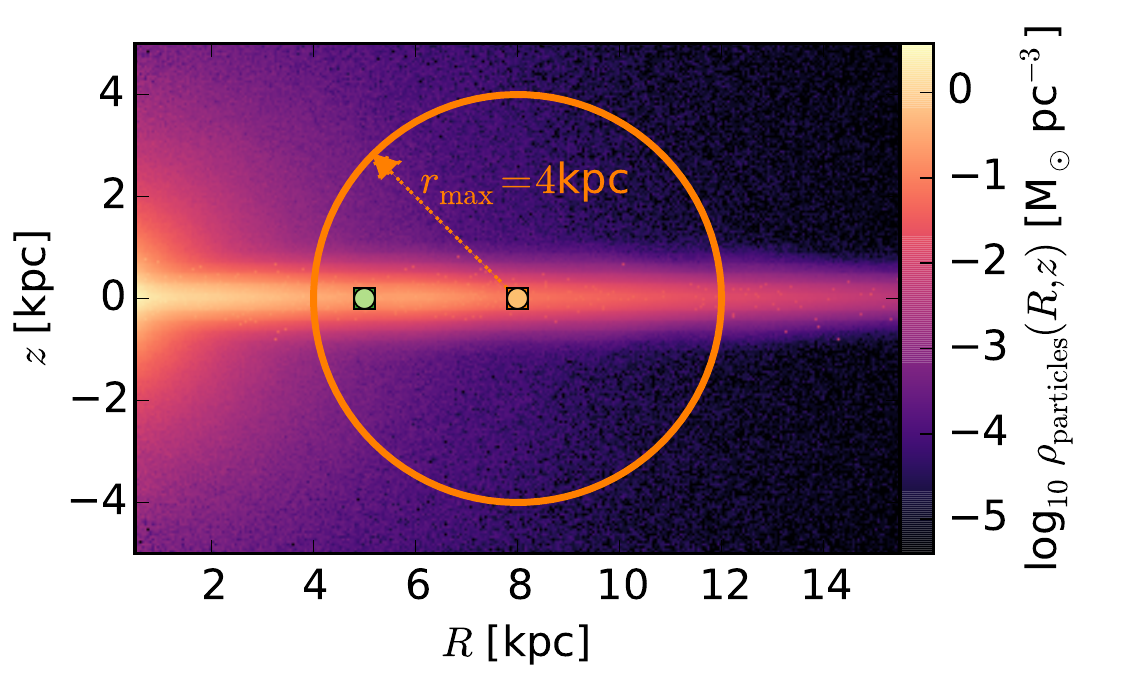}}
\caption{Simulation snapshot by \citet{2013ApJ...766...34D}. Shown are the surface mass density (in the $(x,y)$-plane, panel \ref{fig:simulation_xy}) and mass density (in the $(R,z)$-plane, panel \ref{fig:simulation_Rz}) of the ``star'' particles belonging to disk, bulge and giant molecular clouds (the dark matter halo in this simulation is static and analytic and not shown here). Overplotted are the disk's scale length $R_\text{s}=2.5~\text{kpc}$ (see Section \ref{sec:simulation_description}) and the radii at which we center our test survey volumes in this investigation, $R_0=8~\text{kpc}$ and $R_0 = 5~\text{kpc}$ (see Section \ref{sec:survey_volume_data}). The centers of the different survey volumes are marked with a square, if the survey volume is centered on a spiral arm (\texttt{S8} and \texttt{\texttt{S5}}), or with a circle, if the volume is centered on an inter-arm region (\texttt{I8} and \texttt{\texttt{I5}}). The coordinates are summarized in Table \ref{tbl:volume_positions}. The orange circle with radius $r_\text{max}=4~\text{kpc}$ marks the survey volume in which we conduct the analysis discussed in detail in Section \ref{sec:results_part1}.}
\label{fig:simulation}
\end{figure}

\subsection{Description of the galaxy simulation snapshot} \label{sec:simulation_description}

The high-resolution $N$-body simulation snapshot of a disk galaxy by \citet{2013ApJ...766...34D}, which we use in this work, was carried out with the \texttt{GADGET-3} code, and set up in the manner described in \citet{2005MNRAS.361..776S}. In this simulation, overdensities with properties similar to giant molecular clouds induced four prominent spiral arms---and therefore a non-axisymmetric sub-structure---via the swing amplification mechanism. This galaxy simulation was also investigated by \citet{2013ApJ...766...34D} (their Figure 8, top left panel) and \citet{2015ApJ...808L...8D} (their Figure 2, top left panel). For details see \citet{2013ApJ...766...34D}, here we summarize the essential characteristics.

The simulation has a gravitationally evolving stellar disk within a static/rigid analytic dark matter (DM) halo.

The analytic halo follows a \citet{1990ApJ...356..359H} profile
\begin{equation}
\rho_\text{dm}(r) = \frac{M_\text{dm}}{2\pi} \frac{a_\text{dm}}{r (r+a_\text{dm})^3} \label{eq:dm_hernquist}
\end{equation}
with total halo mass $M_\text{dm} = 9.5\times 10^{11} ~\text{M}_\odot$ and scale length $a_\text{dm} = 29~\text{kpc}$.

The disk consists of $10^8$ ``disk star'' particles, each having a mass of $\sim370 ~\text{M}_\odot$, and 1000 ``giant molecular cloud'' particles with mass $\sim9.5\times 10^{5} ~\text{M}_\odot$. The initial vertical mass distribution of the stars in the disk is specified by the profile of an isothermal sheet with a radially constant scale height $z_\text{s,init}$, i.e.,
\begin{equation}
\rho_*(R,z) = \frac{M_*}{4\pi z_\text{s,init} R_\text{s}^2} \text{sech}^2 \left( \frac{z}{z_\text{s,init}}\right) \exp \left(- \frac{R}{R_\text{s}} \right), \label{eq:sech_disk}
\end{equation}
with total disk mass $M_* = 0.04 M_\text{dm} = 3.8\times 10^{10}~\text{M}_\odot$. The scale-length $R_\text{s}$ is assumed to be $2.5~\text{kpc}$ and $z_\text{s,init}=0.1 R_\text{s}$. In this model the disk fraction within $2.2$ scale lengths is 50\% of the total mass, leading to a formation of approximately four arms \citep{2015ApJ...808L...8D} (see Figure \ref{fig:surf_dens_fourier_b}).

The bulge consists of $10^7$ ``bulge star'' particles with mass $\sim950 ~\text{M}_\odot$ and they are distributed following a spherical Hernquist profile analogous to Equation \eqref{eq:dm_hernquist}, with total mass $M_\text{bulge}=0.01 M_\text{dm} = 9.5\times 10^9~\text{M}_\odot$ and scale length $a_\text{bulge}=0.1 R_\text{s}=0.25~\text{kpc}$.

The initial velocity setup of the ``disk star'' particles assumes for simplicity Gaussian velocity dispersion profiles \citep{2005MNRAS.361..776S}.

The simulation snapshot which we are using in this work has evolved from these initial conditions in isolation for $\sim 250~\text{Myr}$, which corresponds to approximately one orbital period at $R\sim8~\text{kpc}$. The mass density of simulation particles (without the DM halo) at this snapshot time is shown in Figure \ref{fig:simulation}. The ``molecular cloud perturbers'', which caused the formation of the four pronounced spiral arms, can be seen in Figure \ref{fig:simulation} as small overdensities in the disk. The spherical bulge and very flattened disk are shown in Figure \ref{fig:simulation_Rz}.

We have confirmed that the gravitational center of the particles corresponds to the coordinate origin.

\subsection{Data selection and survey volume} \label{sec:survey_volume_data}

\begin{deluxetable}{clccc}[!htbp]
\tabletypesize{\scriptsize}
\tablecaption{Vantage points within the galaxy simulation snapshot around which we center survey volumes of radius $r_\text{max}$. \label{tbl:volume_positions}}
\tablewidth{0pt}
\tablehead{
\colhead{name} & \colhead{position} & \colhead{$R_0$ [kpc]} & \colhead{$\phi_0$ [degrees]} & \colhead{legend}}
\startdata
\tableline
\texttt{S8} & on spiral arm & 8 & 5 & \tikzsquare[fill=darkorange]{6pt}\\
\texttt{I8} & in inter-arm region & 8 & -15 & \tikzcircle[fill=brightorange]{3pt}\\
\texttt{S5} & on spiral arm & 5 & 60 & \tikzsquare[fill=darkgreen]{6pt}\\
\texttt{I5} & in inter-arm region & 5 & 0 & \tikzcircle[fill=brightgreen]{3pt}
\enddata
\tablecomments{All volumes are centered on $z_0=0$ in the plane of the disk, and $\phi_0$ is measured counter-clockwise from the positive $x$-coordinate axis.}
\end{deluxetable}

The selection function of all-sky surveys like Gaia, that are only limited by the brightness of the tracers, are contiguous and---when ignoring anisotropic effects like dust obscuration---spherical in shape. For simplicity we will use spherical survey volumes centered on different vantage points, and with sharp edges at a distance $r_\text{max}$ around it (see Equation \eqref{eq:selection_function}), which corresponds to a magnitude cut for stellar tracers all having the same luminosity. 

\begin{figure}[!htbp]
\centering
 \subfigure[Surface density along $\phi$ at {$R_0= [5,8] ~ \text{kpc}$}. \label{fig:surf_dens_fourier_a}]{\includegraphics[width=0.7\columnwidth]{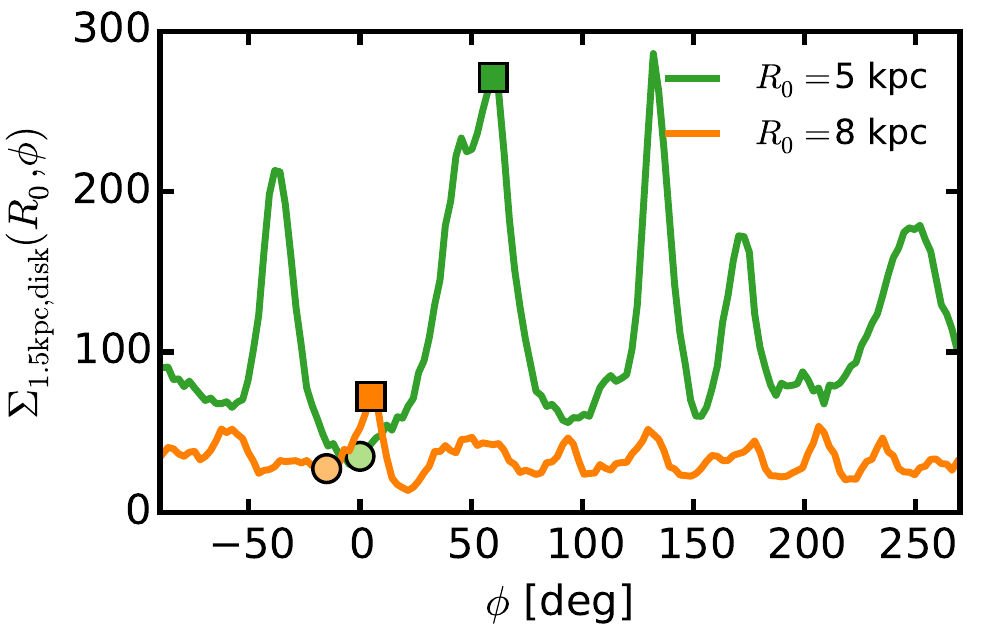}}
  \subfigure[Spiral strength from Fourier mode analysis. \label{fig:surf_dens_fourier_b}]{\includegraphics[width=0.7\columnwidth]{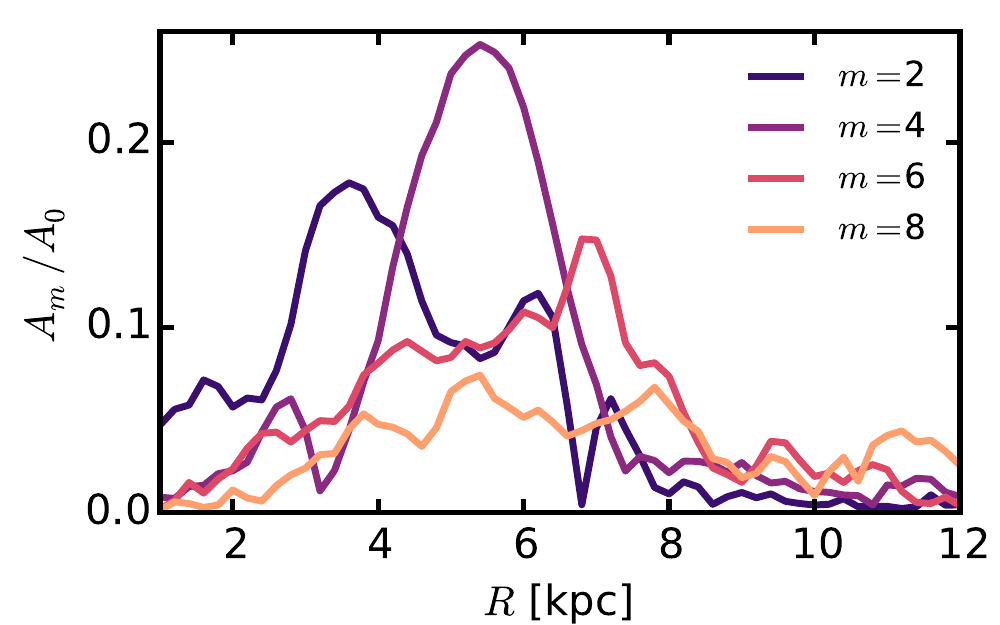}}
\caption{Demonstrating the spiral arm strength at different radii. Panel \ref{fig:surf_dens_fourier_a} shows the surface density along the azimuth angle at the radii $R_0$ on which we center our survey volumes. We also mark the corresponding $\phi_0$ from Table \ref{tbl:volume_positions}. The difference between the surface density at \texttt{S5} (\tikzsquare[fill=darkgreen]{6pt}) and \texttt{I5} (\tikzcircle[fill=brightgreen]{3pt}) is 200\% of the mean surface density at $R_0$; for \texttt{S8} (\tikzsquare[fill=darkorange]{6pt}) and \texttt{I8} (\tikzcircle[fill=brightorange]{3pt}) the difference is 130\%. Panel \ref{fig:surf_dens_fourier_b} shows the Fourier model amplitudes for $m=2,4,6,8$  calculated as $A_m/A_0 = |\sum_l M_l \exp \left(i m \phi_l \right) | / \sum_l M_l$ for all disk particles at a given radius with mass $M_l$ and azimuth position $\phi_l$. As can be seen, the simulation has overall four strong spiral arms, dominating between $R=4~\text{kpc}$ and $R=7~\text{kpc}$. Inside of that there are two, and outside of that six or more arms. As the particle density increases with smaller radius, most tracers in the analysis will come from regions with only a few strong spiral arms.}
\label{fig:surf_dens_fourier}
\end{figure}

\begin{figure}[!htbp]
\centering
\includegraphics[width=0.7\columnwidth]{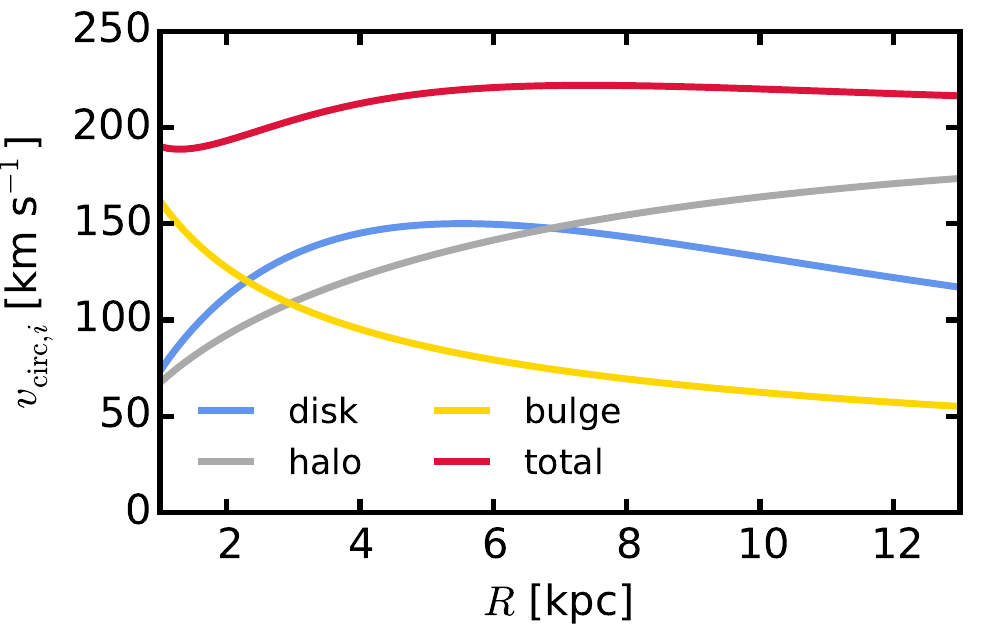}
\caption{Circular velocity curve of the \texttt{DEHH-Pot}, i.e., the symmetrized best fit to the $N$-body simulation, and its disk, halo and bulge components. The rotational support at $2.2$ scale lengths is $(v_{\rm circ,disk}/v_{\rm circ,total})^2 \sim 47\%$. This demonstrates that the simulation is a disk-dominated spiral galaxy.}
\label{fig:DEHH_vcirc_decomposed}
\end{figure}

Figure \ref{fig:simulation} illustrates the different survey volume positions analyzed in this study. We selected volumes with $r_\text{max}=[0.5,1,2,3,4,5]~\text{kpc}$ centered on a spiral arm (\texttt{S}) and on an inter-arm region (\texttt{I}) at both the equivalent of the solar radius, $R_0=8~\text{kpc}$, and at $R_0=5~\text{kpc}$, where the disk strongly dominates (see Figure \ref{fig:DEHH_vcirc_decomposed}), and the spiral arms are more pronounced than at $R_0=8~\text{kpc}$ (see Figure \ref{fig:surf_dens_fourier}). The exact positions of the vantage points \texttt{S8}, \texttt{I8}, \texttt{S5}, and \texttt{I5} are summarized in Table \ref{tbl:volume_positions}.

From within each volume we drew $N_*=20,000$ random ``disk star'' particles, and used their phase-space positions $(\vect{x}_i,\vect{v}_i)$ within the simulated galaxy's rest-frame as data. 

To make the data sample more realistic, one would actually have to add measurement uncertainties, especially to the distances from the survey volume's central vantage point and the proper motions measured from there. We decided not to include measurement uncertainties: Firstly, their effect on \RM{} modeling has been already investigated in Paper I, and we found that the measurement uncertainties of the last data release of Gaia should be small enough to not significantly disturb the modeling. Secondly, in this study we want to isolate and investigate the deviations of the data from axisymmetry and the assumed potential and DF model independently of other effects.

\begin{deluxetable}{lll}[!htbp]
\tabletypesize{\scriptsize}
\tablecaption{Best fit parameters of the \texttt{DEHH-Pot}. \label{tbl:DEHH-Pot}}
\tablewidth{0pt}
\startdata
\tableline
circular velocity & $v_\text{circ}(R_\text{\sun})$ & $222~\text{km s}^{-1}$ \\
disk scale length & $R_\text{s}$ & $2.5~\text{kpc}$ \\
disk scale height & $z_\text{s}$ & $0.17~\text{kpc}$ \\
halo fraction & $f_\text{halo}$ & $0.54$\\
halo scale length & $a_\text{halo}$ & $29~\text{kpc}$ \\
bulge mass & $M_\text{bulge}$ & $0.95 \times 10^{10}~\text{M}_\odot$\\
bulge scale length & $a_\text{bulge}$ & $0.25~\text{kpc}$
\enddata
\tablecomments{The \texttt{DEHH-Pot} is introduced in Section \ref{sec:DEHH-Pot}, and we use it as the global best fit symmetrized potential model for the simulated galaxy. The halo fraction, $f_\text{halo}$, and circular velocity at the ``solar'' radius, $v_\text{circ}(R_\text{\sun})$, which scales the total mass of the model, are defined in Equations \eqref{eq:fhalo} and \eqref{eq:circvel}, with $R_\text{\sun}=8~\text{kpc}$.}
\end{deluxetable}

\pagebreak
\subsection{True symmetrized potential} \label{sec:DEHH-Pot}

For a galaxy with pronounced spiral arms, an axisymmetric model matter distribution \emph{per se} cannot reproduce the true matter distribution globally. We therefore obtain an ``overall best fit symmetrized'' potential model from the distribution of particles to be able (i) to quantify the non-axisymmetries in the simulation snapshot better and (ii) to compare how close our axisymmetric \RM{} results can get to it. 

We derive this model by fitting axisymmetric analytical functions to the density distribution of each of the galaxy components' particles. The bulge and halo follow Hernquist profiles by construction (see Section \ref{sec:simulation_description}).

The disk in this simulation snapshot deviates from its initial conditions in Equation \eqref{eq:sech_disk}: after $250~\text{Myr}$ pronounced spiral arms have formed, also causing some in-plane heating. Except of an overdensity around $R\sim6~\text{kpc}$ (see Figure \ref{fig:4kpc8Spiral_vcirc_surfdens_b} in Section \ref{sec:4kpc8Spiral_potential}), the overall radial surface density profile (i.e., the azimuthal average) did not change by much, and appears smooth and exponential. We therefore chose a double exponential disk model to fit the particle distribution in the disk. The fit assumes the total disk mass to be known and be equal to the total mass of all disk particles. The best fit double exponential disk profile is found by maximizing the likelihood for all disk particles to be drawn from this axisymmetric density profile. In this way the fit does not depend on binning choices and is driven by the number and location of the stars---analogous to our \RM{} procedure. 

\begin{figure*}[!htbp]
\centering
  \subfigure[Non-axisymmetries in the snapshot. \label{fig:spiral_arm_DeltaS_a}]{\includegraphics[height=4cm]{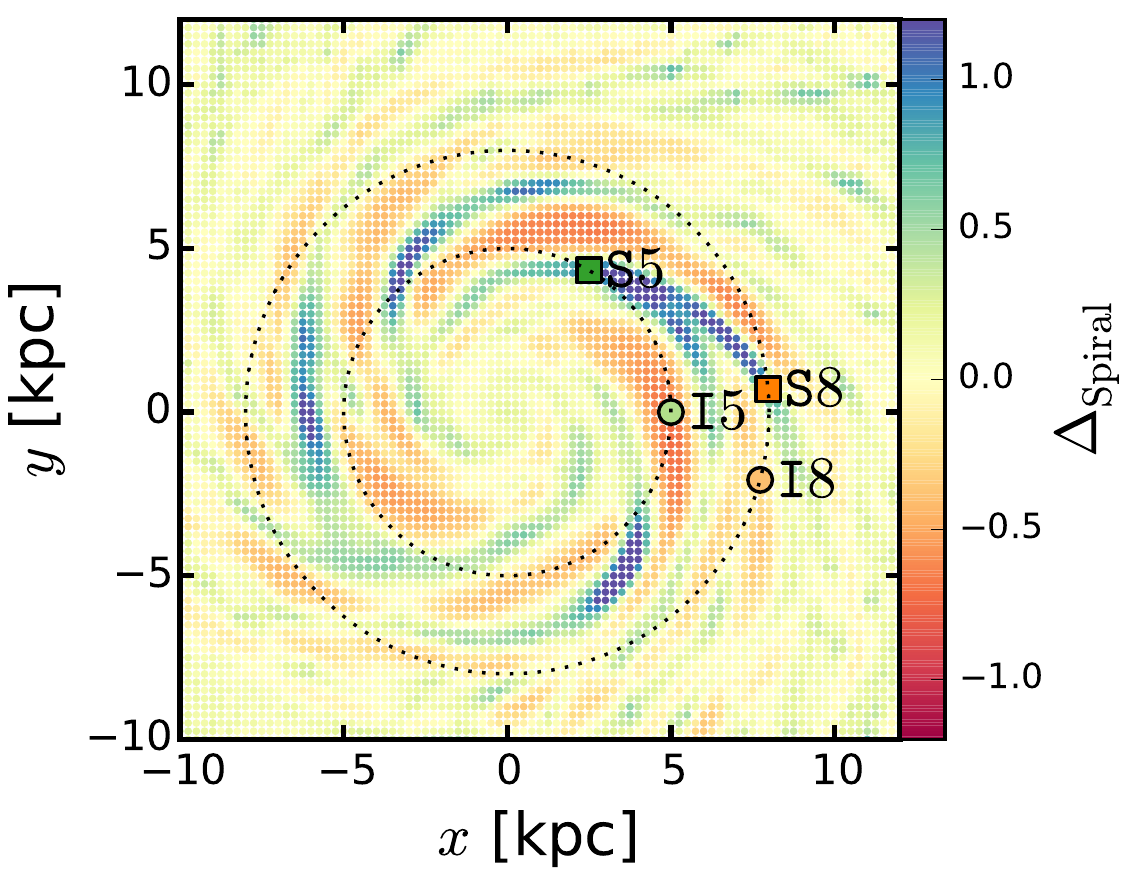}}
  \subfigure[Histogram of $\Delta_{\text{Spiral}}$.\label{fig:spiral_arm_DeltaS_b}]{\includegraphics[height=4cm]{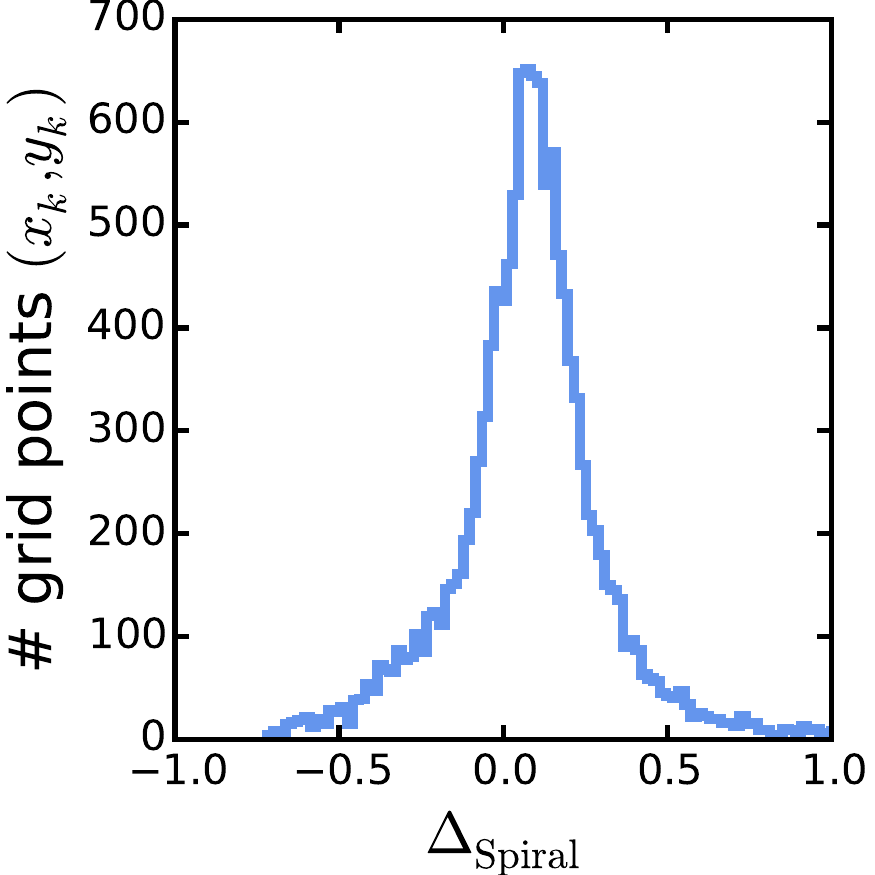}}
  \subfigure[Spiral arm dominance.\label{fig:spiral_arm_DeltaS_c}]{\includegraphics[height=4cm]{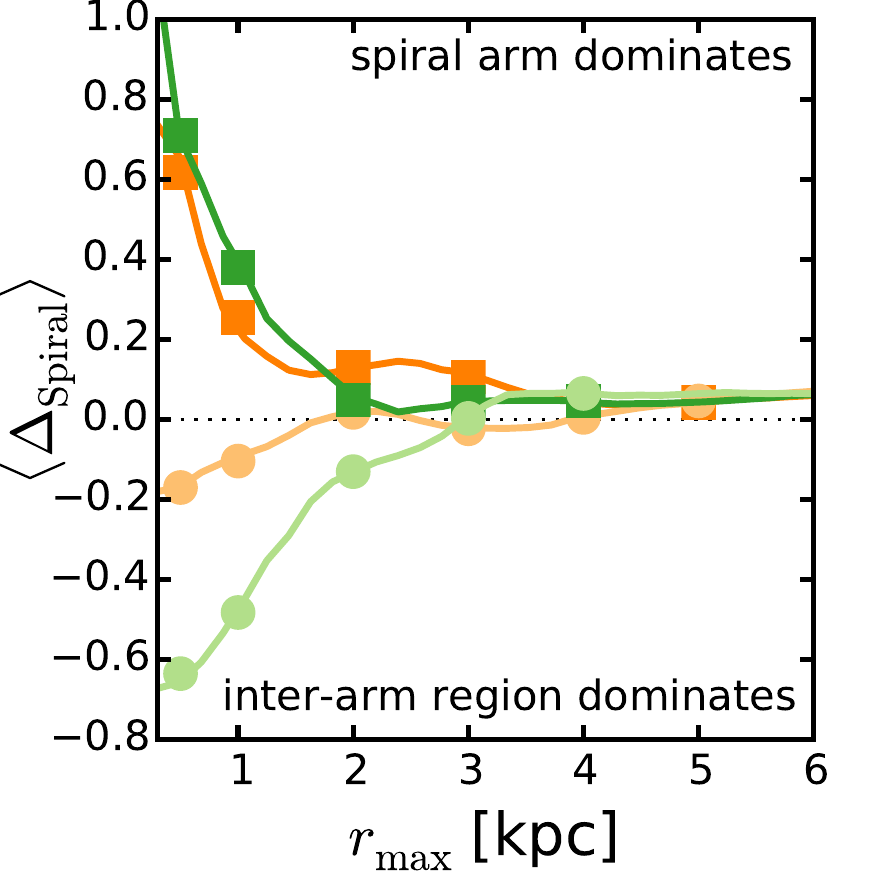}}
  \subfigure[Relative spiral contrast.\label{fig:spiral_arm_DeltaS_d}]{\includegraphics[height=4cm]{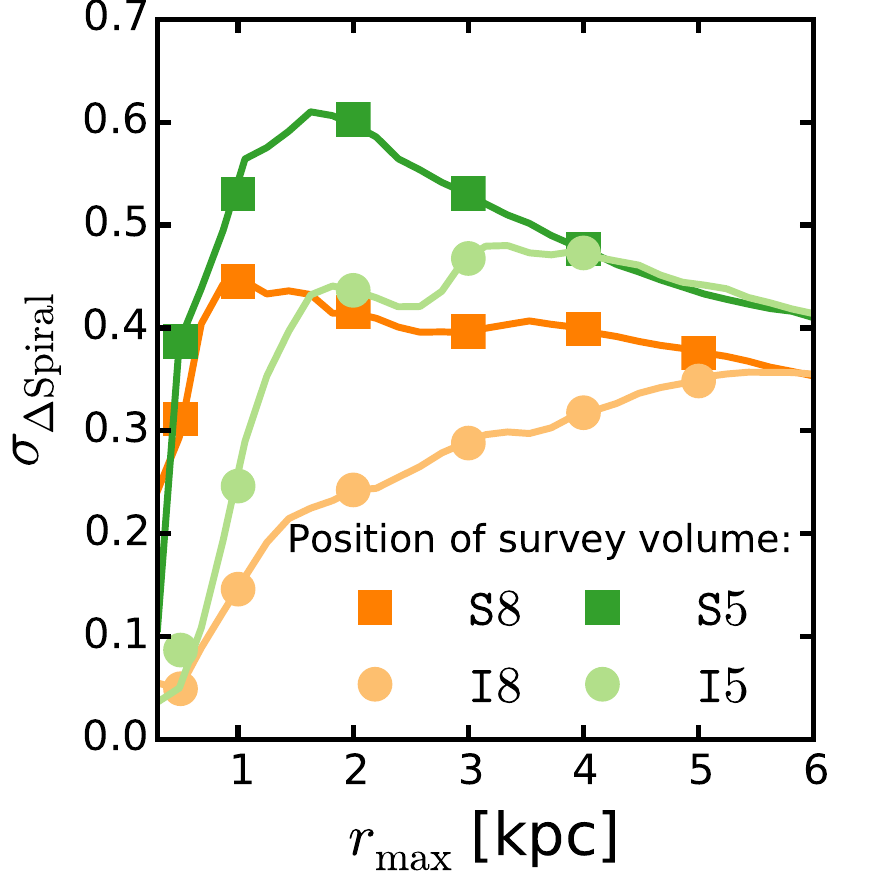}}
\caption{Dominance and contrast of the spiral arms. Panel \ref{fig:spiral_arm_DeltaS_a} shows the local spiral strength $\Delta_{\text{Spiral}}$ (calculated according to Equation \eqref{eq:DeltaS_definition} as described in Section \ref{sec:spiral_arm_DeltaS}) at regular grid points $(x_k,y_k)$ with bin width $0.25~\text{kpc}$. Marked are the centroids of the four test survey volumes of this study analogous to Figure \ref{fig:simulation}. The histogram in panel \ref{fig:spiral_arm_DeltaS_b} demonstrates the number of different $\Delta_\text{Spiral}$ values in the region $x,y \in [-14,14]~\text{kpc}$. The panels \ref{fig:spiral_arm_DeltaS_c} and \ref{fig:spiral_arm_DeltaS_d} then show the dominance and relative contrast of spiral arms and inter-arm regions within each survey volume, depending on the volumes' size, $r_\text{max}$, and position (color-coded). As measure for the spiral dominance we use the mean $\langle \Delta_\text{Spiral} \rangle$, and for the relative spiral contrast the standard deviation $\sigma_{\Delta\text{Spiral}}$, calculated on the basis of all $\Delta_{\text{Spiral}}$ measurements within the given survey volume. We chose two volumes in which the spiral arms dominate, and two in which an inter-arm region dominates. The dominance and contrast of spiral arms and inter-arm regions is stronger at $R_0=5~\text{kpc}$ than at $R_0=8~\text{kpc}$. Also, inter-arm regions appear larger and smoother than spiral arms, as already inside a small volume centered on a spiral arm the contrast is quite large. The larger the volume the more does the overall effect of spiral arms and inter-arm regions average out.}
\label{fig:spiral_arm_DeltaS}
\end{figure*}

The best fit parameters for this reference potential, to which we will refer as the \texttt{DEHH-Pot} (Double-Exponential disk + Hernquist halo + Hernquist bulge) in the remainder of this work, are given in Table \ref{tbl:DEHH-Pot}. As can be seen in Figure \ref{fig:4kpc8Spiral_dens_vcirc_surfdens} in Section \ref{sec:4kpc8Spiral_potential} below, the \texttt{DEHH-Pot} fits the overall true density distribution very well, with the exception of $z\sim0$ where the particle distribution is not as cuspy as the exponential disk. Figure \ref{fig:DEHH_vcirc_decomposed} shows the circular velocity curve of the \texttt{DEHH-Pot}, and its decomposition into disk, halo, and bulge contribution. The disk clearly dominates between $R\sim2~\text{kpc}$ and $R\sim7~\text{kpc}$.

\subsection{Quantifying the strength of spiral arms} \label{sec:spiral_arm_DeltaS}

Depending on the size and position of the survey volume, spiral arms and inter-arm regions dominate the stellar distribution within the volume to different degrees. To quantify the strength of the spiral arms, we introduce the quantity
\begin{equation}
\Delta_\text{Spiral} (x_k,y_k) \equiv \frac{\Sigma_{\text{1.5kpc,disk},T}(x_k,y_k)}{\Sigma_{\text{1.5kpc,disk},S}(x_k,y_k)} -1\label{eq:DeltaS_definition}
\end{equation}
where $\Sigma_{\text{1.5kpc,disk},\alpha}$ is the true surface density of the disk component of the simulation snapshot ($\alpha=T$ for ``true''), or of the symmetrized snapshot model \texttt{DEHH-Pot} in Section \ref{sec:DEHH-Pot} ($\alpha=S$ for ``symmetrized''),
\begin{equation}
\Sigma_{\text{1.5kpc,disk},\alpha}(x_k,y_k) \equiv \int_{-1.5~\text{kpc}}^{1.5~\text{kpc}} \rho_{\text{disk},\alpha}(x_k,y_k,z) \ \diff z.
\end{equation}
$(x_k,y_k)$ are the coordinates of regular grid points with spacing $\delta=0.25~\text{kpc}$.\footnote{We average the particle surface density of the true simulation potential over area element sizes of $\delta \times \delta$ around $(x_k,y_k)$, when calculating $\Sigma_{\text{1.5kpc,disk},T}(x_k,y_k)$.} $(x_c=R_{0,c} \cdot \cos \phi_{0,c},y_c=R_{0,c} \cdot \sin \phi_{0,c},z_c=0)$ is the position of the survey volume's center within the simulation, with $c\in\left\{ \texttt{S8},\texttt{I8},\texttt{S5},\texttt{I5}\right\}$ and $(R_{0,c},\phi_{0,c})$ given in Table \ref{tbl:volume_positions}. We consider all $n \simeq \pi r_\text{max}^2/\delta^2$ values of $\Delta_\text{Spiral} (x_k,y_k)$ inside a given survey volume of radius $r_\text{max}$ around position $c$ and calculate the mean and standard deviation,
\begin{eqnarray}
\langle \Delta_\text{Spiral} \rangle &\equiv & \frac 1n \sum_{k=1}^n \Delta_\text{Spiral}(x_k,y_k)\label{eq:mean_DeltaS}\\
\sigma_{\Delta\text{Spiral}} &\equiv & \sqrt{\frac 1n \sum_{k=1}^{n} \left[ \Delta_\text{Spiral}(x_k,y_k) - \langle \Delta_\text{Spiral} \rangle \right]^2}\label{eq:std_DeltaS}\\
 \text{with } && (x_k-x_c)^2 + (y_k-y_c)^2 \leq r_\text{max}^2.
\end{eqnarray}
These quantities tell us if and how much a spiral arm or an inter-arm region dominates the survey volume ($\langle \Delta_\text{Spiral} \rangle > 0$ for spiral arms, $\langle \Delta_\text{Spiral} \rangle < 0$ for inter-arm regions) and how large the relative contrast between spiral arms and inter-arm regions is ($\sigma_{\Delta\text{Spiral}}$). For example, volumes will have a smaller relative spiral contrast $\sigma_{\Delta\text{Spiral}}$, if they are either small and sitting completely within an inter-arm region, or if they are large volumes that contain---in addition to some spiral arms and depleted inter-arm regions---large areas of unperturbed disk.

Figure \ref{fig:spiral_arm_DeltaS} shows $\Delta_\text{Spiral}$ as function of $(x_k,y_k)$, a histogram over all $\Delta_{\text{Spiral},k}$ within the galaxy, and $\langle \Delta_\text{Spiral} \rangle$ and $\sigma_{\Delta\text{Spiral}}$ calculated for all test survey volumes in this work (see Section \ref{sec:survey_volume_data}), depending on position and size.\footnote{When considering the whole galaxy or a large survey volume, $\langle \Delta_\text{Spiral} \rangle$ in Figures \ref{fig:spiral_arm_DeltaS_b} and \ref{fig:spiral_arm_DeltaS_c} is not exactly at $0$, but slightly larger ($<0.05$). We account this small bias to the different functional forms of the \texttt{DEHH-Pot} and the initial axisymmetric disk in the simulation, Equation \eqref{eq:sech_disk}. This bias will however not affect our results.}

\subsection{Comparison of the $N$-body simulation to the MW} \label{sec:comparison_with_MW}

The $N$-body simulation in this work is not a perfect match to the MW. But to be suitable for this study it is most important that it satisfies our main requirements: it is a disk-dominated spiral galaxy with a very high number of star particles and strong spiral arms. To set the context, we discuss in the following the differences between the simulation at hand and what we know about the MW.

\paragraph{Bar.} The MW has a central bar; our simulation does not. The Galactic bar can introduce non-axisymmetries around the co-rotation and Lindblad resonances (\citealt{2000AJ....119..800D,2001A&A...373..511F,2003AJ....125..785Q,2005AJ....130..576Q,2010MNRAS.409..145S,2010ApJ...722..112M,2017MNRAS.466L.113M}; see also review by \citealt{2011MSAIS..18..185G}). Bar effects are consequently not part of this study and remain a possible source of uncertainty in \RM{}. We suspect however that its influence is modest outside of the bar region, except close to the outer Lindblad resonance.

\paragraph{Mass components.} The MW's dark halo is estimated to have a mass $M_{\text{dm},200,\text{MW}}\approx10^{12}~\text{M}_\odot$ \citep{2016ARA&A..54..529B}; the simulation's halo is only slightly less massive with $M_{\text{dm},200}\approx7\times 10^{11}~\text{M}_\odot$. The stellar bulge mass of $M_\text{bulge}=9.5\times 10^9~\text{M}_\odot$ in this simulation is a bit smaller than the estimated stellar mass of the MW bulge with $M_\text{bulge,MW} = (1.4-1.7) \times 10^{10}~\text{M}_\odot$ \citep{2015MNRAS.448..713P}. The total stellar mass of the MW is estimated to be $M_\text{stars,MW} =(5\pm1)\times 10^{10}~\text{M}_\odot$ \citep{2016ARA&A..54..529B,2013ApJ...779..115B}, consistent with the total baryonic mass in the simulation of $M_\text{stars}=4.75\times 10^{10}~\text{M}_\odot$. The fraction of bulge mass to total stellar mass of the MW is $M_\text{bulge,MW}/M_\text{stars,MW}= 0.3\pm0.06$; in our simulation it is $0.25$.

\paragraph{Disk.} The best estimate for the MW's thin disk scale length from combining several measurements in the literature is $R_\text{s} = 2.6 \pm 0.5~\text{kpc}$ \citep{2016ARA&A..54..529B}. \citet{2013ApJ...779..115B}, for example, found a stellar disk scale length of $R_\text{s}=2.15\pm0.14~\text{kpc}$. The disk of this $N$-body simulation has a similar scale length, $R_\text{s} = 2.5~\text{kpc}$. The stellar disk is however thinner than in the MW. \citet{2008ApJ...673..864J} found scale heights $300~\text{pc}$ and $900~\text{pc}$ (with $20\%$ uncertainty) for the thin and thick disk of the MW, respectively; \citet{2012ApJ...753..148B}, who considered the disk as a superposition of many exponential MAPs, measured scale heights from $\approx200~\text{pc}$ up to $1~\text{kpc}$, continuously increasing with the age of the sub-population. In our simulation there is only a very thin stellar disk component with scale height $z_\text{s}=170~\text{pc}$, and no gas and thick disk component as compared to the MW. This discrepancy does, however, not affect the objective of this study. The thick disk has a much higher velocity dispersion and is less prone to spiral perturbations. When we will apply \RM{} to real MW data, the gas disk can be included as an additional component in the mass model, analogous to \citet{2013ApJ...779..115B}, and if needed also the thick disk. By slicing data according to MAPs, the thick disk will be accounted for implicitly in the tracer selection.

\paragraph{Disk fraction.} The strength of the perturbations in the disk depends on the disk fraction (e.g., \citealt{2015ApJ...808L...8D}). The total disk mass in the simulation might be only 4\% of the halo mass, but within $2.2R_\text{s}$ it is already 50\% of the total mass. It is still under debate if the MW disk is maximal (i.e., rotational support of the disk at $2.2R_\text{s}$ is $\sim55-90\%$; \citealt{1997ApJ...483..103S, 2008gady.book.....B}, \S 6.3.3). \citet{2013ApJ...779..115B} found, for example, a maximum disk with rotational support $\left( v_\text{circ,disk}/v_\text{circ,total}\right)^2=(69\pm6)\%$ at $2.2R_\text{s}$ with $R_\text{s}=2.15\pm0.14~\text{kpc}$. The \texttt{DEHH-Pot} (and therefore our $N$-body model) has $\left( v_\text{circ,disk}/v_\text{circ,total}\right)^2\approx47\%$ at $2.2R_\text{s}$ and is therefore slightly sub-maximal (see Figure \ref{fig:DEHH_vcirc_decomposed}). The decomposed rotation curve in Figure \ref{fig:DEHH_vcirc_decomposed} is qualitatively similar to the MW models by \citet{2011MNRAS.414.2446M} (their Figure 5) and by \citet{2016A&A...593A.108B} (their Figure 5, left panels, model MI). Their disks dominate between $R\approx2.5-8.5~\text{kpc}$ and $R\approx5-9.5~\text{kpc}$, respectively, while our simulation's disk dominates between $R\approx2-7~\text{kpc}$. We account for this slight difference by also drawing mock data sets from regions around $R_0=5~\text{kpc}$ (see Section \ref{sec:survey_volume_data}).

\paragraph{Number of spiral arms.} The exact number of spiral arms in the MW is still under debate. The distribution of star-forming regions traced, e.g., by HII regions \citep{1976A&A....49...57G}, maser sources \citep{2009ApJ...700..137R,2014ApJ...783..130R}, and young massive stars \citep{2014MNRAS.437.1791U}, suggest that the MW has four major spiral arms (see also \citet{2008AJ....135.1301V,2014AJ....148....5V} and references therein). Observations in the infrared, e.g., of old red-clump giant stars in the Spitzer/GLIMPSE survey \citep{2009PASP..121..213C} or of stellar NIR-emission by the COBE satellite \citep{2001ApJ...556..181D} indicate that the MW is a grand-design two-armed spiral. One hypothesis is that the four-armed spiral observed in young stars and gas is due to the response of the gas to the two-armed spiral in old stars \citep{2000A&A...358L..13D,2004MNRAS.350L..47M}. Our simulation is overall a four-armed spiral galaxy, with the mode $m=4$ dominating between $R=4-7~\text{kpc}$, $m=2$ at smaller and $m\geq6$ at larger radii (see Figure \ref{fig:surf_dens_fourier_b}).

\paragraph{Strength of spiral arms.} As demonstrated in Figure \ref{fig:surf_dens_fourier_a} the spiral arms introduce strong peak-to-peak differences in the stellar surface density (e.g., $\sim200\%$ at $R=5~\text{kpc}$). The disk dominates inside $R=8~\text{kpc}$ (see Figure \ref{fig:DEHH_vcirc_decomposed}), and in Section \ref{sec:4kpc8Spiral_potential}, Figure \ref{fig:4kpc8Spiral_forces}, we will see that the spiral arms introduce relative perturbations in the total gravitational forces of up to $30\%$. Figure \ref{fig:DF_velres}, which we will discuss in Section \ref{sec:4kpc8Spiral_DF}, suggests an excess of stars with radial velocities up to $50~\text{km s}^{-1}$ in our simulated galaxy as compared to an axisymmetric model. This can be compared to \citet{2014ApJ...783..130R}, who measured, for example, that typical peculiar non-circular motions in the MW spiral arms were around $10-20~\text{km s}^{-1}$ for $R \gtrsim 4~\text{kpc}$. \citet{2012MNRAS.425.2335S} and \citet{2015ApJ...800...83B} found velocity fluctuations of the same order in the solar neighborhood. We therefore expect the strength of the perturbation to the total potential due to the spiral arms in this simulation---especially inside $R=8~\text{kpc}$ where most of our mock data is drawn from---to be similar or even stronger than those in the MW.

\section{RoadMapping modeling} \label{sec:RoadMapping}
In this section we summarize the mathematical ingredients of \RM{}, and motivate the DF and potential model that we are going to fit to the data. \RM{} makes extensive use of the \texttt{galpy} python library by \citet{2015ApJS..216...29B}\footnote{The \texttt{galpy} python package by \citet{2015ApJS..216...29B} can be downloaded from \url{http://github.com/jobovy/galpy}.}. For full details on the \RM{} machinery see also Paper I.

\pagebreak

\subsection{Likelihood} \label{sec:likelihood}
As already laid out in Section \ref{sec:survey_volume_data}, we use as data the 6D $(\vect{x}_i,\vect{v}_i)$ coordinates of $N_*$ stars within a spherical survey volume. The corresponding, purely spatial selection function $\text{SF}(\vect{x})$ is
\begin{equation}
\text{SF}(\vect{x}) \equiv \begin{cases} 1 &\mbox{if } \left| \vect{x}-\vect{x}_0 \right| \leq r_\text{max} \\
0 & \mbox{otherwise} \end{cases}, \label{eq:selection_function}
\end{equation}
with $\vect{x}_0 = (R_0,\phi_0,z_0=0)$ from Table \ref{tbl:volume_positions}.

Given a parametrized axisymmetric potential model $\Phi(R,z)$ with parameters $p_\Phi$, the probability that the $i$-th star is on an orbit with the actions 
\begin{equation}
\vect{J}_i \equiv \vect{J}[\vect{x}_i,\vect{v}_i \mid p_\Phi],
\end{equation}
is proportional to the given orbit distribution function $\text{DF}(\vect{J})$ with parameters $p_\text{DF}$,
\begin{equation}
\text{DF}(\vect{J}_i \mid p_\text{DF}) \equiv \text{DF}(\vect{x}_i,\vect{v}_i \mid p_\Phi,p_\text{DF}).
\end{equation}

The joint likelihood of a star being within the survey volume and on a given orbit is therefore
\begin{eqnarray}
\mathscr{L}_i &\equiv& \mathscr{L}(\vect{x}_i,\vect{v}_i \mid p_\Phi,p_\text{DF}) \nonumber\\
&=& \frac{\text{DF}(\vect{x}_i,\vect{v}_i\mid p_\phi, p_\text{DF}) \cdot \text{SF}(\vect{x_i})}{\int \text{DF}(\vect{x},\vect{v}\mid p_\Phi, p_\text{DF}) \cdot \text{SF}(\vect{x}) \ \Diff3 x \Diff3 v}.
\end{eqnarray}
The details of how we numerically evaluate the likelihood normalization to a sufficiently high precision are discussed in Paper I.\footnote{In the terminology of Paper I we use the high numerical accuracy $N_x = 20$, $N_v = 28$, $n_\sigma = 5.5$ to calculate the likelihood normalization, or in other words, to evaluate the spatial and velocity integrals over the qDF within the survey volume.}

In the scenario considered in this paper it can happen that there are a few ($\sim 1$ in 20,000) stars entering the catalog that are for some reason on rather extreme orbits, e.g., moving radially directly towards the center. These kinds of orbits do not belong to the set of orbits that we classically expect to make up an overall smooth galactic disk. To avoid that such single stars with very low likelihood have a strong impact on the modeling we employ here a simple strategy to ensure a robust likelihood,
\begin{equation}
\mathscr{L}_i \longrightarrow \max \left( \mathscr{L}_i, \epsilon \times \text{median}(\mathscr{L})\right),
\end{equation}
where $\epsilon = 0.001$ for $N_*=20,000$ stars and $\text{median}(\mathscr{L})$ is the median of all the $N_*$ stellar likelihoods $\mathscr{L}_i$ with the given $p_\Phi$ and $p_\text{DF}$. This robust likelihood method was not used in Paper I. For future applications to real MW data an outlier model similar to the one in \citet{2013ApJ...779..115B} could be added: a stellar outlier distribution with constant spatial number density and velocities following a broad Gaussian.

Following Paper I, we assume for now uninformative flat priors on the model parameters $p_\Phi$ and $p_\text{DF}$ and find the maximum and width of the posterior probability function $pdf(p_\Phi,p_\text{DF} \mid \text{data}) \propto \prod_{i=1}^{N_*} \mathscr{L}_i \times prior(p_\Phi,p_\text{DF})$ using a nested-grid approach and then explore the full shape of the $\pdf$ using a Monte Carlo Markov Chain (MCMC)\footnote{We use the MCMC software \texttt{\texttt{emcee}} by \citet{2013PASP..125..306F}.}. Full details on this procedure are given in Paper I.

\begin{deluxetable*}{lll}[!htbp]
\tabletypesize{\scriptsize}
\tablecaption{Best fit \texttt{MNHH-Pot} and qDF parameters as recovered from the \RM{} analysis of a survey volume with $r_\text{max}=4~\text{kpc}$ centered on a spiral arm at $R_0=8~\text{kpc}$ (position \texttt{S8}). \label{tbl:MNHHdiffSph2_4kpc8Spiral}}
\tablewidth{0pt}
\startdata
\tableline
circular velocity at $R_\text{\sun}=8~\text{kpc}$& $v_\text{circ}(R_\text{\sun})$ & $(223.0 \pm 0.1) ~\text{km s}^{-1}$\\
Miyamoto-Nagai disk scale length & $a_\text{disk}$ & $(3.62^{+ 0.06}_{- 0.05}) ~\text{kpc}$\\
Miyamoto-Nagai disk scale height & $b_\text{disk}$ & $(0.26 \pm 0.02) ~\text{kpc}$\\
halo fraction at $R_\text{\sun}=8~\text{kpc}$ & $f_\text{halo}$ & $(0.53 \pm 0.02)$\\
halo scale length & $a_\text{halo}$ & $(21 \pm 2) ~\text{kpc}$\\
bulge mass & $M_\text{bulge}$ & $0.95 \times 10^{10}~\text{M}_\odot$ (fixed)\\
bulge scale length & $a_\text{bulge}$ & $0.25~\text{kpc}$ (fixed)\\
\tableline
qDF tracer scale length & $h_R$ & $(3.34^{+ 0.05}_{- 0.04} ) ~\text{kpc}$\\
qDF radial velocity dispersion & $\sigma_{R,0}$ & $(15.91 \pm 0.08) ~\text{km s}^{-1}$\\
qDF vertical velocity dispersion & $\sigma_{z,0}$ & $(14.0^{+ 0.2}_{- 0.1}) ~\text{km s}^{-1}$\\
qDF radial velocity dispersion scale length & $h_{\sigma,R}$ & $(4.6 \pm 0.5)~\text{kpc}$\\
qDF vertical velocity dispersion scale length & $h_{\sigma,z}$ & $(5.65^{+ 0.06}_{- 0.07}) ~\text{kpc}$
\enddata
\tablecomments{The bulge mass and scale length were fixed in the analysis to their true values, see Sections \ref{sec:simulation_description} and \ref{sec:potential_model}.}
\end{deluxetable*}

\subsection{Distribution function model} \label{sec:DF_model}

The most simple action-based orbit DF is the quasi-isothermal DF (qDF) introduced by \citet{2010MNRAS.401.2318B} and \citet{2011MNRAS.413.1889B}, which has been a successful ingredient in Paper I and many disk modeling approaches \citep{2013ApJ...779..115B,2014MNRAS.445.3133P,2015MNRAS.449.3479S}. The exact functional form of the qDF$(J_R,L_z,J_z \mid p_\text{DF})$ is given, for example, in \citet{2011MNRAS.413.1889B}, or in Equations (2)-(4) of Paper I.

The qDF is expressed in terms of actions, frequencies, and scaling profiles for the radial stellar tracer density $n(R_g)$, and velocity dispersion profiles $\sigma_z(R_g)$ and $\sigma_R(R_g)$. The latter are functions of the guiding-center radius $R_g$, i.e., the radius of a circular orbit with given angular momentum $L_z$ in a given potential. We set the scaling profiles to
\begin{eqnarray}
n(R_g \mid p_\text{DF}) &\propto& \exp\left(-\frac{R_g}{h_R} \right)\\
\sigma_R(R_g \mid p_\text{DF}) &=& \sigma_{R,0} \times \exp\left(- \frac{R_g-R_\text{\sun}}{h_{\sigma,R}} \right)\label{eq:sigmaRRg}\\
\sigma_z(R_g \mid p_\text{DF}) &=& \sigma_{z,0} \times \exp\left(- \frac{R_g-R_\text{\sun}}{h_{\sigma,z}} \right)\label{eq:sigmazRg}.
\end{eqnarray}
The free model parameters of the qDF are
\begin{equation}
p_\text{DF} \equiv \left\{ \ln h_R, \ln \sigma_{R,0}, \ln \sigma_{z,0}, \ln h_{\sigma,R}, \ln h_{\sigma,z}\right\}. \label{eq:qDF_parameters}
\end{equation}

In an axisymmetric potential superimposed with non-axisymmetric perturbations the actions are not exact integrals of motion. But if one simply considers action-angles as phase-space coordinates, a DF that is a function of the actions only might still be a good model for $\text{DF}(\vect{x},\vect{v},t)= \text{DF}(\vect{J},\vect{\theta},t)$ at a given point in time and if the system is well-mixed in phase.

We motivate the use of the qDF as specific action-based DF for the simulation snapshot in this work as follows: There is no stellar abundance or age information in the simulation. We therefore cannot define stellar sub-populations, as we normally would for the MW (see Paper I and \citealt{2013ApJ...779..115B}). However, the disk of the galaxy simulation was originally set up as a single axisymmetric flattened particle population whose density decreases exponentially with radius (see Section \ref{sec:simulation_description}). This is actually very similar to the stellar distribution generated by a single qDF (see, e.g., \citealt{2013MNRAS.434..652T}). As all particles in the disk have evolved for the same $\sim 250~\text{Myr}$ since its axisymmetric set-up, we can consider the disk essentially as a mono-age population. All of this motivates us therefore to use one single qDF to model the whole disk.

Locally, the current particle distribution in the snapshot at hand might be dominated by non-axisymmetries which evolved later in the simulation. We have no indication if for small survey volumes the qDF is still a good model for the data. We will use it anyway---to see how far we can get with the simplest model possible and to test if actions are still informative in this case.

\subsection{Potential model} \label{sec:potential_model}

In all \RM{} analyses in this work we will fit an axisymmetric gravitational potential model consisting of a (fixed and known) Hernquist bulge, a free Hernquist halo and a free Miyamoto-Nagai disk \citep{1975PASJ...27..533M},
\begin{equation}
\Phi_\text{disk}(R,z) = - \frac{GM}{\sqrt{R^2+(a_\text{disk}+\sqrt{z^2+b_\text{disk}^2})^2}}, \label{eq:MN-disk}
\end{equation}
where $a_\text{disk}$ and $b_\text{disk}$ are the equivalents of a disk scale length and scale height. Using Hernquist profiles for halo and bulge is motivated by our knowledge of the snapshot galaxy, and we fix the bulge's total mass and scale length to the true values (see Section \ref{sec:simulation_description} and Table \ref{tbl:MNHHdiffSph2_4kpc8Spiral}). As the bulge contribution to the total radial force at $R_\text{\sun}\equiv 8~\text{kpc}$ is only $\sim9-10\%$, this will not give the modeling an unfair advantage. The free model parameters of the halo are the halo scale length $a_\text{halo}$ and the halo fraction, i.e., the relative halo-to-disk contribution to the radial force at $R_\text{\sun}$, defined as
\begin{equation}
f_\text{halo} \equiv \left. \frac{F_{R,\text{halo}}}{F_{R,\text{disk}} + F_{R,\text{halo}}} \right|_{\stackrel{R=R_\text{\sun}}{z=0}}.\label{eq:fhalo}
\end{equation}
As a parameter that scales the total mass of the galaxy model we use the circular velocity at the ``solar'' radius $R_\text{\sun}$,
\begin{equation}
v_\text{circ}(R_\text{\sun}=8~\text{kpc}) \equiv \left. \sqrt{ R \frac{\partial \Phi}{\partial R} }\right|_{\stackrel{R=R_\text{\sun}}{z=0}} . \label{eq:circvel}
\end{equation}
The total set of free potential model parameters is therefore
\begin{equation}
p_\Phi \equiv \left\{ v_\text{circ}(R_\text{\sun}),a_\text{disk},b_\text{disk},a_\text{halo},f_\text{halo}\right\}.
\end{equation} 
We will call this potential model the \texttt{MNHH-Pot} (Miyamoto-Nagai disk + Hernquist halo + Hernquist bulge) in the remainder of this work.

To estimate the stellar actions $\vect{J}=(J_R,L_z,J_z)$ in the axisymmetric \texttt{MNHH-Pot}, we use the \emph{St\"{a}ckel fudge} algorithm by \citet{2012MNRAS.426.1324B} with fixed focal length $\Delta=0.45$, and interpolate the actions on a grid \citep{2012MNRAS.426.1324B,2015ApJS..216...29B}. We made sure that the accuracy of the parameter estimates are not degraded by interpolation errors.\footnote{For the action interpolation grid following \citet{2015ApJS..216...29B}, we use $R_\text{max}=40~\text{kpc}$, $n_E=70$, $n_\psi=40$, $n_{L_z}=50$ in their terminology.}

\begin{figure*}[!htbp]
  \centering
  \subfigure[DF density residuals.\label{fig:DF_densres}]{\includegraphics[width=0.32\linewidth]{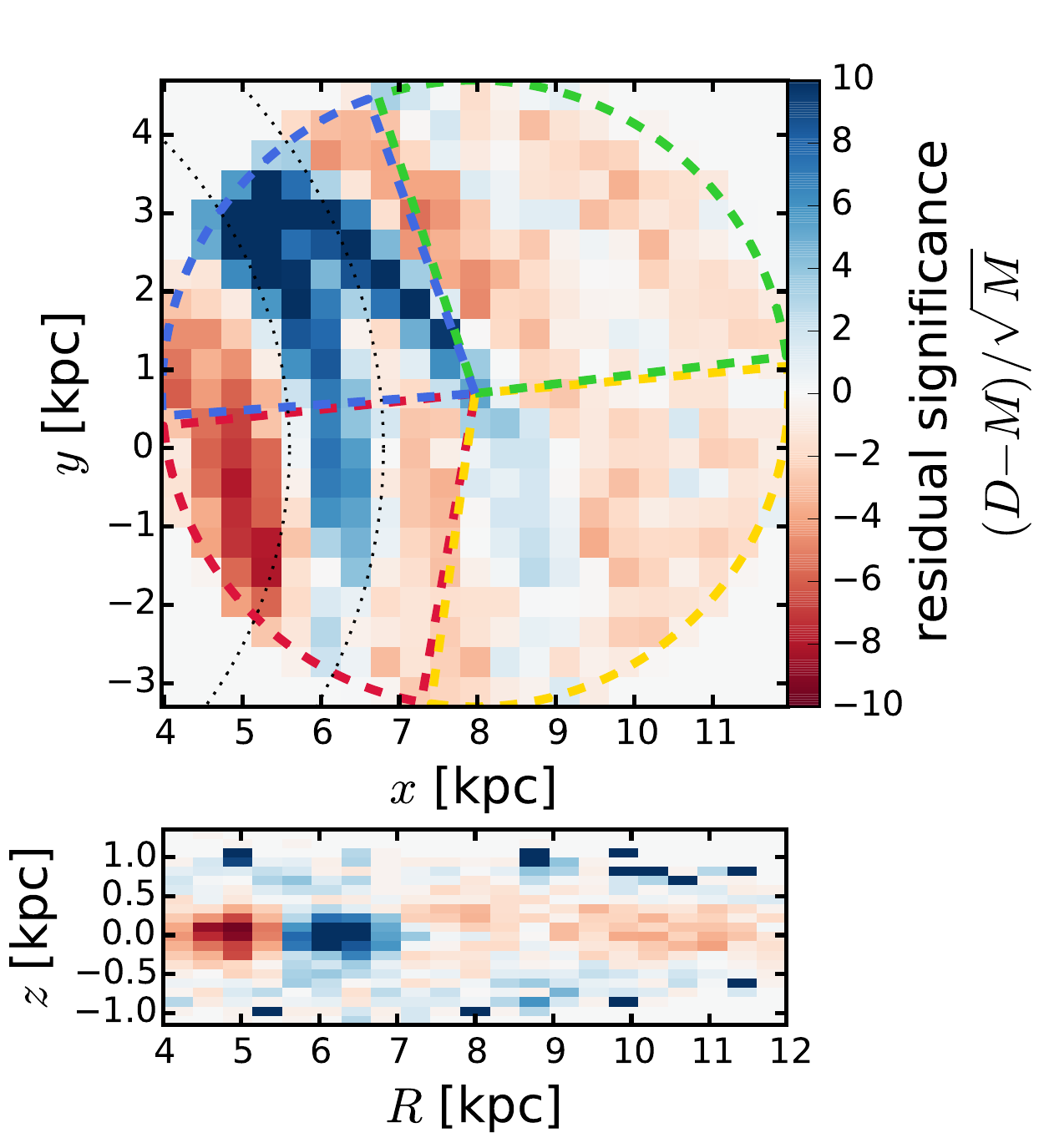}}
  \subfigure[DF density profiles.\label{fig:DF_densprof}]{\includegraphics[width=0.32\linewidth]{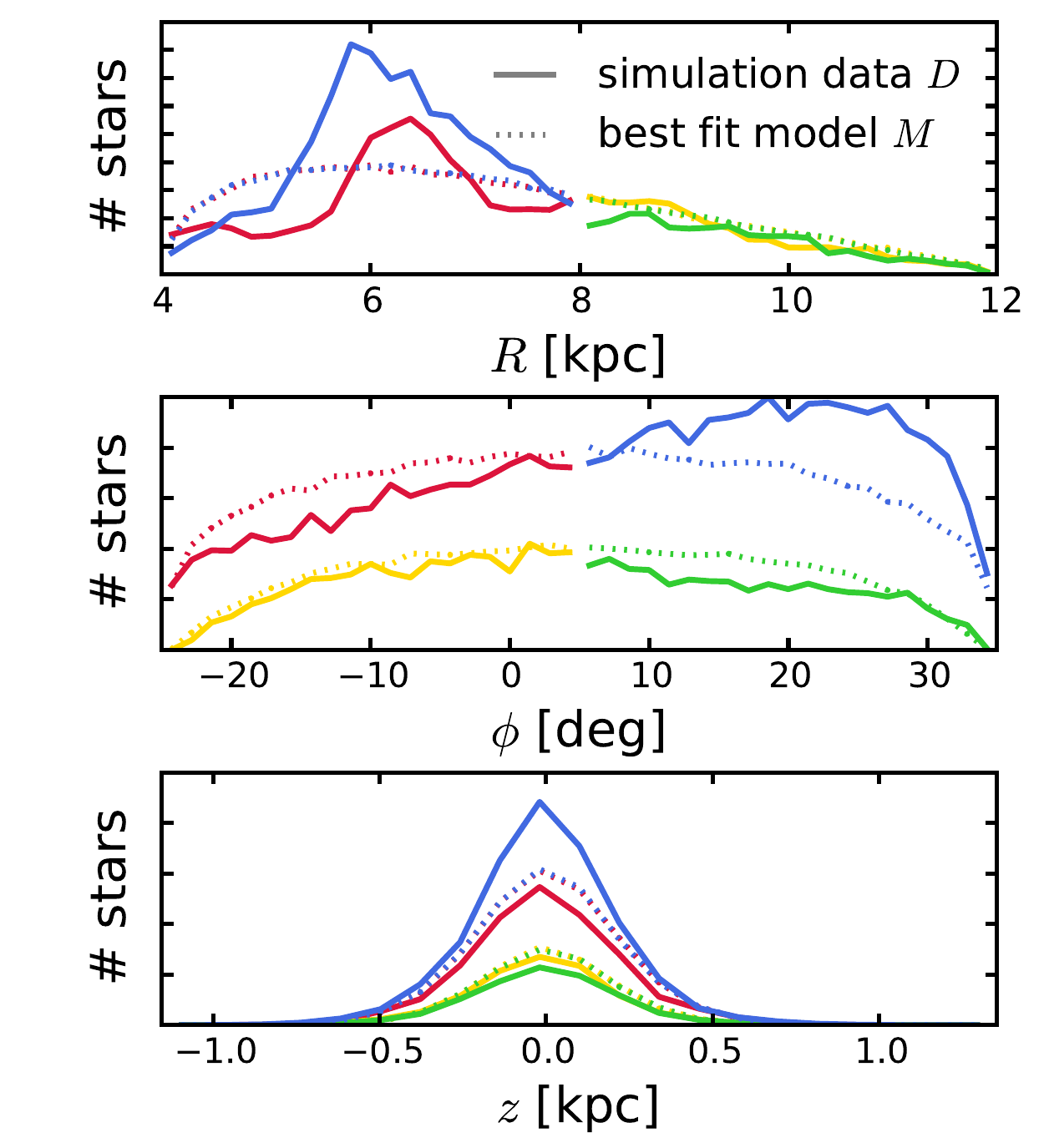}}
  \subfigure[DF velocity residuals.\label{fig:DF_velres}]{\includegraphics[width=0.32\linewidth]{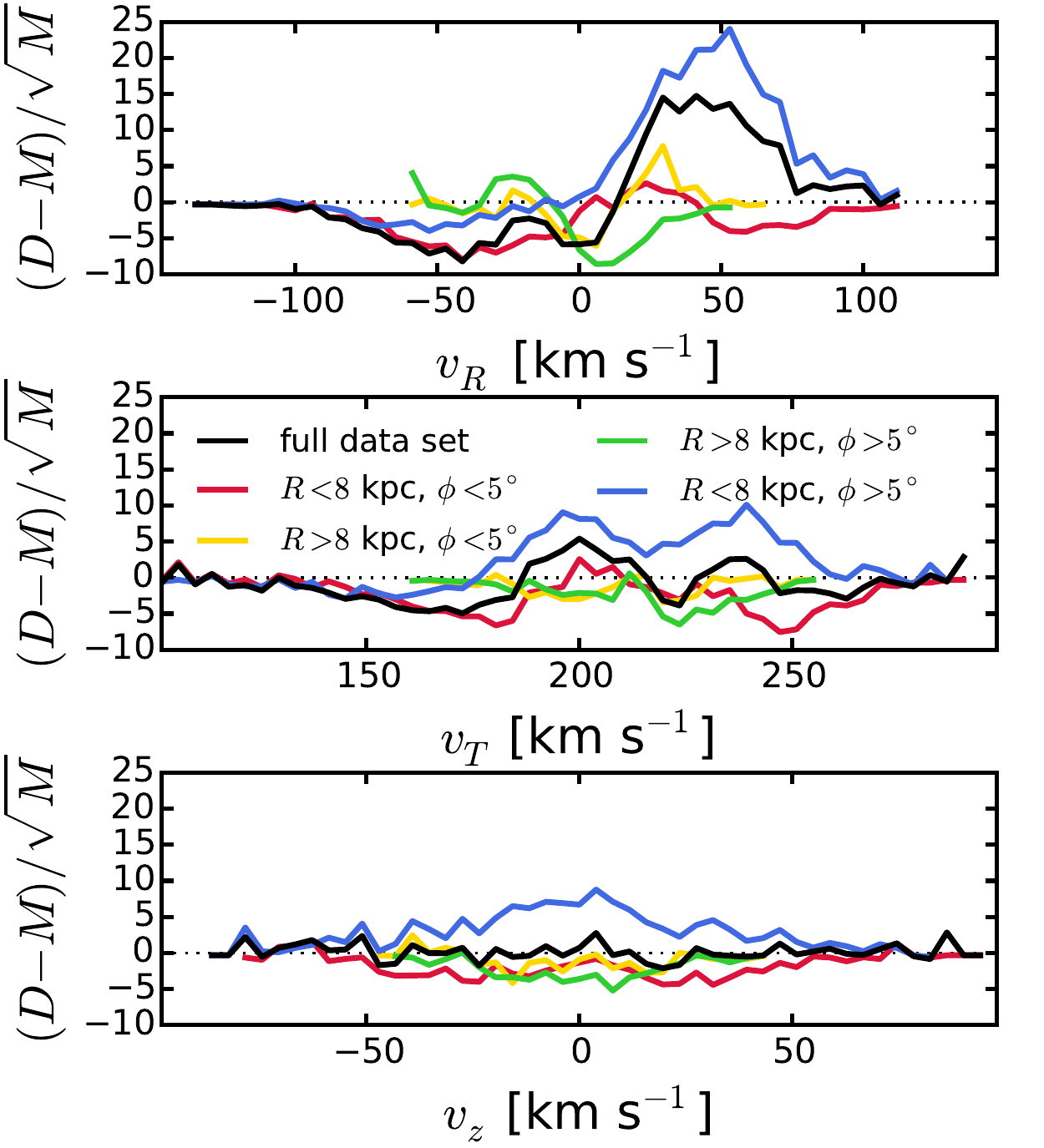}}
  \caption{Comparison of the true and best fit stellar DF$(\vect{x},\vect{v})$ in position-velocity space. The true DF$(\vect{x},\vect{v})$ (i.e., the data $D$) is the distribution of all the stars in the data set drawn from the simulation snapshot (with $N_*=20,000$ and $r_\text{max}=4~\text{kpc}$ centered on position \texttt{S8}). The best fit DF$(\vect{x},\vect{v})$ (i.e., the model $M$) is generated by MC sampling of the best fit qDF$(\vect{J})$ in the best fit potential model from \RM{} in Table \ref{tbl:MNHHdiffSph2_4kpc8Spiral}, given the known selection function. Panel \ref{fig:DF_densres} shows the spatial density residual significance $(D-M)/\sqrt{M}$ of the projection to the $(x,y)$ and $(R,z)$ plane (the MC sampled $M$ is the expected number of stars per bin, and $\sqrt{M}$ the expected error due to Poisson statistics). In the $(x,y)$ panel the following regions are marked: $R<8~\text{kpc},\phi>5^\circ$ (blue), $R<8~\text{kpc},\phi<5^\circ$ (red), $R>8~\text{kpc},\phi>5^\circ$ (green), $R>8~\text{kpc},\phi<5^\circ$ (yellow). Panels \ref{fig:DF_densprof} and \ref{fig:DF_velres} show the density profiles and velocity residual significance along each of the 6D phase-space coordinates separately for each of the four spatial regions. The blue region is very much dominated by the non-axisymmetric spiral arm. For the yellow region the axisymmetric single-qDF model is a good description. Overall the qDF is a good average axisymmetric model for the data. (In panel \ref{fig:DF_densres} we overplot the radii $R_\text{spiral} \in [5.6,6.8]~\text{kpc}$ as black dotted lines to mark the approximate extent of the stronger spiral arm, to compare it with Figure \ref{fig:4kpc8Spiral_actions}.)}
  \label{fig:4kpc8Spiral_DF_comparison}
\end{figure*}

Galaxy disks in general, as well as the simulated disk in this work, have exponential radial density profiles. A single Miyamoto-Nagai disk is more massive at large radii than an exponential disk (see e.g., \citealt{2015MNRAS.448.2934S}). By construction, the \texttt{DEHH-Pot} introduced in Section \ref{sec:DEHH-Pot} is therefore better suited to reproduce the overall density distribution in the simulation than the \texttt{MNHH-Pot} (as we will see in Figure \ref{fig:4kpc8Spiral_dens_vcirc_surfdens} in the next section). However, the closed form expression of the Miyamoto-Nagai potential in Equation \eqref{eq:MN-disk} has the crucial advantage of allowing much faster force and therefore action calculations. In addition, by using a potential model where we already know that it is not be the optimal model for the galaxy's disk, we challenge \RM{} even further.

\section{Results} \label{sec:results}

At the core of this work is a suite of 22 data sets consisting of the phase-space coordinates of stellar tracer particles, drawn from the spiral galaxy simulation snapshot introduced in Section \ref{sec:simulation_description}. Each data set comes from a different survey volume within the galaxy's disk (see Section \ref{sec:survey_volume_data}). We modeled all data sets with \RM{} as described in Section \ref{sec:RoadMapping}, by fitting to it a single qDF (see Section \ref{sec:DF_model}) and the potential model \texttt{MNHH-Pot} (introduced in Section \ref{sec:potential_model}). This resulted in 22 independent measurements of the simulated galaxy's potential and DF.

We present our results in two steps. In Section \ref{sec:results_part1} we look at one of these \RM{} models in detail. In Section \ref{sec:results_part2} we then compare all 22 \RM{} results, and discuss their differences in the context of spiral arms.

\subsection{An axisymmetric galaxy model from \RM{}} \label{sec:results_part1}

In this section we will discuss all aspects of a \RM{} model for one single data set. This data set has $N_*=20,000$ stars that were drawn from the spherical volume with $r_\text{max}=4~\text{kpc}$ centered on a spiral arm at the ``solar'' radius $R_0=8~\text{kpc}$. This volume is shown in orange in Figure \ref{fig:simulation} (position \texttt{S8}). We chose this volume because of its position centered on a smaller spiral arm, similar to our Sun being located in the Orion spiral arm. It is a bit larger than our conservative guess for the survey volume size for which we currently expect unbiased potential estimates from the final Gaia data release (DR), $r_\text{max}=3~\text{kpc}$ from the Sun (see Paper I and Section \ref{sec:discussion_choosing_SV}). But improvements in \RM{} with respect to the treatment of measurement errors might ultimately allow to also model larger volumes. This example survey volume ranging from $R=4-12~\text{kpc}$ also has the advantage of being crossed by several spiral arms of different spiral strength (see Figures \ref{fig:simulation_xy} and \ref{fig:surf_dens_fourier}). In this section our goal is now to investigate the ability of the best fit \RM{} model to serve as an overall axisymmetric model for the galaxy.

The parameters of the best fit \texttt{MNHH-Pot} and qDF recovered with \RM{} from this data set are summarized in Table \ref{tbl:MNHHdiffSph2_4kpc8Spiral}. The circular velocity $v_\text{circ}(R_\text{\sun})$ and halo fraction $f_\text{halo}$ are especially well-recovered (compare to Table \ref{tbl:DEHH-Pot}).

\subsubsection{Recovering the stellar distribution} \label{sec:4kpc8Spiral_DF}

The \RM{} fit itself takes place in action space. However, an important sanity check to decide if the fit was successful, is to test if the best fit \RM{} model (i.e., best fit action-based DF in best fit potential and in given selection function) generates a stellar distribution that reproduces the distribution of data points in observable phase-space, $(\vect{x},\vect{v})$. This comparison is shown in Figure \ref{fig:4kpc8Spiral_DF_comparison}. 

We note that the spiral arms introduce very strong non-axisymmetries in the data, both in the spatial and the velocity distribution (especially in $v_R$, where a significantly larger number of stars move outward than inward as compared to an axisymmetric model). We therefore compare the data and fit separately for different spatial regions, $R>8~\text{kpc}$ and $R<8~\text{kpc}$, and $\phi<5^\circ$ and $\phi>5^\circ$. In the region where the spiral arm dominates (blue in Figure \ref{fig:4kpc8Spiral_DF_comparison}) the best fit \RM{} model is actually a very poor model. However, what the model underestimates in the spiral arm, it slightly overestimates in the other regions and is therefore indeed something like a good average model for the overall distribution. The region at $R>8~\text{kpc}$ and $\phi<5^\circ$ (yellow in Figure \ref{fig:4kpc8Spiral_DF_comparison}), where neither the spiral arm nor the inter-arm regions dominate strongly, is especially well-described by the model. 

Overall the qDF appears to be a good model for unperturbed regions of the disk and averages over spiral arms.

\begin{figure*}[!htbp]
\hspace{0.04\textwidth}
\begin{minipage}{0.96\textwidth}
\centering
\subfigure[Equidensity contours, $\rho_\Phi(R,z)$. \label{fig:4kpc8Spiral_density_a}]{\includegraphics[height=5cm]{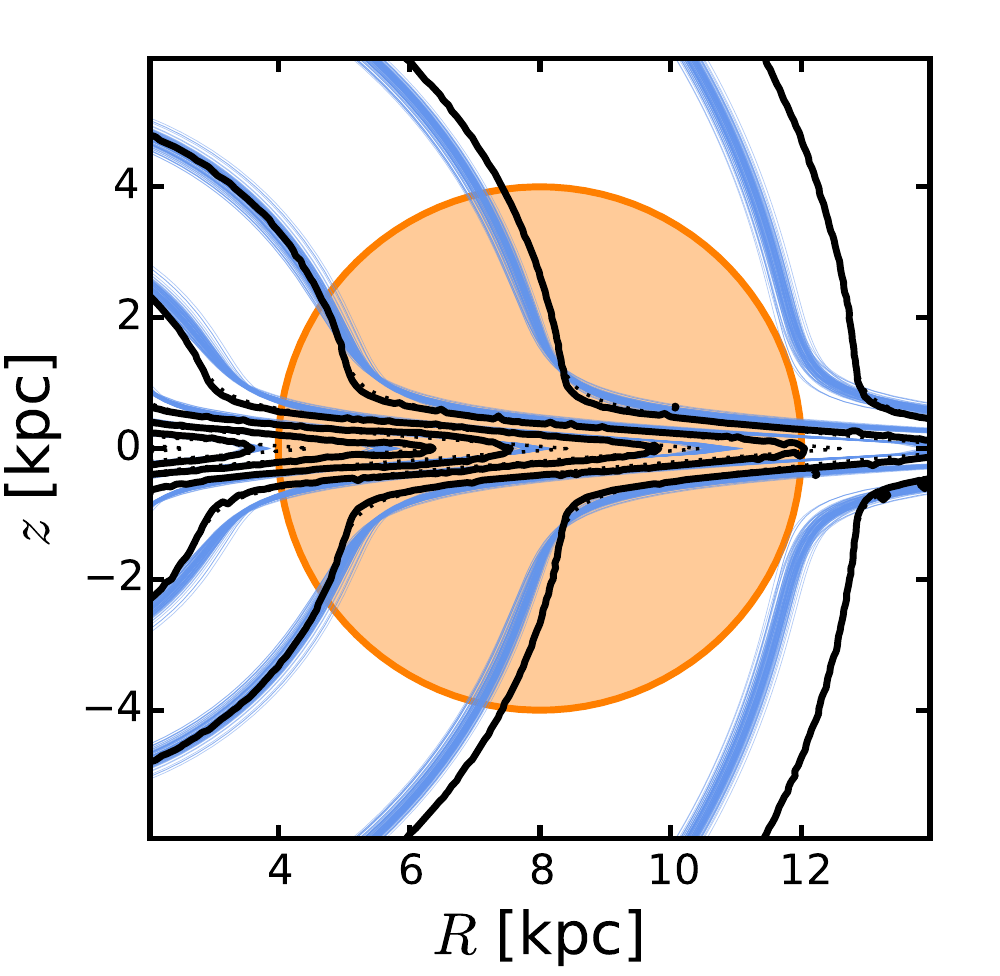}}\hspace{0.01\textwidth}
   \subfigure[Density residuals. \label{fig:4kpc8Spiral_density_b}]{\includegraphics[height=5cm]{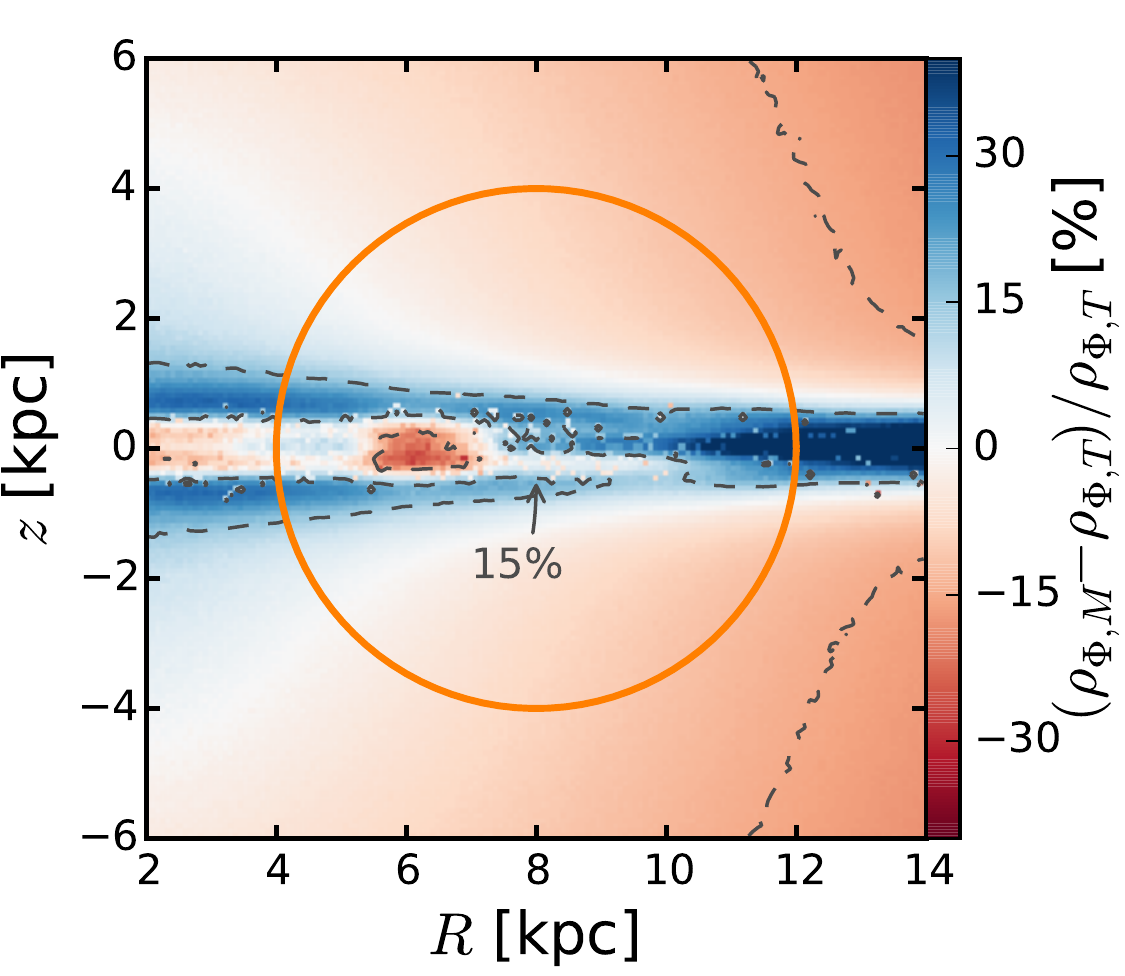}}
 \end{minipage}\\
 \begin{minipage}{\textwidth}
 \centering
  \subfigure[Circular velocity curve. \label{fig:4kpc8Spiral_vcirc_surfdens_a}]{\includegraphics[height=5.5cm]{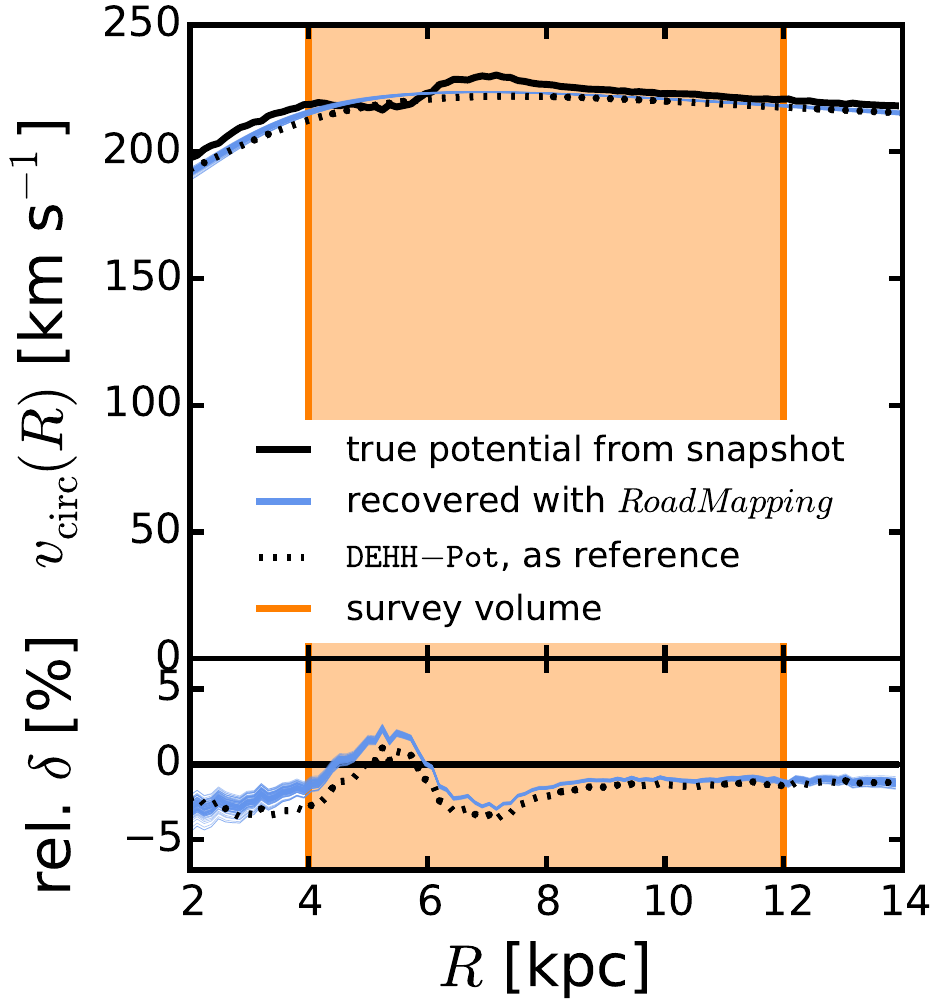}}
    \subfigure[Surface density profile. \label{fig:4kpc8Spiral_vcirc_surfdens_b}]{\includegraphics[height=5.5cm]{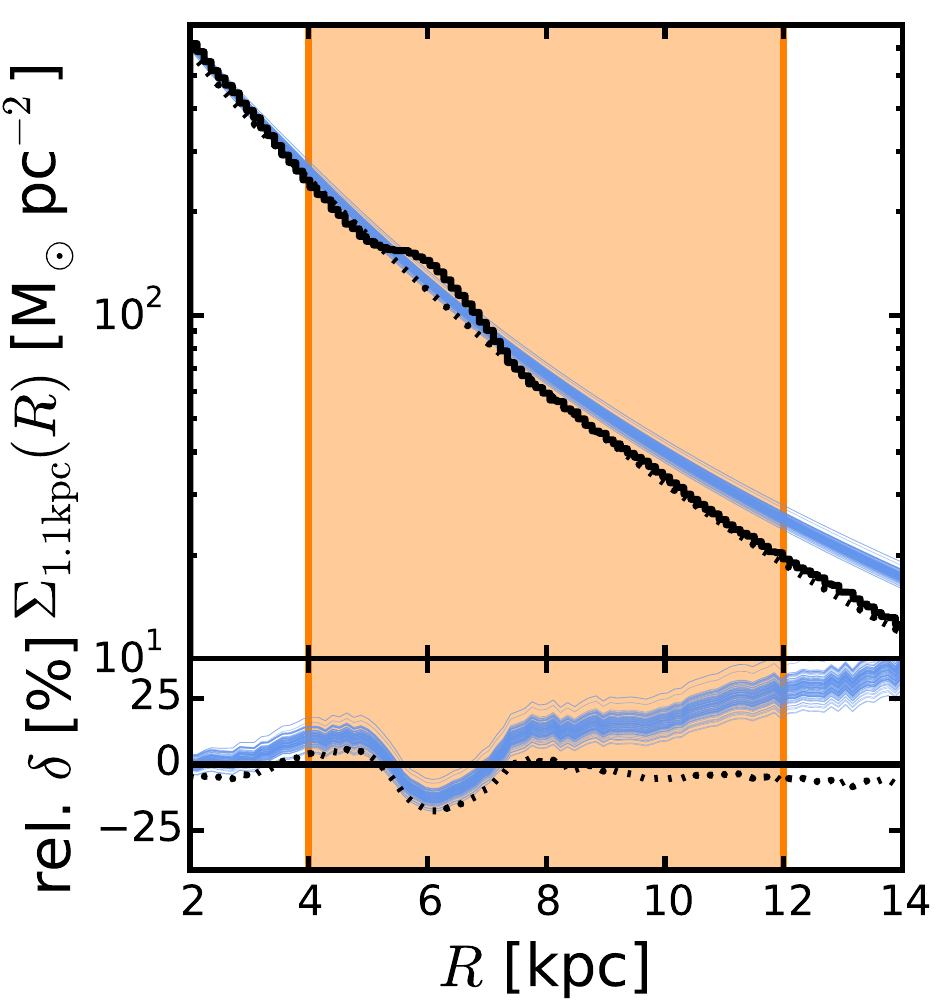}}
    \end{minipage}
    \caption{Comparison of the true and recovered gravitational potential $\Phi$. Panel \ref{fig:4kpc8Spiral_density_a} compares in the $(R,z)$ plane equidensity contours of the overall matter density distribution $\rho_{\Phi}$, generating the potential. Panel \ref{fig:4kpc8Spiral_vcirc_surfdens_a} and \ref{fig:4kpc8Spiral_vcirc_surfdens_b} show the potential's circular velocity curve and radial surface density profile within $|z|=1.1~\text{kpc}$, respectively. We compare the true $\rho_{\Phi}$, $v_\text{circ}$, and $\Sigma_{\rm 1.1kpc}$, of the galaxy simulation (azimuthally averaged over the whole galaxy; black solid lines), with 100 \texttt{MNHH-Pot} potentials drawn from the pdf of the best fit \RM{} model (blue lines). This model was derived from $N_*=20,000$ stars in the spherical survey volume at \texttt{S8} with $r_\text{max}=4~\text{kpc}$. (The extent of the survey volume is marked in orange.) The best fit parameters are given in Table \ref{tbl:MNHHdiffSph2_4kpc8Spiral}, and Panel \ref{fig:4kpc8Spiral_density_b} shows the residuals between the matter density corresponding to the median values in this table, $\rho_{\Phi,\texttt{M}}$, and the true density $\rho_{\Phi,\texttt{T}}$. Overplotted in Panels \ref{fig:4kpc8Spiral_density_a}, \ref{fig:4kpc8Spiral_vcirc_surfdens_a}, and \ref{fig:4kpc8Spiral_vcirc_surfdens_b} is also the reference \texttt{DEHH-Pot} (see Section \ref{sec:DEHH-Pot}; black dotted line). Over wide areas even outside of the survey volume the relative difference between true and recovered density is less than $15\%$. At $R\gtrsim8~\text{kpc}$ and $z\sim0$ it becomes apparent that the chosen potential model cannot perfectly capture the structure of the disk. However, in the plane of the disk and at smaller radii within the survey volume, where most of the stars are located, the model gives good constraints on the density. The circular velocity curve is recovered to less than $5\%$.}
\label{fig:4kpc8Spiral_dens_vcirc_surfdens}
\end{figure*}

\subsubsection{Recovering the gravitational potential} \label{sec:4kpc8Spiral_potential}

As shown in the previous section the best fit \RM{} model seems to reproduce the average stellar phase-space distribution quite well. But is the corresponding potential close to the true potential? 

\begin{figure*}[!htbp]
\centering
  \subfigure[Recovery of the radial forces. \label{fig:4kpc8Spiral_forces_a}]{\includegraphics[width=6cm]{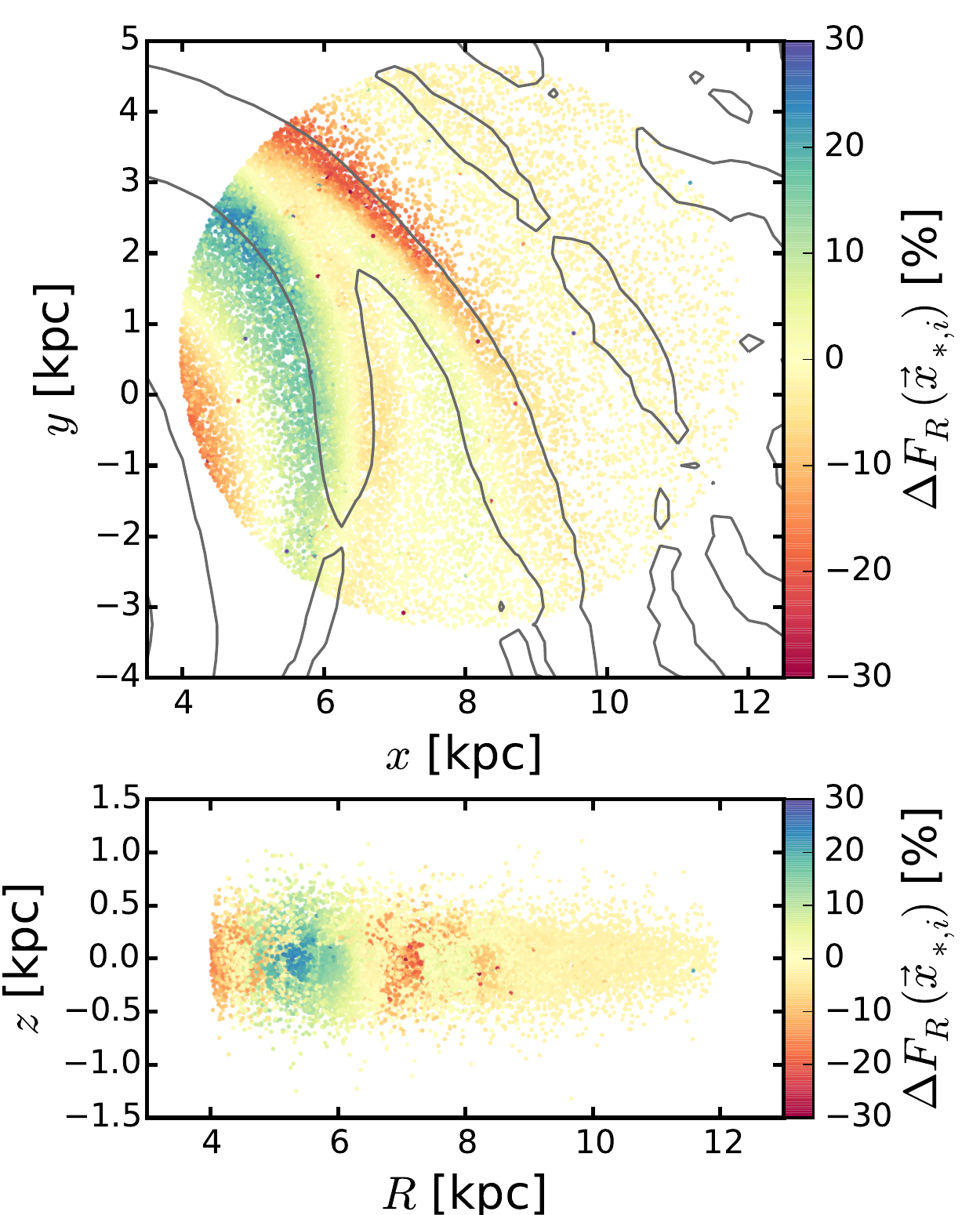}}
  \subfigure[Recovery of the vertical forces. \label{fig:4kpc8Spiral_forces_b}]{\includegraphics[width=6cm]{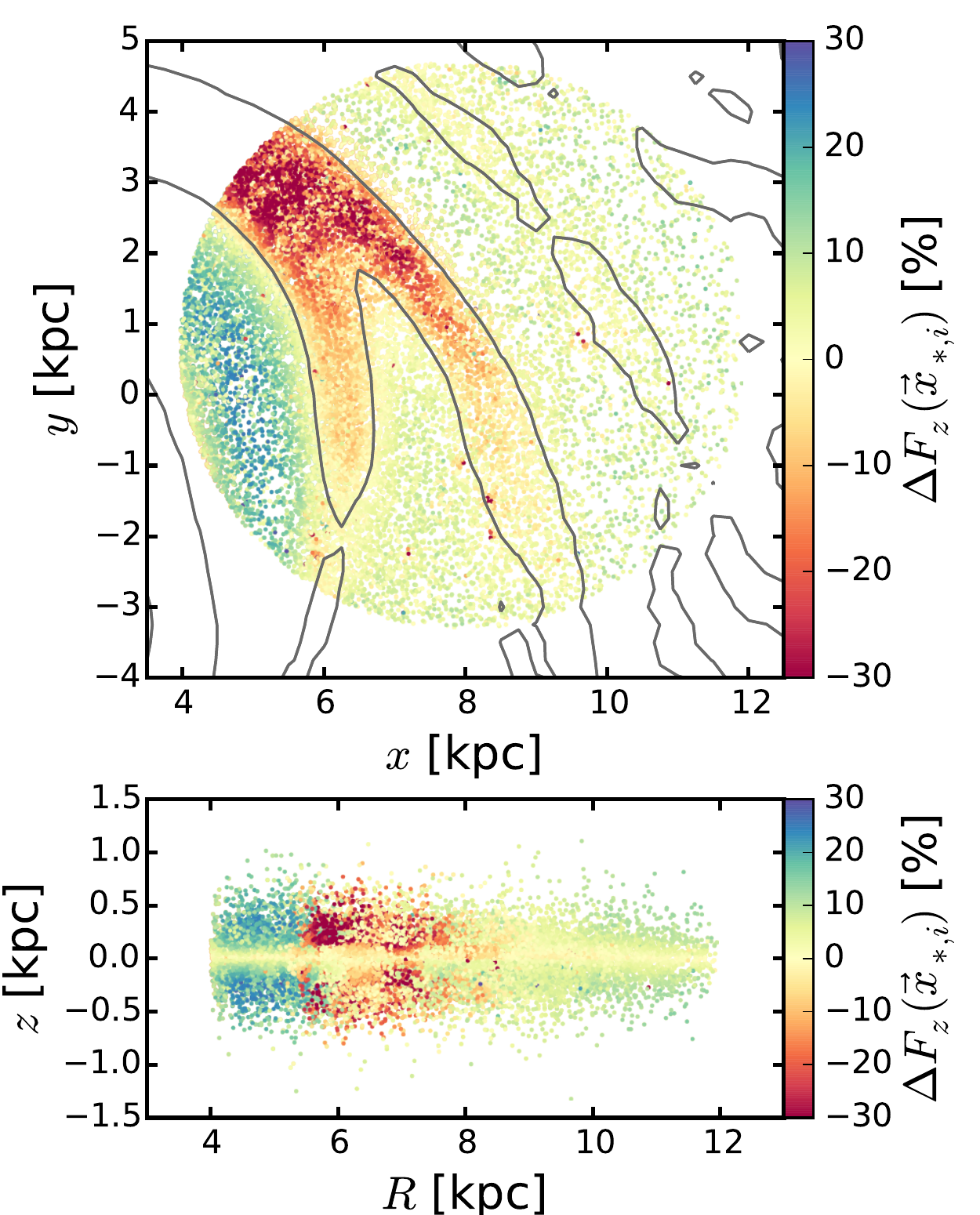}}
\caption{Recovery of the gravitational forces with \RM{}. We compare the true gravitational forces with the forces estimated from the \RM{} best fit potential in Table \ref{tbl:MNHHdiffSph2_4kpc8Spiral} at the $(x,y)$ positions (upper panels) and $(R,z)$ positions (lower panels) of the stars that entered the analysis. In particular, we color-code the positions of the stars according to the radial (panel \ref{fig:4kpc8Spiral_forces_a}) and vertical (panel \ref{fig:4kpc8Spiral_forces_b}) force residuals scaled by a typical force, i.e., we show $\Delta F_R(\vect{x}_{*,i})$ and $\Delta F_z(\vect{x}_{*,i})$ in Equations \eqref{eq:delta_FR}-\eqref{eq:delta_Fz}. The overplotted grey contours correspond to $\Sigma_{\text{1.5kpc,disk},T}/\Sigma_{\text{1.5kpc,disk},S}=1.15$, i.e., the true vs.\ the symmetric disk surface density, and mark the position of the spiral arms. The red dots mark stars for which the best fit model underestimates the (absolute value of the) force. This is the case for the radial force in the leading sides of the spiral arms and the vertical force within the spiral arms, which cannot be reproduced. Blue marks correspond to stars for which the force is overestimated. Overall the radial forces are very well recovered, which is related to the good recovery of the circular velocity curve in Figure \ref{fig:4kpc8Spiral_vcirc_surfdens_a}. There are more problems with the vertical force, which is related to the higher surface densities in spiral arms which slightly biases the overall \RM{} model.}
\label{fig:4kpc8Spiral_forces}
\end{figure*}

Figure \ref{fig:4kpc8Spiral_dens_vcirc_surfdens} compares the true potential from the simulation snapshot (symmetrized by averaging over the whole $\Delta\phi=2\pi$) and the axisymmetric reference \texttt{DEHH-Pot} from Table \ref{tbl:DEHH-Pot} with the best fit \texttt{MNHH-Pot} from the \RM{} analysis. In particular, Figure \ref{fig:4kpc8Spiral_dens_vcirc_surfdens} illustrates the overall matter density distribution, the rotation curve and the surface density profile. Figure \ref{fig:4kpc8Spiral_forces} compares the true and recovered (median) gravitational forces at the position of each star in the data set.

The recovery of density, surface density and circular velocity curve is especially good in the region where most of the stars are located, around $R\sim6~\text{kpc}$ and in the plane of the disk. In large regions inside the survey volume, and even outside, the density is recovered to within 15\%. The circular velocity curve is recovered to within 5\%, which is also approximately the extent of perturbation that the spiral arms cause with respect to a smooth rotation curve. There is however a very small ($<1.5\%$) underestimation of $v_\text{circ}$ at larger radii. We suspect that this bias is introduced by the spiral arms (see discussion in Sections \ref{sec:biases_explained} and \ref{sec:MNdHHinit}). The overall surface density profile is a bit overestimated ($\sim 15\%$) at smaller radii; this is clearly due to the local spiral arm at $R\sim6~\text{kpc}$ with its higher surface density and many stars entering the analysis, which bias the result and which was to be expected. At larger radii ($R\sim10-12~\text{kpc}$) the fit of the local density in the disk and surface density profile starts to flare due to the choice of the Miyamoto-Nagai disk family with its shallow profile (see Section \ref{sec:MNdHHinit}). But again, where most of the stars are located, our \RM{} model is a very good average model for the true galaxy.

The aspect of the potential to which the stellar orbits are actually sensitive is the gravitational forces. In Figure \ref{fig:4kpc8Spiral_forces} we therefore compare the true force that each star in the data set feels (i.e., the radial force, $F_{R,T}(\vect{x}_{*,i})$, and vertical force, $F_{z,T}(\vect{x}_{*,i})$, calculated as the sum of the individual contributions by each particle in the simulation and the analytic DM halo at the position of each star $\vect{x}_{*,i}\equiv (x_i,y_i,z_i)$) with the force that the \RM{} median model predicts for each star ($F_{R,M}(\vect{x}_{*,i})$ and $F_{z,M}(\vect{x}_{*,i})$; $M$ for ``median model''). We scale the difference between truth and model by a typical radial or vertical force at the given radius for which we use
\begin{eqnarray}
F_{R,\text{typ}}(R) &\equiv& v^2_{\text{circ},S}(R) / R\label{eq:FRtyp}\\
F_{z,\text{typ}}(R) &\equiv& F_{z,S}(R,z=z_\text{s}),
\end{eqnarray}
where $v_{\text{circ},S}$ and $F_{z,S}$ are the circular velocity and vertical force evaluated in the ``true symmetric'' reference potential \texttt{DEHH-Pot} in Table \ref{tbl:DEHH-Pot}. As the typical vertical force at a given radius we use $F_{z,S}$ evaluated at the scale height $z_\text{s}=0.17~\text{kpc}$ of the disk. This is of the same order as the true vertical force averaged over all stars at this radius. Figure \ref{fig:4kpc8Spiral_forces} shows therefore
\begin{eqnarray}
\Delta F_R(\vect{x}_{*,i}) \equiv \frac{|F_{R,M}(R_i,z_i)| - |F_{R,T}(x_i,y_i,z_i)|}{|F_{R,\text{typ}}(R_i)|}\label{eq:delta_FR}\\
\Delta F_z(\vect{x}_{*,i}) \equiv \frac{|F_{z,M}(R_i,z_i)| - |F_{z,T}(x_i,y_i,z_i)|}{|F_{z,\text{typ}}(R_i)|}\label{eq:delta_Fz}
\end{eqnarray}
for each star ($R_i = \sqrt{x_i^2+y_i^2}$). The recovery is as expected: The true vertical force is stronger in the spiral arms (due to the higher surface density) and weaker in the inter-arm regions as compared to the axisymmetric best fit model. The radial force is well recovered where the majority of the stars are located, i.e., in the wide inter-arm regions and in the peaks of the spiral arms. Misjudgments happen in the wings of the spiral arms: The true radial force (i.e., the pull towards the galactic center) is stronger at the outer edge/leading side of the spiral arm because of the additional gravitational pull towards the massive spiral arm, and for the same reason weaker at the inner edge/trailing side. Overall the recovered \RM{} model appears to be a good mean model, averaging over spiral arms and inter-arm regions.

\begin{figure*}[!htbp]
\centering
\includegraphics[width=0.7\textwidth]{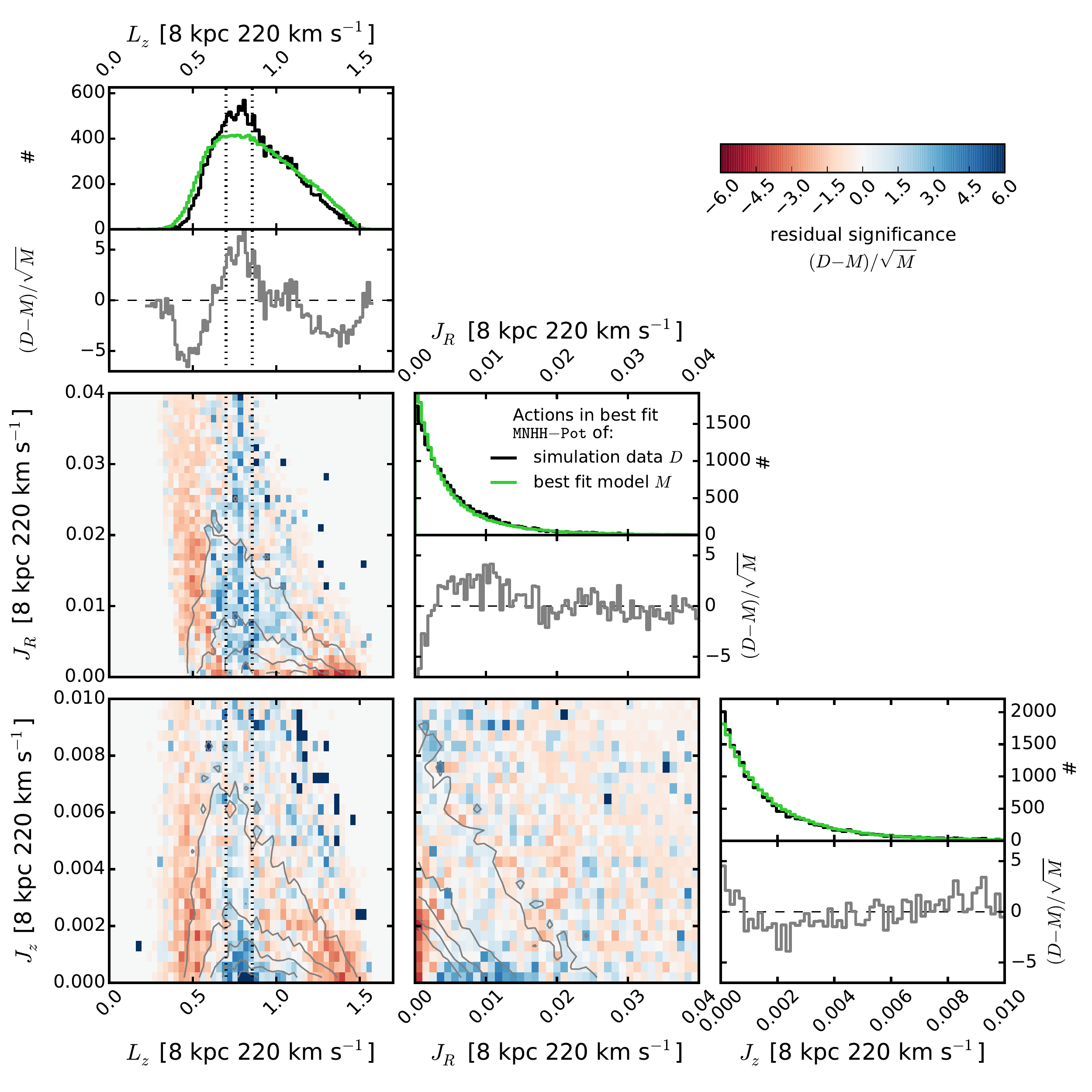}
\caption{Comparison of the stellar action distribution of the data set $D$ used in the analysis and the recovered axisymmetric action distribution $M$ (see Figure \ref{fig:4kpc8Spiral_DF_comparison} for the comparison in configuration space). All actions of the data set and best fit distribution were calculated in the best fit \texttt{MNHH-Pot} in Table \ref{tbl:MNHHdiffSph2_4kpc8Spiral}. The upper panel in each column shows one-dimensional histograms of the $D$ and $M$ distribution of angular momentum, $L_z$, the radial action, $J_R$, and the vertical action, $J_z$. The other panels display the residual significance $(D-M)/\sqrt{M}$, as both one-dimensional and two-dimensional distribution (the model $M$ was constructed by MC sampling the best fit qDF, $\sqrt{M}$ is the expected noise due to Poisson statistics, and $(D-M)/\sqrt{M}$ therefore describes how significant any difference between $D$ and $M$ is). The two-dimensional residuals are overplotted with equidensity contours of the data $D$'s two-dimensional action distribution (grey solid lines). In Figure \ref{fig:DF_densres} we have marked the approximate radial extent of the stronger spiral arm with black dotted lines ($R_\text{spiral} \in [5.6,6.8]~\text{kpc}$); the dotted lines in the $L_z$ distributions in this figure correspond to $L_z = R_\text{spiral} \times v_\text{circ}(R_\text{spiral})$. This comparison gives a first impression of how the approximate action distribution in spiral arms might look.}
\label{fig:4kpc8Spiral_actions}
\end{figure*}

\subsubsection{Recovering the action distribution} \label{sec:4kpc8Spiral_actions}

In Sections \ref{sec:4kpc8Spiral_DF} and \ref{sec:4kpc8Spiral_potential} we have demonstrated the goodness of the fit in the configuration space of the data, and of the recovered gravitational potential. What \RM{} is actually fitting, however, is the distribution in action space. Figure \ref{fig:4kpc8Spiral_actions} compares the data and the model action distribution (generated by the best fit qDF) given the best fit median \texttt{MNHH-Pot} in Table \ref{tbl:MNHHdiffSph2_4kpc8Spiral}. (We use this axisymmetric potential to calculate the actions which lead to the best fit model, and do not attempt to estimate the true actions in the true potential.)

We note that the radial and vertical action distribution fits quite well; the axisymmetric model however contains many more stars on close-to-circular orbits $(J_R \sim 0,J_z \sim 0)$ than the simulation. In the data set there is an excess of stars in the galactic plane $(J_z\sim0)$ that have more eccentric orbits than the axisymmetric model would predict. In Figure \ref{fig:DF_densres} we have marked the radial extent of the stronger spiral arm with dotted lines ($R_\text{spiral} \in [5.6,6.8]~\text{kpc}$), and overplotted the corresponding angular momenta $L_z = R_\text{spiral} \times v_\text{circ}(R_\text{spiral})$ in Figure \ref{fig:4kpc8Spiral_actions}. This serves as a rough estimate for the region in action space where we expect the stars of this spiral arm to be located. It is again obvious that this spiral arm contains (i) more stars in general and (ii) more stars with eccentric orbits $(J_R>0)$ which are (iii) mostly located close to the plane $(J_z\sim0)$, as compared to the axisymmetric model. All of this confirms our expectations for orbits in a spiral arm. 

One of the open tasks that the Galactic dynamical modeling community faces is the description of the orbit distribution of spiral arms. The above exercise of comparing the data and the model actions in a best-fit axisymmetric potential should therefore be performed for any future application to data in the Milky Way as well. It could help to learn more about the approximate orbits that stars move on in real spiral arms, and how spiral arms perturb axisymmetric action DFs.

\subsubsection{Calibrating the method by modeling a snapshot without spiral arms} \label{sec:MNdHHinit}

To better understand the effect of spiral arms on the \RM{} modeling in the previous sections, we also performed a calibration test run. For that we applied \RM{} to a mock data set drawn from the same spherical survey volume as in Section \ref{sec:results_part1} (centered at \texttt{S8} with maximum radius $r_\text{max}=4~\text{kpc}$), but from the initial axisymmetric snapshot of the galaxy simulation ($t=0~\text{Myr}$), in which no spiral arms had evolved yet.

The spatial tracer distribution was very well recovered by \RM{}, as well as the distribution of tracers in $v_R$ and $v_z$. There were some deviations in the $v_T$ distribution between data and best fit model, however, with our model predicting more asymmetric drift. We attribute this to the simple initial setup of the disk's stellar velocities, which does not follow a physical distribution function but only assumes a triaxial Gaussian velocity distribution \citep{2005MNRAS.361..776S}. We expect the modeling in this calibration run to be slightly biased by this. Other than that, this confirms our expectation that the qDF is indeed a good model for the unperturbed disk, and that the deviations between data and model in Figure \ref{fig:4kpc8Spiral_DF_comparison} were purely due to the spiral arms.

The surface density profile generated by the best fit potential was a perfect fit inside of $R\sim7~\text{kpc}$, but it started flaring further out. This was expected, because the Miyamoto-Nagai disk is known to be more massive at larger radii than an exponential disk \citep{2015MNRAS.448.2934S}. Also, if the fit is driven by the majority of stars, the fit is expected to be better at smaller radii with its higher tracer number density. The same flaring in the surface density also showed up in Figure \ref{fig:4kpc8Spiral_dens_vcirc_surfdens}. Overall our chosen potential model will therefore systematically bias vertical forces at large radii to be too strong as compared to the truth.

The recovered potential parameters for the initial snapshot, $a_\text{disk}=(3.73 \pm 0.04) ~\text{kpc}$, $b_\text{disk}=(0.34 \pm 0.03) ~\text{kpc}$, $f_\text{halo}=(0.52 \pm 0.03)$, and $a_\text{halo}=(24 \pm 2) ~\text{kpc}$, are consistent with the spiral arm-affected measurements in Table \ref{tbl:MNHHdiffSph2_4kpc8Spiral} to within 2, 4, 1, and 1.5 times the statistical error, respectively. While the halo scale length in this calibration run is also consistent with the truth within 3 times the error, we expect it to be underestimated to at least partly account for the Miyamoto-Nagai disk being too massive at large radii. As we will see later (in Section \ref{sec:parameter recovery} and Figure \ref{fig:model_parameters}) we seem to need an even larger survey volume to have enough radial coverage to constrain the halo scale length properly.

\begin{figure*}[!htbp]
\centering
	\subfigure[$r_\text{max}=2~\text{kpc}$. \label{fig:2kpcSuite}]{\includegraphics[width=0.6\textwidth]{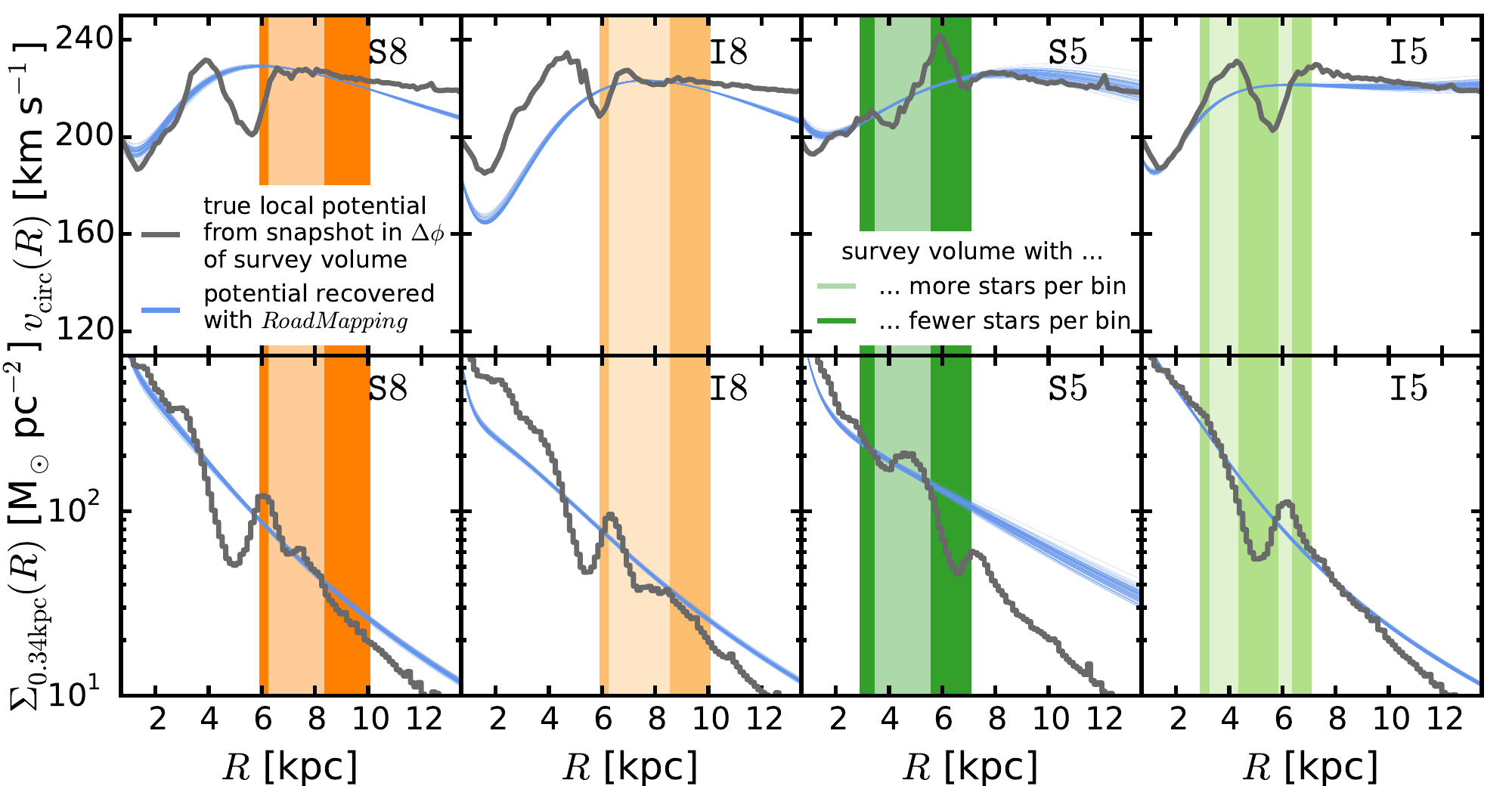}} %
    \subfigure[$r_\text{max}=3~\text{kpc}$. \label{fig:3kpcSuite}]{\includegraphics[width=0.6\textwidth]{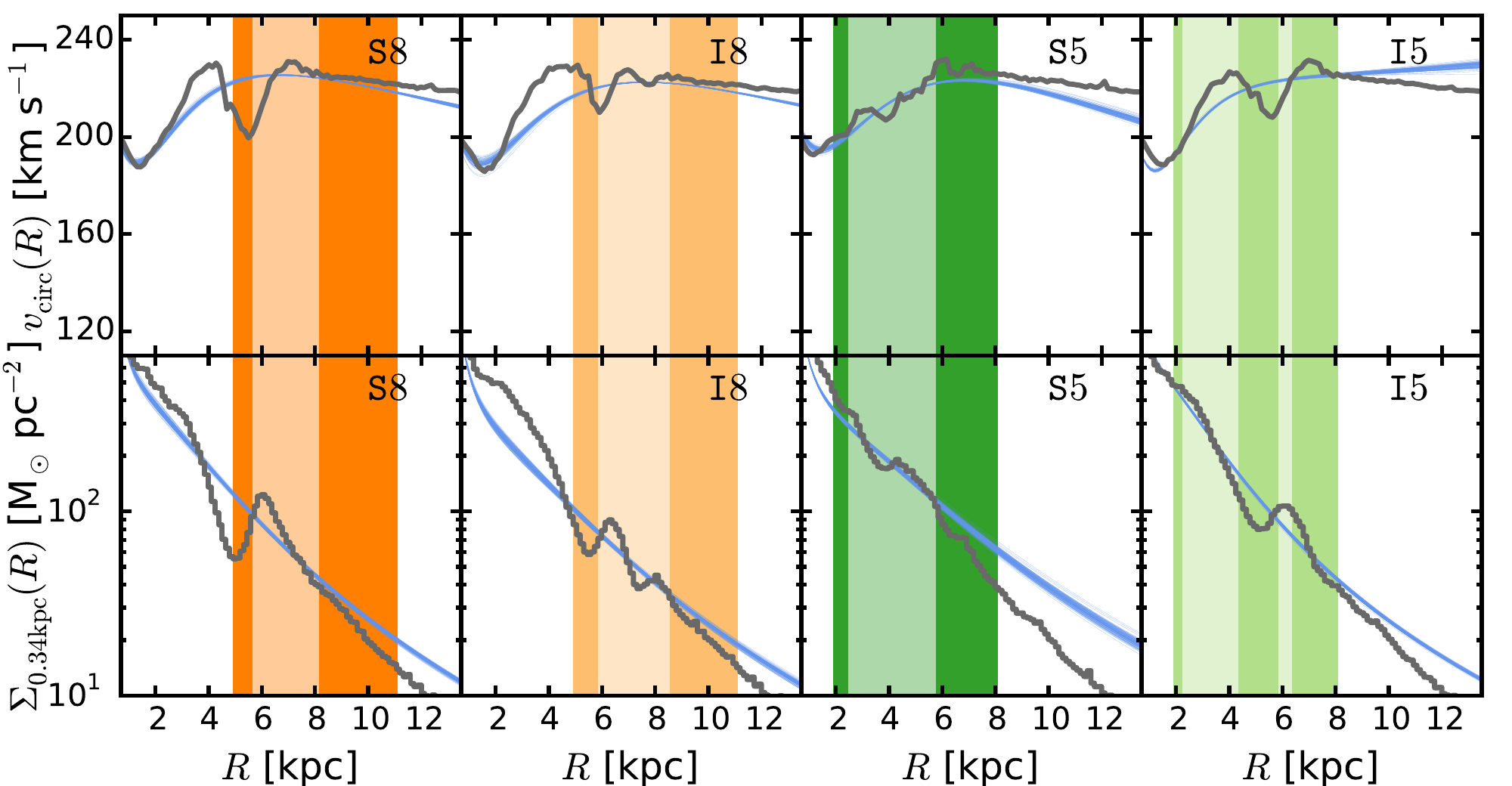}}
\caption{Comparison of the true local circular velocity curve and surface density within $|z| \leq 2 z_\text{s} = 0.34~\text{kpc}$ with the recovered \RM{} models from survey volumes of size $r_\text{max}=2~\text{kpc}$ and $r_\text{max}=3~\text{kpc}$. The blue lines show the \RM{} potential models recovered from these data sets (each line is one of 100 potentials drawn from the full $pdf$ sampled with the MCMC). The grey curves show the true profiles as derived from the galaxy simulation snapshot, averaged over the angular wedge $\phi_0\pm\arcsin(r_\text{max}/R_0)$ that encloses the corresponding survey volume (see Table \ref{tbl:volume_positions} for all $R_0$ and $\phi_0$ values). In other words, we show the true profiles only for the region of the spiral galaxy that was actually probed by the data. The orange and green colored regions mark the radial extent of the survey volumes. We sorted the stars of each data set into radial bins of size $\Delta R = 200~\text{kpc}$. The radial bins with a higher than average number of stars are marked with a lighter shade of the corresponding color, and the bins with a lower than average number with a darker shade. It turns out that the constraints of highest accuracy and precision are always where most of the stars are located---within the survey volume and in particular at the peak of the distribution.}
    \label{fig:vcirc_surfdens_suite_medium}
\end{figure*}

\begin{figure*}[!htbp]
\centering
	\subfigure[$r_\text{max}=500~\text{pc}$. \label{fig:500pcSuite}]{\includegraphics[width=0.6\textwidth]{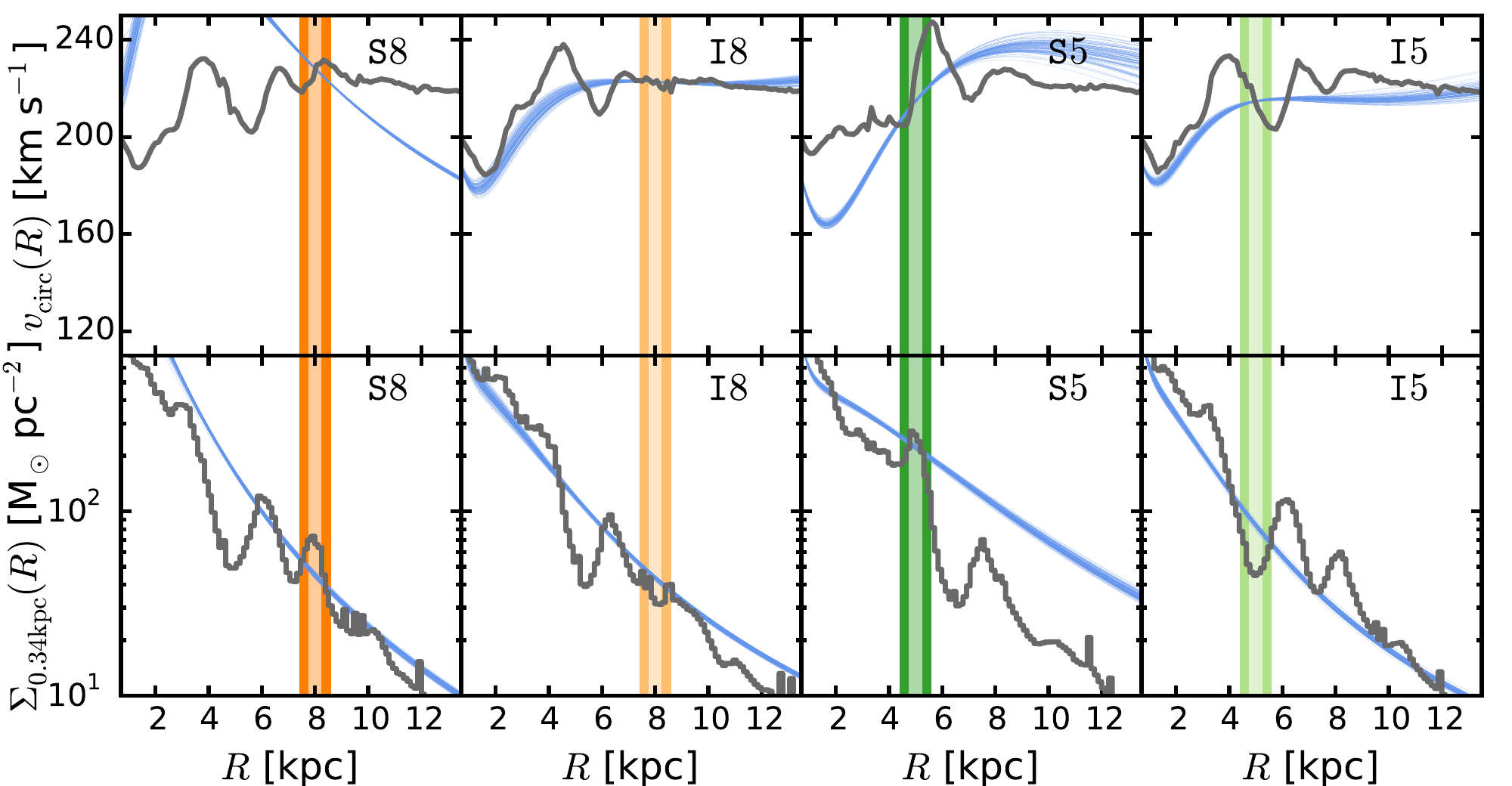}} %
	\subfigure[$r_\text{max}=1~\text{kpc}$. \label{fig:1kpcSuite}]{\includegraphics[width=0.6\textwidth]{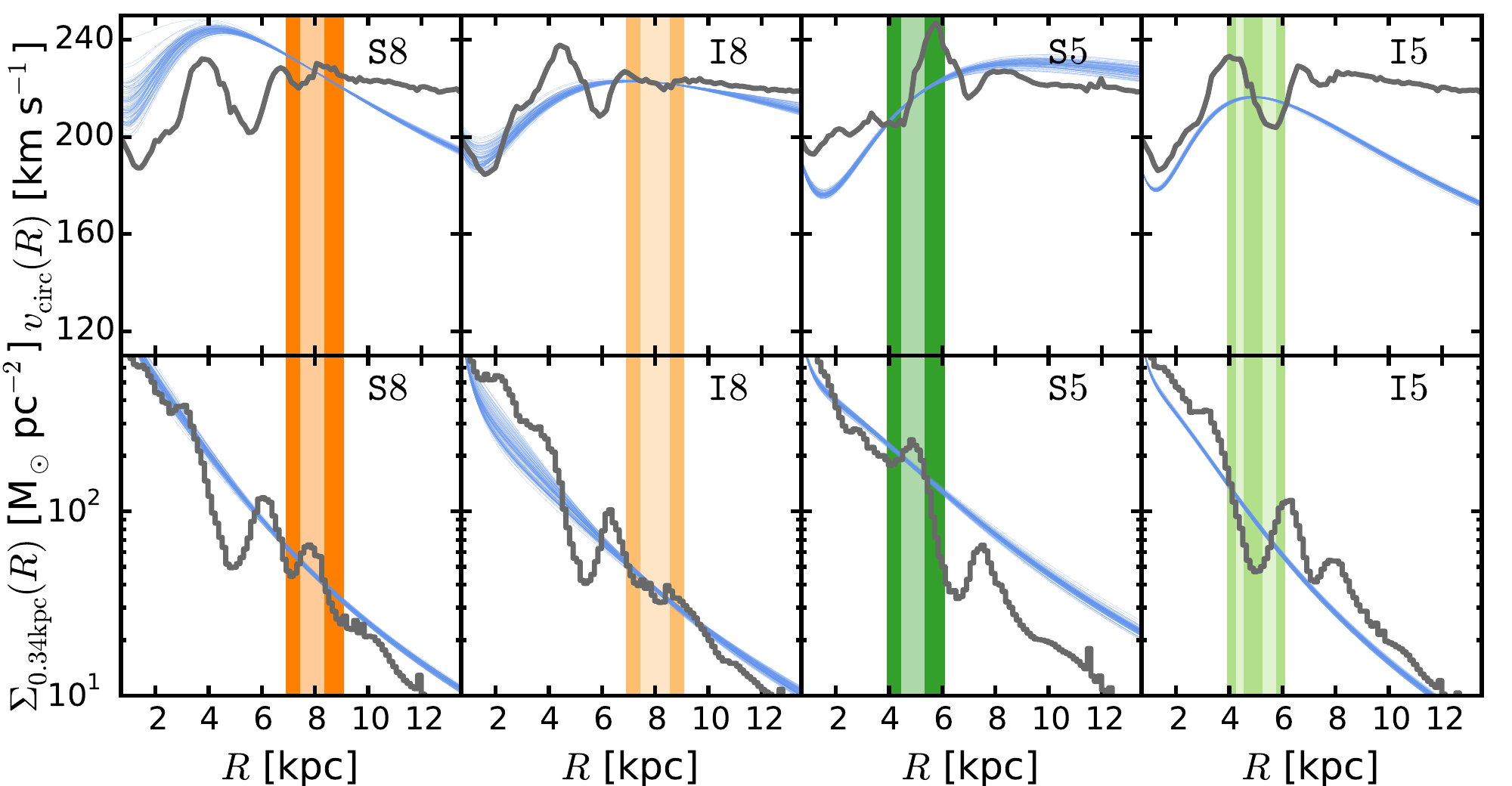}} 
	\caption{Same as Figure \ref{fig:vcirc_surfdens_suite_medium}, but for all small survey volumes with $r_\text{max}=500~\text{pc}$ and $r_\text{max}=1~\text{kpc}$. }
\label{fig:vcirc_surfdens_suite_small}
\end{figure*}

\begin{figure*}[!htbp]
\centering
	\subfigure[$r_\text{max}=4~\text{kpc}$. \label{fig:4kpcSuite}]{\includegraphics[width=0.6\textwidth]{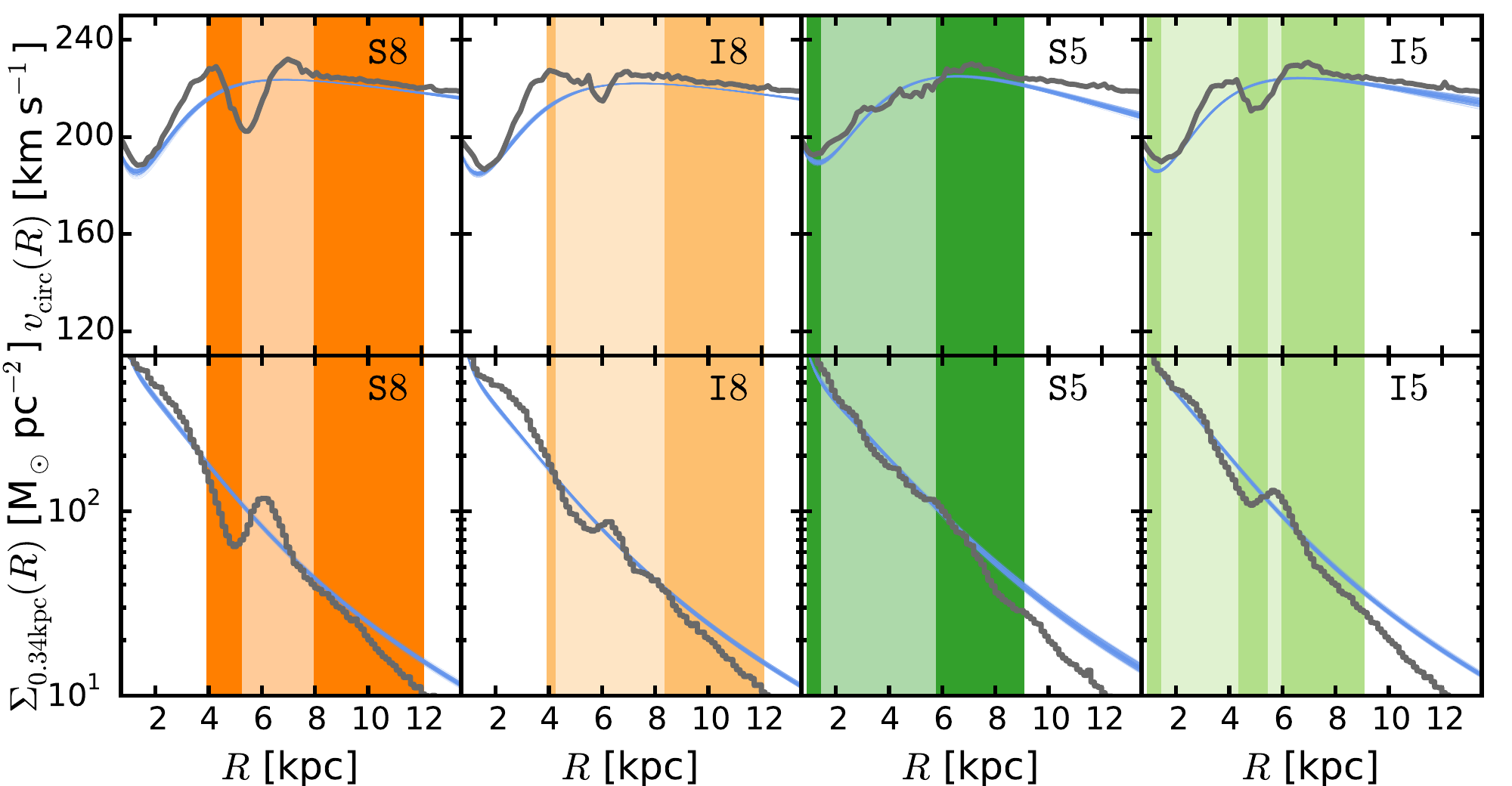}} %
	\subfigure[$r_\text{max}=5~\text{kpc}$. \label{fig:5kpcSuite}]{\includegraphics[width=0.6\textwidth]{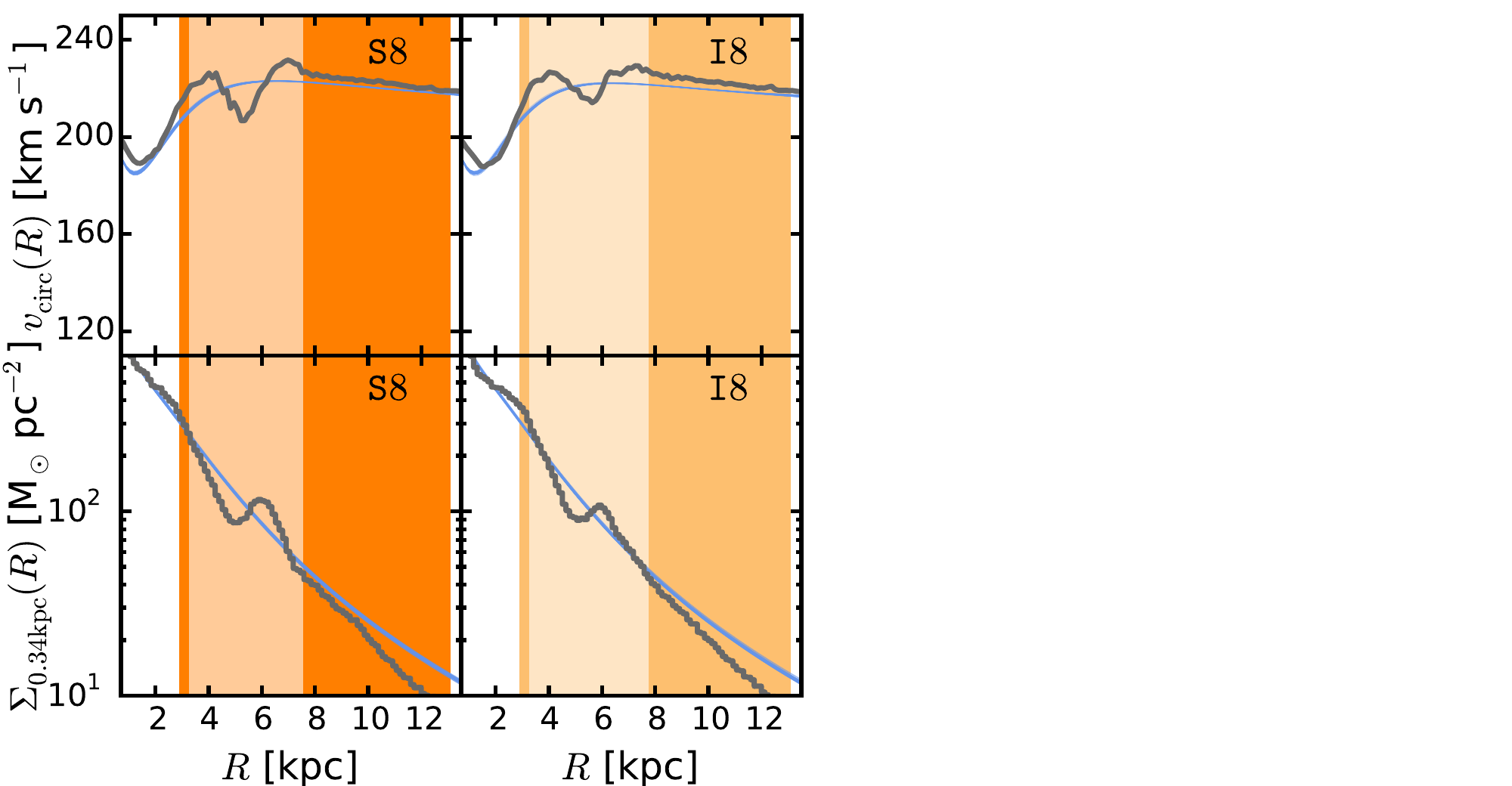}} 
	\caption{Same as Figure \ref{fig:vcirc_surfdens_suite_medium}, but for all big survey volumes with $r_\text{max}=4~\text{kpc}$ and $r_\text{max}=5~\text{kpc}$. }
\label{fig:vcirc_surfdens_suite_big}
\end{figure*}

Compared to the truth, the circular velocity curve derived from the initial snapshot was underestimated by $\sim2.5\%$; $v_\text{circ}(R_\text{\sun})=(219.36 \pm 0.08) ~\text{km s}^{-1}$. For the snapshot with spiral arms (see Figure \ref{fig:4kpc8Spiral_dens_vcirc_surfdens}) we observe an underestimation of only $\sim1.5\%$, so the bias in the initial snapshot might be related to its unphysical setup of the $v_T$ velocities. One the one hand, because this underestimation is present for both snapshots, it could also be a systematic bias due to the chosen Miyamoto-Nagai disk model. On the other hand, the bias also shows up in the \texttt{DEHH-Pot}'s circular velocity curve in Figure \ref{fig:4kpc8Spiral_dens_vcirc_surfdens}. The \texttt{DEHH-Pot} has a more realistic shape and was found as a direct fit to the spiral arm affected disk particle distribution. The $\sim1.5\%$ underestimate is therefore most likely introduced by the spiral arms (see Section \ref{sec:biases_explained}).

Because the simulation immediately develops spiral arms after the initial snapshot, there was no snapshot that was still axisymmetric, yet already in a dynamical steady state.

\subsection{The influence of spiral arms in \RM{} modeling} \label{sec:results_part2}

In the previous section we showed that for a large survey volume ($r_\text{max}=4~\text{kpc}$) \RM{} can construct a good average axisymmetric potential (and DF) model for a galaxy with spiral arms. In the following we want to investigate how this modeling success depends on the position and the size of the survey volume within the galaxy and with respect to the spiral arms.

\subsubsection{A suite of data sets drawn from spiral arms and inter-arm regions} \label{sec:suite}

To investigate a range of data sets affected in different proportions by spiral arms, we center our test survey volumes at the positions marked in Figures \ref{fig:simulation} and \ref{fig:spiral_arm_DeltaS} (see also Table \ref{tbl:volume_positions}) and consider volume sizes with $r_\text{max} \in [0.5,1,2,3,4,5]~\text{kpc}$ for $R_0 = 8~\text{kpc}$ and $r_\text{max} \in [0.5,1,2,3,4]~\text{kpc}$ for $R_0 = 5~\text{kpc}$ (to avoid the galactic center). As demonstrated in Figure \ref{fig:spiral_arm_DeltaS} the spiral arm strength is very different in these test volumes. Each data set that we draw from the simulation contains a random selection of $N_*=20,000$ stars inside the given spherical volume and we fit a single qDF and \texttt{MNHH-Pot} to it. 

\pagebreak

\subsubsection{Recovering the circular velocity curves and surface density profiles} \label{sec:circvel_surfdens}

It turns out that \RM{} is successful in finding reasonable and even very good best fit potential models for each one of the 22 test data sets independent of size and location---given the data and limitations of the model. To illustrate this and to make this encouraging result immediately obvious, we show the circular velocity curves and surface density profiles of all analyses in Figures \ref{fig:vcirc_surfdens_suite_medium}-\ref{fig:vcirc_surfdens_suite_big}. 

In contrast to Figure \ref{fig:4kpc8Spiral_dens_vcirc_surfdens}, these Figures only show the true profiles for the region within the galaxy where the data comes from: averaged over the angular wedge covering the radial extent of the survey volume, $\Delta \phi = \phi_0 \pm \arcsin(r_\text{max}/R_0)$, and within $|z| \leq 2 z_\text{s} = 0.34~\text{kpc}$, i.e., twice the scale height of the disk, which contains most of the disk mass. This is the matter distribution in which the stars are currently moving, and therefore the potential to which the modeling should be sensitive. In Figures \ref{fig:vcirc_surfdens_suite_medium}-\ref{fig:vcirc_surfdens_suite_big} we also mark the survey volume and the radial bins of size $\Delta R =200~\text{pc}$ with the highest number of stars. 

Even though the curves vary extremely between the individual data sets, it becomes very obvious that it is indeed the regions in which the majority of stars is located that drives the \RM{} fit. Furthermore, whether this region is dominated by a spiral arm or an inter-arm region, and even if this region is only as small as $r_\text{max}=500~\text{pc}$, \RM{} indeed constrains the local potential where most of the stars of the data set are located. Also, the constraints are not only most accurate but also most precise in these regions.

Only in the two volumes with $r_\text{max}=[0.5,1]~\text{kpc}$ at position \texttt{S8} \RM{} has some difficulties fitting the circular velocity curve; the model expects a flat or falling rotation curve and is presented with a steeply rising rotation curve due to the spiral arm dominating the region. But given that \RM{} recovers a good average surface density profile and the circular velocity at least at the center of the small volume, the fit is still quite successful. 

In an application to real data in the MW we would also have the possibility to impose some informative prior information on the potential shape (e.g., on the rotation curve), to avoid very unrealistic results (see also discussion in Section \ref{sec:discussion_sun_location}).

\subsubsection{Discussion of the model parameter recovery} \label{sec:parameter recovery}

\begin{figure}[!htbp]
\centering
\includegraphics[width=\columnwidth]{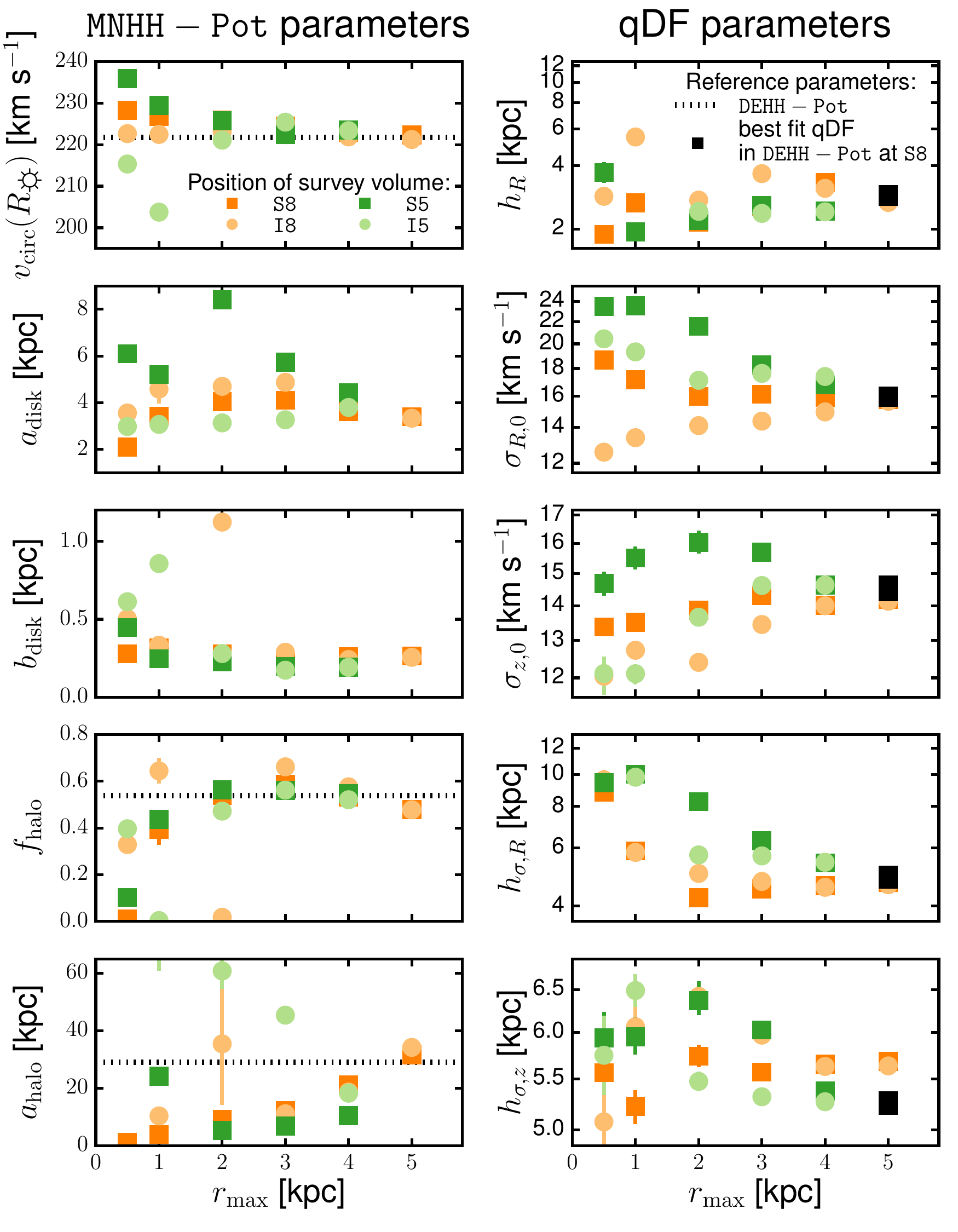}
\caption{Overview of the model parameter estimates (\texttt{MNHH-Pot} parameters on the left, qDF parameters on the right) recovered with \RM{} from 22 different data sets. All data sets were drawn from the same simulation snapshot, but from survey volumes at different positions in the galaxy (color-coded) and of different sizes ($r_\text{max}$ as indicated on the $x$-axis). Note that all five qDF parameters are shown here on a logarithmic scale, because \RM{} uses a logarithmically flat prior for them in the fit (see Equation \eqref{eq:qDF_parameters}). The black dotted line shows the known model parameters from the reference potential \texttt{DEHH-Pot} in Table \ref{tbl:DEHH-Pot} (the Miyamoto-Nagai disk parameters $a_\text{disk}$ and $b_\text{disk}$ are related but not directly comparable to an exponential disk scale length and height). The black squares denote the qDF parameters we recovered by fixing the potential to the \texttt{DEHH-Pot}, centering a survey volume with $r_\text{max}=5~\text{kpc}$ on the spiral arm at $R_0=8~\text{kpc}$ (position \texttt{S8}), and fitting the qDF only. A survey volume with a radial coverage as large as $r_\text{max}=5~\text{kpc}$ is required to properly recover all ``true'' model parameters. For smaller volumes there seem to be quite large deviations between truth and model; that these recovered parameters still all correspond to successful \RM{} fits to the data is discussed in Section \ref{sec:parameter recovery}.}
\label{fig:model_parameters}
\end{figure}

Figures \ref{fig:vcirc_surfdens_suite_medium}-\ref{fig:vcirc_surfdens_suite_big} in the previous Section have illustrated how well the potential is recovered by \RM{}. Figure \ref{fig:model_parameters} compares the potential and qDF parameters found with \RM{} to the parameters of the reference \texttt{DEHH-Pot} from Table \ref{tbl:DEHH-Pot}. And at first glance there appear to be several discrepancies. In the following we will discuss the deviations and explain why each set of parameters still corresponds to a good fit to the data.

Overall the statistical random errors on the parameter recovery are very small for $N*=20,000$ and possible systematic errors dominate. There are only a few exceptions ($r_\text{max}=500~\text{pc}$, $r_\text{max}=1~\text{kpc}$ at \texttt{I5}, $r_\text{max}=2~\text{kpc}$ at \texttt{I8}), which we will discuss later.

We will first consider the parameters of the gravitational potential (left column in Figure \ref{fig:model_parameters}): All volumes recover $v_\text{circ}(R_\text{\sun})$ within a few $\text{km s}^{-1}$; in the largest volumes---where the circular velocity curve is probed over several $\text{kpc}$---the estimate is the most accurate. The halo fraction $f_\text{halo}$ of the radial force at the ``solar'' radius $R_\text{\sun}$ is very well recovered, especially for $r_\text{max}\gtrsim 2~\text{kpc}$. The estimate that we get for the best fit Miyamoto-Nagai disk scale height $b_\text{disk}$ seems to be also approximately independent of the size of the volume. We can even recover the true halo scale length $a_\text{halo}$, however only for a volume as large as $r_\text{max}=5~\text{kpc}$. The models at $r_\text{max}=500~\text{pc}$ appear to be too small to constrain the halo at all, and the MCMCs diverged completely for this parameter. Smaller volumes that underestimate $a_\text{halo}$ get slightly larger estimates for the disk scale length $a_\text{disk}$ and the overall radial density slope is then probably closer to the truth, even if the individual parameters are not. Outliers can often be explained by having a look at the data: The large disk scale length recovered from the $r_\text{max}=2~\text{kpc}$ volume at \texttt{S5}, for example, mirrors the comparably flat matter distribution caused by two spiral arms close together and dominating the volume (see Figure \ref{fig:2kpcSuite} and the large $\sigma_{\Delta\text{Spiral}}$ for this analysis in Figure \ref{fig:spiral_arm_DeltaS_d}). 

The right column of Figure \ref{fig:model_parameters} compares the recovered qDF parameters for the different survey volumes with the qDF parameters we got from fixing the potential model to the \texttt{DEHH-Pot} and fitting the qDF only in a $r_\text{max}=5~\text{kpc}$ volume at \texttt{S8}. Even though the qDF parameters for small volumes are widely different for different positions within the galaxy, they all approach the values recovered with the \texttt{DEHH-Pot} for larger volumes. There seems, therefore, to be an overall best-fit qDF describing the average tracer distribution in the galaxy's disk. The only difference is in the $h_{\sigma,z}$ parameter, where the models fitting a \texttt{MNHH-Pot} recover a slightly larger value than the models using the known \texttt{DEHH-Pot}. The suspected reason is that the Miyamoto-Nagai disk flares at larger radii as compared to the double exponential-disk (see Figure \ref{fig:4kpc8Spiral_dens_vcirc_surfdens}), which leads to a less-steep radial decline in the vertical forces, and therefore mean vertical orbital energies $\langle E_z \rangle \sim \nu \times J_z$, and therefore to a slightly longer $h_{\sigma,z}$ scale length. In general, volumes centered on spiral arms have larger velocity dispersion parameters $\sigma_{R,0}$ and $\sigma_{z,0}$ as compared to volumes at the same radius $R_0$ but centered on an inter-arm region. And the volumes at $R_0=5~\text{kpc}$ with their stronger spiral arms have larger velocity dispersions than those at $R_0=8~\text{kpc}$---which is what we expect. Most volumes recover similar tracer scale lengths $h_R\sim2.5\pm0.5~\text{kpc}$ close to the known disk scale length $R_\text{s}$. Only the volumes centered on the inter-arm region at $R_0=8~\text{kpc}$ (position \texttt{I8}) recover much longer $h_R$. This might be related to the fact that volumes at \texttt{I8} are dominated by an especially extended inter-arm region. The volumes at \texttt{I5} with $r_\text{max}=[0.5,1]~\text{kpc}$ were not able to constrain the tracer scale length at all because of the unfortunate position between the rising density wings of two strong spiral arms (see Figure \ref{fig:vcirc_surfdens_suite_small}).

There are a few survey volumes for which the recovered parameters show some peculiarities: The models from volumes with $r_\text{max}=1~\text{kpc}$ at \texttt{I5} and $r_\text{max}=2~\text{kpc}$ at \texttt{I8} reject the DM halo completely, i.e., $f_\text{halo}=0$. The corresponding halo scale lengths $a_\text{halo}$ are therefore unconstrained,\footnote{For these analyses and the $r_\text{max}=500~\text{pc}$ analyses the fit could not constrain $a_\text{halo}$ and the MCMC was diverging. We had to stop the MCMC after some time, so the $a_\text{halo}$ might in truth be even less constrained than shown in Figure \ref{fig:model_parameters}.} while the corresponding disk scale heights $b_\text{disk}$ are grossly overestimated to account for the missing contribution of the spherical halo. We have investigated the reason for this fitting result and found that for the way in which the spiral arms affect the circular velocity curve in these volumes, the recovered models with unusual radial profiles are indeed a better description for the data (see Figure \ref{fig:2kpcSuite} and \ref{fig:1kpcSuite}). Also, while most analyses average the vertical forces radially over the spiral arms (see Figures \ref{fig:4kpc8Spiral_forces}, lower right panel), for these analyses the averaging happens vertically, i.e., at approximately one scale height above the plane where the model's vertical forces are equally well recovered at all radii (in spiral arms and between), while at small and large $|z|$ the model is bad. \RM{} therefore also found a good average fit model for the stars in these volumes. 

Overall we find that if the volume is large enough to average over several spiral arms and inter-arm regions, an unlucky positioning with respect to the spiral arms does not lead to strong biases in the parameter recovery. We stress again that for particularly large volumes, $r_\text{max}=5~\text{kpc}$, we were able to recover all model parameters, including the halo scale length $a_\text{halo}$.

\subsubsection{Recovering the local gravitational forces} \label{sec:local_grav_forces}

\begin{figure}[!htbp]
\centering
  \includegraphics[width=\columnwidth]{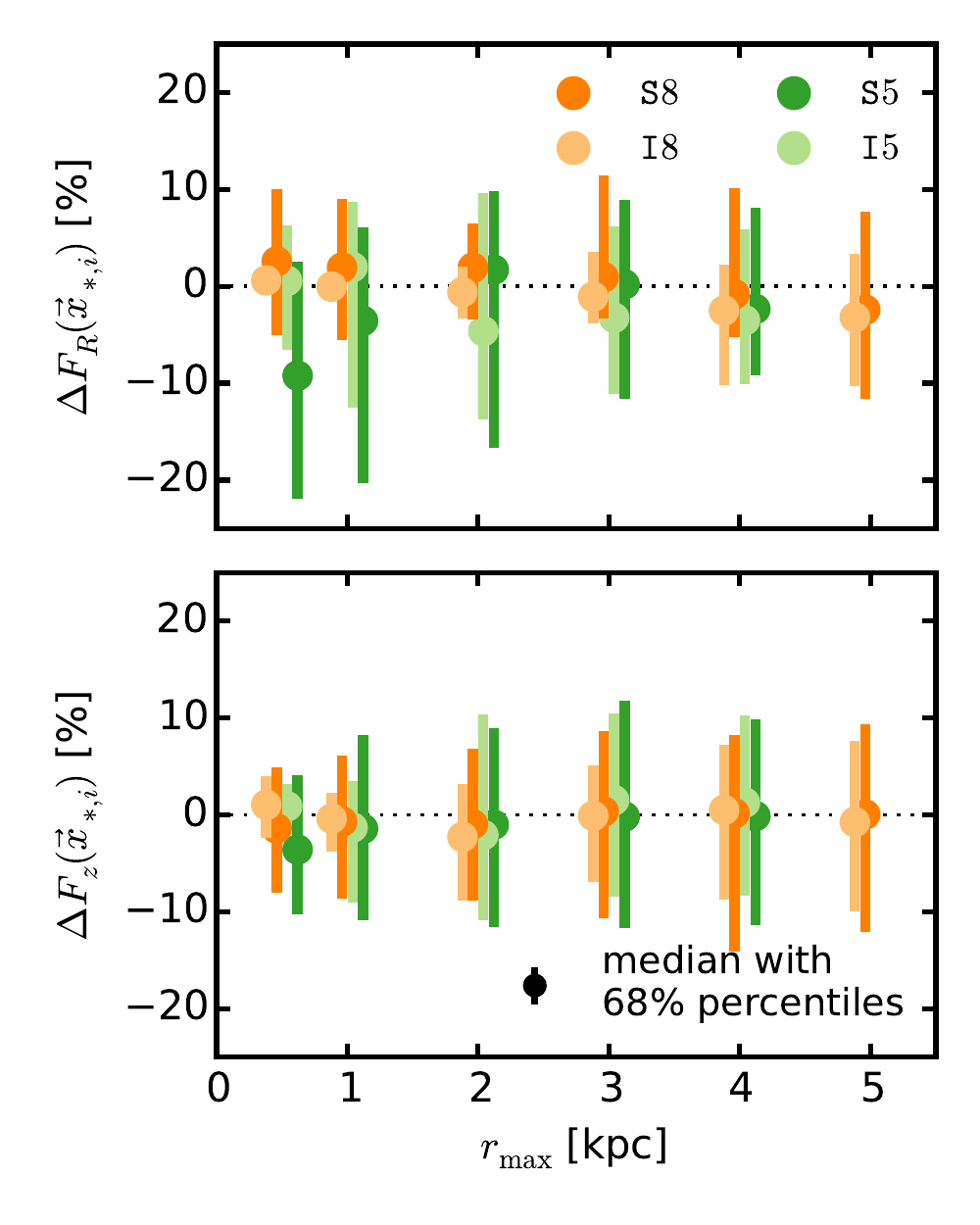}
  \caption{Accuracy of the radial (upper panel) and vertical (lower panel) gravitational forces recovered with \RM{} from the suite of data sets introduced in Section \ref{sec:suite} for the ensemble of stars in each data set. The $x$-axis denotes the radial size $r_\text{max}$ of the survey volume belonging to each data set. (For presentation purposes we added a small offset $\ll 1$ to $r_\text{max}$ on the $x$-axis.) The $y$-axis shows the distribution of force residuals, $\Delta F_{R}(\vect{x}_{*,i})$ and $\Delta F_{z}(\vect{x}_{*,i})$ in Equations \eqref{eq:delta_FR}-\eqref{eq:delta_Fz}, at the positions of the stars $x_{*,i}$ that entered the analysis. In particular we show here for each distribution of $\Delta F(\vect{x}_{*,i})$ the median as a dot with the [$16th$,$84th$] percentile range as a bar. We find that the forces are very well recovered at the positions of the stars independent of the size of the volume.}
\label{fig:forces_bias_a}
\end{figure}

In the previous section we found that the potential and qDF parameters recovered from different survey volumes can be quite different. While the differences can be explained qualitatively, it is not yet clear how good the corresponding potential constraints actually are in a quantitative sense. To test this we calculate again $\Delta F_R(\vect{x}_{*,i})$ and $\Delta F_z(\vect{x}_{*,i})$ from Equations \eqref{eq:delta_FR}-\eqref{eq:delta_Fz} at the position of each star $\vect{x}_{*,i}$ in each data set (analogous to Figure \ref{fig:4kpc8Spiral_forces}). From the corresponding histograms of number of stars vs. $\Delta F$ we derive the median and the $16th$ and $84th$ percentiles ($1\sigma$ range) and show them in Figure \ref{fig:forces_bias_a}. We chose this diagnostic because the forces at the positions of the stars are the quantities of the potential to which our modeling is sensitive.

The important key result from Figure \ref{fig:forces_bias_a} is that we get very close to recovering the true forces $\Delta F(\vect{x}_{*,i}) \lesssim 10\%$ at the positions of the majority of stars in the survey volume, no matter how large or small the survey volume is. On average, the force recovery is also unbiased\footnote{An exception are the radial forces for small volumes strongly dominated by spiral arms (e.g., at \texttt{S5}). The small systematic bias in $F_R$ is discussed in detail in Section \ref{sec:biases_explained}.} for the ensemble of stars.

\begin{figure}[!htbp]
\centering
\includegraphics[width=\columnwidth]{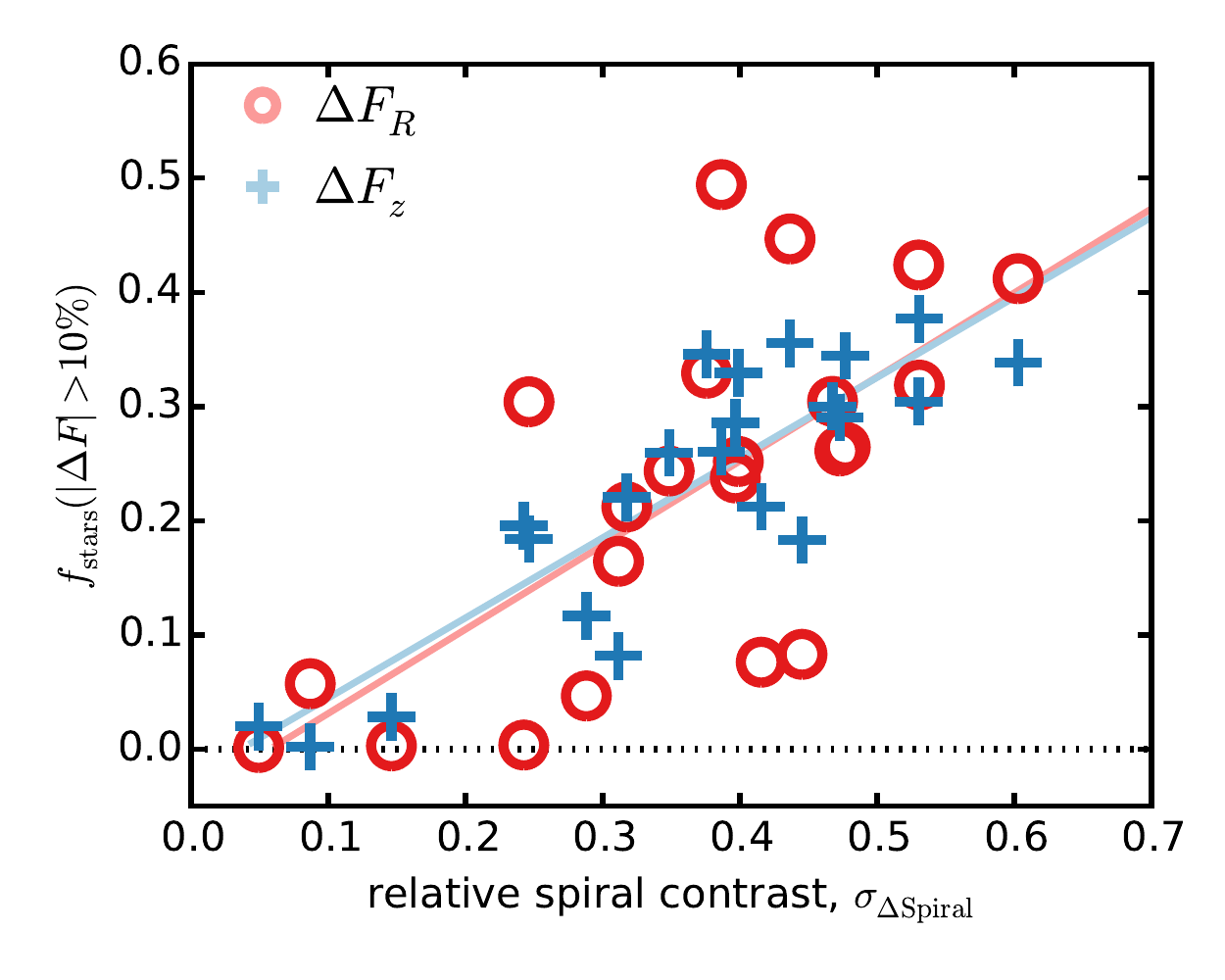}
\caption{Influence of the spiral arm contrast on the recovery of the gravitational forces at the positions of the stars $x_{*,i}$ that entered the \RM{} analysis. Each circle/cross pair corresponds to one of our 22 data sets. The relative spiral contrast on the $x$-axis is quantified as $\sigma_{\Delta\text{Spiral}}$ calculated within each survey volume according to Equation \eqref{eq:std_DeltaS} in Section \ref{sec:spiral_arm_DeltaS}. On the $y$-axis the fraction of stars ($f_\text{stars}$) in each data set is shown for which the radial (red circles) and vertical (blue crosses) force residual calculated from Equations \eqref{eq:delta_FR}-\eqref{eq:delta_Fz} is larger than 10\% (i.e., at $0$ all stars have good force measurements, at $1$ everything went wrong). The red and blue lines are linear fits to the radial and vertical force residual fraction, respectively, and are guides to the eye that show the clear and expected trend that in volumes with smaller spiral arm contrast, where comparably fewer stars are located in spiral arms, the axisymmetric best-fit model can recover the true gravitational forces also for more stars. On average, the radial and vertical forces are equally well-recovered at a given spiral contrast.}
\label{fig:std_DeltaS_vs_frac10_stars}
\end{figure}

Figure \ref{fig:forces_bias_a} also contains some subtle clues that suggest that the quality of the force recovery could be correlated with the position of the data set with respect to the spiral arms. We investigate this further by relating in Figure \ref{fig:std_DeltaS_vs_frac10_stars} the local force recovery, i.e., the distribution of $\Delta F_R(\vect{x}_{*,i})$ and $\Delta F_z(\vect{x}_{*,i})$ for each data set, to the relative spiral contrast within the respective survey volume, $\sigma_{\Delta \text{Spiral}}$ (Equation \eqref{eq:std_DeltaS}; see also Figure \ref{fig:spiral_arm_DeltaS_d}). Figure \ref{fig:std_DeltaS_vs_frac10_stars} shows that the average fraction of stars for which the recovery of the radial or vertical force is bad (i.e., larger than 10\%) increases with increasing spiral contrast $\sigma_{\Delta\text{Spiral}}$. This is as expected: Volumes in which the steep gradient in surface density around a strong spiral arm ($|\Delta_{\text{Spiral},k}| > 0$) is not balanced by larger areas with less perturbations ($\Delta_{\text{Spiral},k} \sim 0$) have (i) a large relative spiral contrast $\sigma_{\Delta\text{Spiral}}$, and (ii) a large relative number of stars affected by the non-axisymmetric kinematics of the spiral arms. And for these individual stars, the axisymmetric \RM{} model is less successful in recovering the correct forces. (But as we saw in Figure \ref{fig:forces_bias_a}, the ensemble average is even in these cases unbiased.)

Interestingly, and even though there is some scatter, the force recovery at a given $\sigma_{\Delta\text{Spiral}}$ is on average very similar for the radial and vertical forces (compare the linear fits in Figure \ref{fig:std_DeltaS_vs_frac10_stars}). This means that \RM{} attempts to fit both the radial and vertical forces at the positions of the stars, and is not particularly sensitive to just one of them.

As we saw in Figure \ref{fig:spiral_arm_DeltaS_d}, the spiral contrast $\sigma_{\Delta \text{Spiral}}$ increases for the different test volume positions approximately in this order: \texttt{I8} $\longrightarrow$ \texttt{S8} $\longrightarrow$ \texttt{I5} $\longrightarrow$ \texttt{S5}. From Figure \ref{fig:std_DeltaS_vs_frac10_stars} it follows that this is also the order in which the accuracy of the force recovery decreases. (We did not include this piece of additional information in Figure \ref{fig:std_DeltaS_vs_frac10_stars}, but it can be seen in Figure \ref{fig:forces_bias_a}, especially for the smaller volumes.)

\subsubsection{Extrapolating the gravitational potential model}

\begin{figure}[!htbp]
\centering
  \includegraphics[width=\columnwidth]{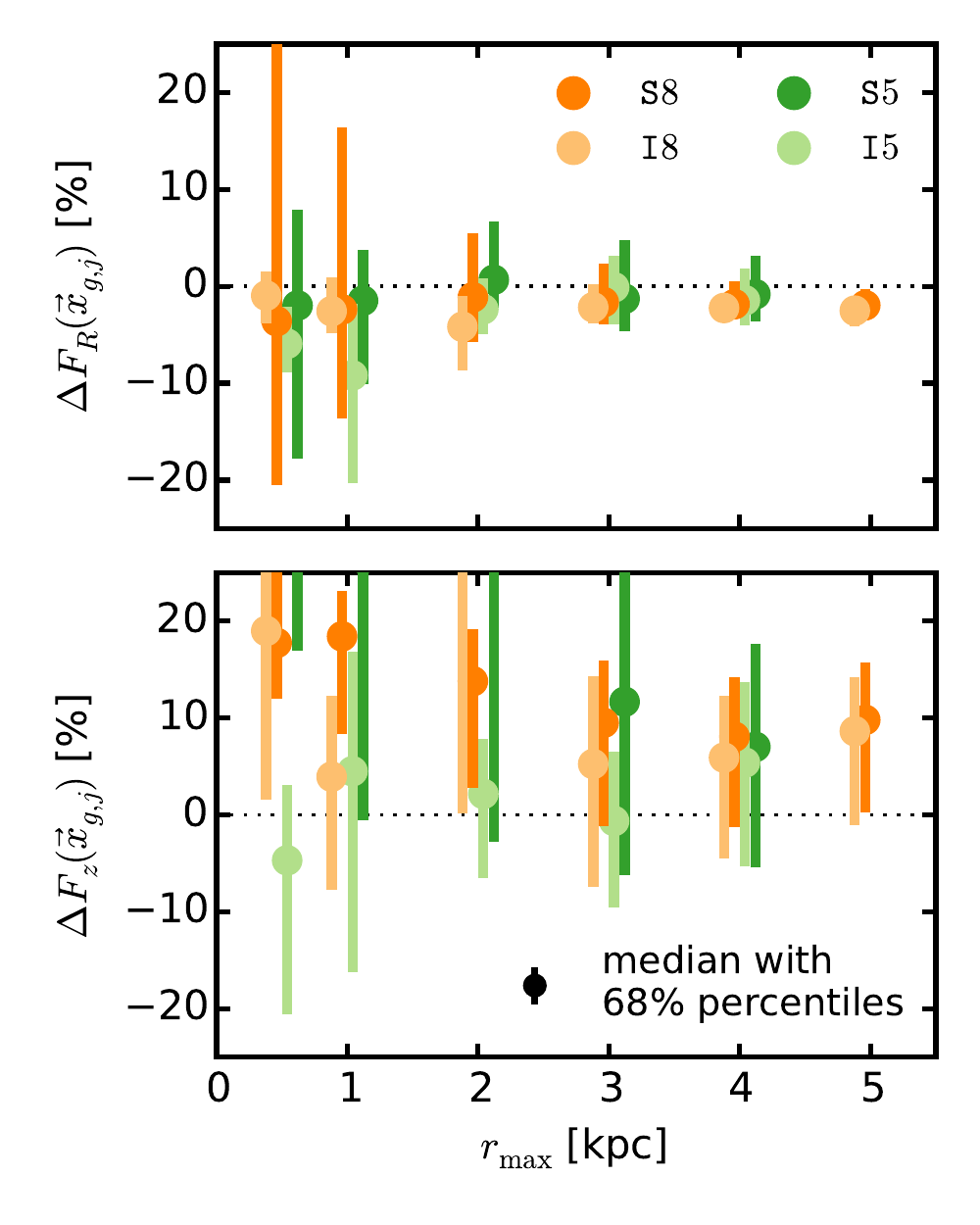}
  \caption{Extrapolability of a \RM{} potential model. This figure is similar to Figure \ref{fig:forces_bias_a} and shows the radial (upper panel) and vertical (lower panel) gravitational force residuals. But instead of calculating the residuals at the positions of the stars, we evaluated them here at regular grid points $\vect{x}_{g,j}$ in a large cylinder of $r_\text{max}=5~\text{kpc}$ and height $|z|=1.5~\text{kpc}$ around each survey volumes' center: $\Delta F_{R}(\vect{x}_{g,j})$ and $\Delta F_{z}(\vect{x}_{g,j})$ in Equations \eqref{eq:delta_FR_grid}-\eqref{eq:delta_Fz_grid}. This demonstrates how well the model can be extrapolated out to $5~\text{kpc}$, i.e., how close to the truth the corresponding volume averaged model is. The $x$-axis shows the radial size $r_\text{max}$ of the survey volume from which the model was derived. The dot and error bars denote the median and [$16th$,$84th$] percentile range for each distribution of $\Delta F(\vect{x}_{g,j})$. The extrapolability works better for the radial forces than for the vertical forces. We account the systematic overestimation of the vertical forces to the spiral arms and the flaring of the disk model at large radii (see Section \ref{sec:biases_explained}). We find that we need at least a survey volume of $r_\text{max}=3~\text{kpc}$ to get a potential with a reasonable extrapolability.}
\label{fig:forces_bias_b}
\end{figure}

Also, it is interesting to see how well the extrapolation of a recovered potential describes the overall gravitational potential of the galaxy. To investigate the extrapolability, we introduce another diagnostic which uses a cylindrical grid centered on the respective positions in Table \ref{tbl:volume_positions}, always having a radius of $r_\text{max}=5~\text{kpc}$ and a height of $z=1.5~\text{kpc}$ both above and below the plane. In the $(x,y)$ plane the regular grid points have a distance of $0.25~\text{kpc}$ and in $z$ they have a distance of $0.125~\text{kpc}$ to better sample the thin disk (we throw out grid points close to the galactic center with $R<0.125~\text{kpc}$, however). We then evaluate at the position $\vect{x}_{g,j} \equiv (x_j,y_j,z_j)$ of each regular grid point the force residuals
\begin{eqnarray}
\Delta F_R(\vect{x}_{g,j}) \equiv \frac{|F_{R,M}(R_j,z_j)| - |F_{R,T}(x_j,y_j,z_j)|}{|F_{R,\text{typ}}(R_j)|} \label{eq:delta_FR_grid}\\
\Delta F_z(\vect{x}_{g,j}) \equiv \frac{|F_{z,M}(R_j,z_j)| - |F_{z,T}(x_j,y_j,z_j)|}{|F_{z,\text{typ}}(R_j)|},\label{eq:delta_Fz_grid}
\end{eqnarray}
analogous to Equations \eqref{eq:FRtyp}-\eqref{eq:delta_Fz}. The two panels in Figure \ref{fig:forces_bias_b} show the [$16th$,$84th$] percentile range and the median of the grid points' distribution in $\Delta F_R(\vect{x}_{g,j})$ and $\Delta F_z(\vect{x}_{g,j})$.

\begin{figure}[!htbp]
\centering
\includegraphics[width=\columnwidth]{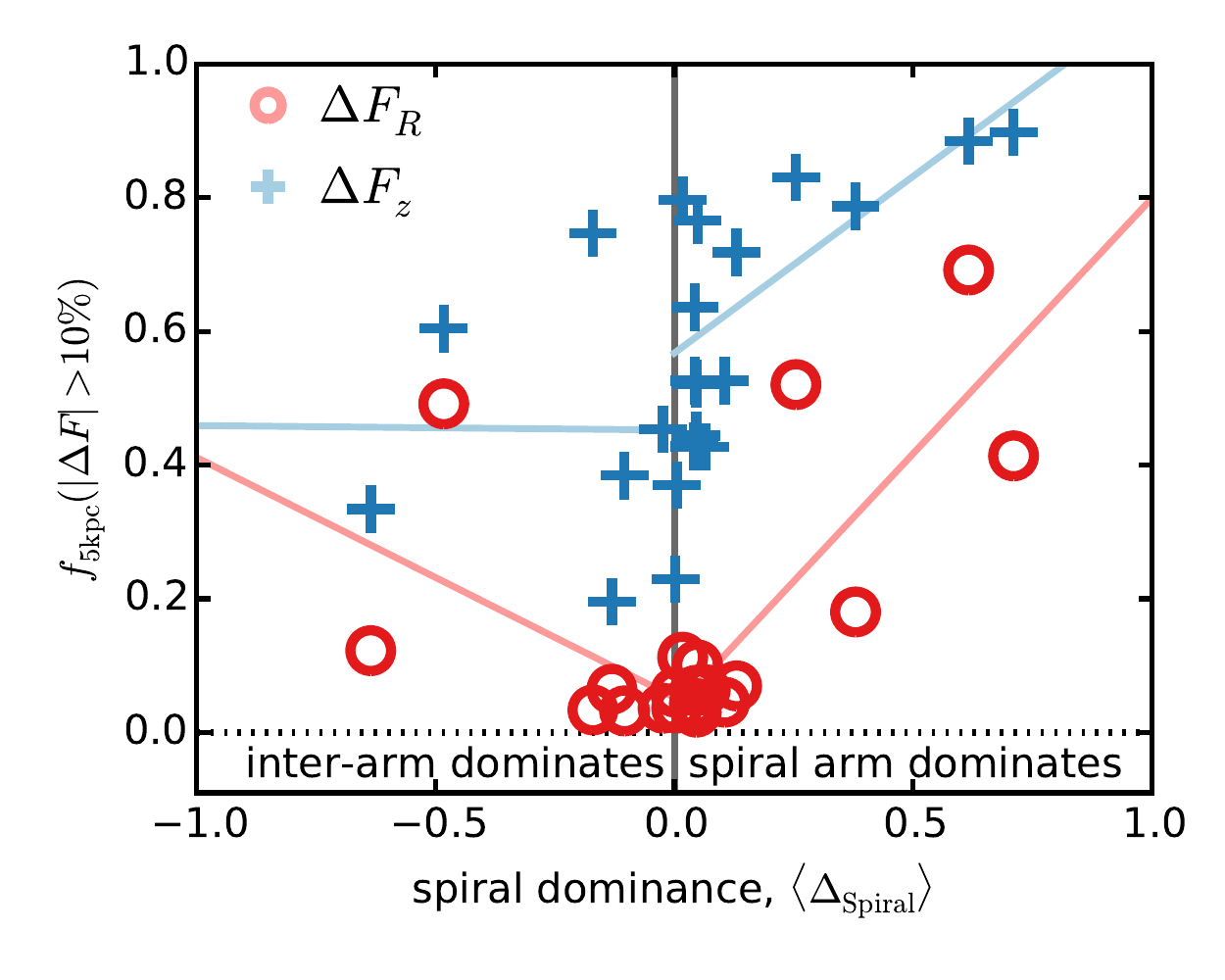}
\caption{Influence of the spiral arm dominance on the extrapolability of the \RM{} potential models. Each circle/cross pair corresponds to one of our 22 \RM{} analyses. How much a spiral arm dominates within a given survey volume is quantified by $\langle \Delta_\text{Spiral} \rangle$ from Equation \eqref{eq:mean_DeltaS} and shown on the $x$-axis. $\langle \Delta_\text{Spiral} \rangle\sim0$ means spiral arms and inter-arm regions dominate equally in the survey volume. The more negative $\langle \Delta_\text{Spiral} \rangle$ is, the more dominated by an inter-arm region is the survey volume. The $y$-axis shows the volume fraction within which the radial (red circles) and vertical (blue crosses) gravitational force residuals are larger than 10\%. In particular, we extrapolate each potential model within a cylindrical volume with $r_\text{max}=5~\text{kpc}$ and $|z| \leq 1.5~\text{kpc}$, centered on the respective survey volume positions given in Table \ref{tbl:volume_positions}. $f_\text{\rm 5kpc}(|\Delta F| > 10\%)$ therefore quantifies how well a \RM{} potential model can be extrapolated out to $5~\text{kpc}$ from the center of the survey volume. At $f_\text{\rm 5kpc}(|\Delta F| > 10\%) \sim 0$ the extrapolation works best. The red and blue lines are linear fits and only serve as guide to the eye. This figure demonstrates that the extrapolability of the model gets better the less a spiral arm dominates the data set. For the dominance of inter-arm regions this trend is less pronounced. Overall, the radial force can be much better extrapolated than the vertical force, and models from data sets centered on inter-arm regions can be more reliably extrapolated than those centered on spiral arms.}
\label{fig:mean_DeltaS_vs_frac10_grid}
\end{figure}

First, we find that the radial forces are overall very well predicted, especially when derived from large survey volumes. There is however an overestimation of $\sim5-20\%$ in the vertical forces (depending on volume size and position) which is induced by the spiral arms and is partly also due to a systematic error caused by the choice of potential model (see explanation below in Section \ref{sec:biases_explained}).

Second, the constraints we get on the spatially averaged forces inside $r<5~\text{kpc}$ are almost as good when derived from a survey volume of $r_\text{max}=3~\text{kpc}$ as compared to survey volumes of $r_\text{max}=4$ or $5~\text{kpc}$. If we had to decide between a $r_\text{max}=3~\text{kpc}$ volume with good data quality and a larger volume with worse data quality, we would lose nothing in terms of extrapolability when using the smaller volume (only the halo scale length might not be as well constrained, see Figure \ref{fig:model_parameters}).
 
Third, there are further indications in Figure \ref{fig:forces_bias_b} that the position of the survey volume with respect to the spiral arms matters for the force recovery. 

In Figure \ref{fig:mean_DeltaS_vs_frac10_grid} we stress that even more by relating the extrapolability, i.e., the volume-averaged distribution of force residuals, $\Delta F_R(\vect{x}_{g,j})$ and $\Delta F_z(\vect{x}_{g,j})$, to the dominance of the spiral arm $\langle \Delta_\text{Spiral} \rangle$ in the survey volume (Equation \eqref{eq:mean_DeltaS}; see also Figure \ref{fig:spiral_arm_DeltaS_c}). We derive the fraction of grid points $\vect{x}_{g,j}$ in the reference cylinder for each data set/\RM{} model with forces that are misjudged by more than 10\%, and plot it against $\langle \Delta_\text{Spiral} \rangle$. Positive or negative $\langle \Delta_\text{Spiral}\rangle$ quantify how much the spiral arms or the inter-arm regions dominate the corresponding survey volume, respectively. A low fraction of grid points with $|\Delta F| > 10\%$ means that the extrapolation of the potential model works well.

Firstly, we note that there is a clear trend that the extrapolability gets worse if a spiral arm strongly dominated the survey volume from which the potential constraint was derived ($\langle \Delta_\text{Spiral} \rangle \gg 0$). The same trend can be seen for the dominance of inter-arm regions ($\langle \Delta_\text{Spiral} \rangle \ll 0$), but it is weaker and less clear.

Secondly, we note again that $F_z(\vect{x}_{g,j})$ is predicted less good than $F_R(\vect{x}_{g,j})$. The reason for this is laid out in Section \ref{sec:biases_explained}. 

\begin{figure}[!htbp]
\centering
    \includegraphics[width=\columnwidth]{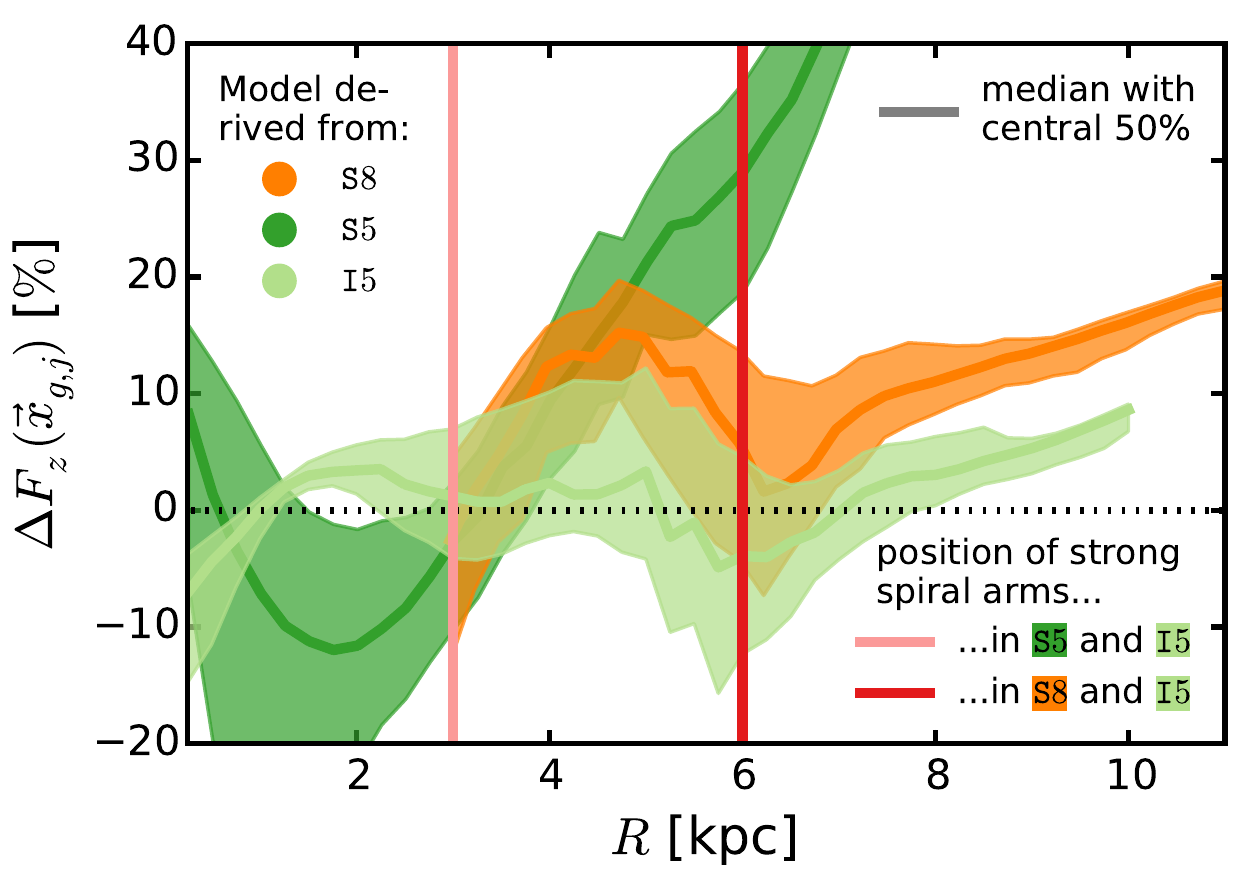}
\caption{Examples for the extrapolability of a \RM{} gravitational potential model as function of galactocentric radius, $R$. In particular, we show the vertical force residuals $\Delta F_z(\vect{x}_{g,j})$ from Equation \eqref{eq:delta_Fz_grid} for the \RM{} potential models derived from three data sets with $r_\text{max}=2~\text{kpc}$ at positions \texttt{S8}, \texttt{S5}, and \texttt{I5} (color-coded; see also Figure \ref{fig:simulation}). The colored bands show the distribution (median with the central $50\%$ percentiles) of $\Delta F_z$'s of all regular grid points $\vect{x}_{g,j}$ within a distance $r\leq5~\text{kpc}$ and $|z|\leq1.5~\text{kpc}$ from the survey volumes center at a given $R$. This figure demonstrates the origin of the bias in the vertical force prediction, which we found in Figure \ref{fig:forces_bias_b}. As the fit is driven by the excess of stars that feel stronger vertical forces in the spiral arms, e.g., at $R\sim3~\text{kpc}$ or $R\sim6~\text{kpc}$, we get the vertical force right at these radii. We consequently overestimate it in the inter-arm regions. In addition, the assumed gravitational potential model family, \texttt{MNHH-Pot} with a Miyamoto-Nagai, disk flares outside of $R\sim8~\text{kpc}$ as compared to the true exponential disk.}
\label{fig:Fg_vs_R}
\end{figure}

The main result of Figure \ref{fig:mean_DeltaS_vs_frac10_grid} is, however, the following: The extrapolability of models derived from data sets drawn from survey volumes centered on inter-arm regions appears to be in general better than that of data sets centered on spiral arms. We suspect that the reason for this is that the stellar distribution between spiral arms is smoother, more extended, and closer to the overall axisymmetric average model, such that the potentials recovered from these volumes have real predictive power for a much larger volume.

\subsubsection{Biases in the potential recovery caused by the spiral arms} \label{sec:biases_explained}

What are the reasons for the biases that we observe in Figure \ref{fig:forces_bias_b}? 

The peak of the distribution in $\Delta F_R(\vect{x}_{g,j})$ (and also in $\Delta F_R(\vect{x}_{*,i})$ in Figure \ref{fig:forces_bias_a}) is slightly ($\sim1.5\%$) biased towards an underestimation of $|F_{R}|$ in our \RM{} models. This bias already showed up in the circular velocity curve in Figure \ref{fig:4kpc8Spiral_dens_vcirc_surfdens} and also in the reference potential model \texttt{DEHH-Pot}. We therefore suspect that this bias is caused by the spiral arms. The line of argument goes like this: Spiral arms are very thin. If a spiral arm crosses the observation volume, both its leading side (at large radii) and its trailing side (at small radii) are also in the volume. Stars on the trailing side feel a lower gravitational pull towards the galaxy center than they would if there was no spiral arm. Because there are in general more stars at smaller radii, the \RM{} fit is slightly biased to reproduce in general slightly weaker radial forces.\footnote{In the special case that the survey volume coincidentally only contains part of a spiral arm---as was the case with the analyses for $r_\text{max}=[0.5,1]~\text{kpc}$ at position \texttt{I5} (see Figure \ref{fig:vcirc_surfdens_suite_small}, right panels)---the fitting behaves differently anyway, as was already discussed in Section \ref{sec:parameter recovery}.}

The peak in the distribution of $\Delta F_z(\vect{x}_{g,j})$ is strongly biased towards an overestimation of $|F_{z}|$ by the \RM{} model. We illustrate the reason for this in Figure \ref{fig:Fg_vs_R}, where we show how $\Delta F_z(\vect{x}_{g,j})$ varies as a function of $R$ for three example data sets with $r_\text{max}=2~\text{kpc}$. One reason for this bias is a relic of the assumed \texttt{MNHH-Pot} disk model family, which flares outside of $R\sim8~\text{kpc}$ (see Section \ref{sec:potential_model}). The vertical forces in this region are therefore much higher in the model than in the true galaxy with its exponential disk. The main reason for this overestimation of $F_z$ comes however directly from the spiral arms: There are much more stars in the spiral arms than in the inter-arm regions, and the stars in the spiral arm feel stronger vertical forces because of the higher surface mass density. The \RM{} fit is driven by the majority of stars and the best fit model therefore predicts in general higher vertical forces. As expected, the overestimation of $F_{z}$ in Figure \ref{fig:forces_bias_b} is especially strong ($\sim 20 \%$) for small survey volumes dominated by spiral arms, while small volumes dominated by an inter-arm region result in much better estimates for the spatially averaged $F_z(\vect{x}_{g,j})$ ($\sim5\%$ bias). Large volumes lie somewhere in between (bias of $\sim10\%$).

Why do the biases show up in different strength in Figures \ref{fig:forces_bias_a} and \ref{fig:forces_bias_b}?

The stellar number asymmetry in the trailing vs.\ leading sides of spiral arms is much smaller than the stellar number asymmetry in the spiral arm vs.\ the inter-arm region. The bias is therefore visible in the distribution of $\Delta F_R(\vect{x}_{*,i})$ (because the $F_R$ recovery is biased only by a few stars, which leads to a bias that is visible for the majority of stars) and not in $\Delta F_z(\vect{x}_{*,i})$ (because the majority of stars bias the fit and we therefore also recover $F_z$ for the majority of stars). The bias becomes particularly pronounced for $\Delta F_z(\vect{x}_{g,j})$ (because the inter-arm regions dominate when averaging spatially which leads to a large average overestimation of $F_z$) and stays small for $\Delta F_R(\vect{x}_{g,j})$ (because trailing and leading sides of spiral arms are similarly important when averaging spatially so it becomes visible that the bias is actually not that big).

\section{Discussion and Outlook} \label{sec:discussion}

\subsection{On the informativeness of an orbit distribution function}

The qDF appears to be very informative. We did expect it to be at least a reasonable model for the overall symmetrized disk of the galaxy simulation, considering its initial set-up as an axisymmetric, exponentially decreasing particle distribution that subsequently evolved as a mono-age population (see Sections \ref{sec:simulation_description} and \ref{sec:DF_model}). In Sections \ref{sec:4kpc8Spiral_DF}, \ref{sec:4kpc8Spiral_actions}, and \ref{sec:MNdHHinit} we demonstrated that the qDF is indeed a good average model for the tracer distribution in a large survey volume---even though the spiral arms did introduce considerable deviations.

We had, however, no indications beforehand of how well the axisymmetric qDF would perform in a small survey volume completely dominated by non-axisymmetric spiral arms. It would not have been surprising if \RM{} had failed. But in Section \ref{sec:circvel_surfdens} it turned out that the potential measurements were reliable even in most of the small volumes with $r_\text{max}=[0.5,1]~\text{kpc}$. And the corresponding qDF parameters were tightly constrained and reasonable as well. 

We deduce that the qDF is indeed flexible and robust enough to work with data affected by non-axisymmetries.

That the corresponding potential constraints were reliable as well leads to the following conclusion: A potential model that does not fit the gravitational forces acting on the stars appears to lead to such an unrealistic orbit/action distribution, that a fit with even such a simple orbit DF as the qDF is impossible. This demonstrates once more how powerful the concept of an orbit DF is.

\subsection{On the restrictiveness of the parametrized potential model}

How much does the choice of potential model matter for the success of the modeling?

We used, on the one hand, a bulge and halo model that reproduces the true bulge and halo better than we can hope to use in reality for the MW. The fact that for all except the largest volumes the true halo scale length is not remotely recovered (and some small volumes even have $f_\text{halo}=0$), and that the contribution of the bulge to the overall potential is small (i.e., the bulge contribution to the total radial force at $R=8~\text{kpc}$ is only $\sim 9-10\%$) remedies this apparent advantage. 

On the other hand, we used a disk model, the Miyamoto-Nagai disk, that we chose purely for its convenient parametric form and of which we know that it is not a good model for the simulation snapshot; especially not for the radial density profile at large radii \citep{2015MNRAS.448.2934S}. As we saw in most figures in this paper this lead to biases in predicting the potential at radii where we have only a few or no stars. But because the spiral arms are such strong perturbations in the overall potential, a better disk model would not give much better results.

It appears that a potential model with a reasonable shape and flexibility (here: disk+bulge+halo structure with 5 free parameters) can do well enough in finding a good fit, both locally for small volumes and overall for large volumes.

This is in agreement with one of our key results of Paper I, where we managed to successfully fit data from a MW-like (but axisymmetric) galaxy model with a bulgeless potential of a restrictive St\"ackel form. This was illustrated in Figure 16 of Paper I. Considering that we used there the same number of stars, $N_*=20,000$, the potential uncertainties were much greater than in the analogous figure of this work, Figure \ref{fig:4kpc8Spiral_density_a}. We believe that this is how \RM{} accounts for an inconvenient potential parametrization---by increasing the uncertainty of the model estimate---which is exactly as it should be.

\subsection{Gaia measurement errors and choosing the survey volume size} \label{sec:discussion_choosing_SV}

Considering measurement uncertainties of distances and proper motions, we found in Paper I that for a survey volume with $r_\text{max} = 3~\text{kpc}$, distance uncertainties of $<10\%$ and proper motion uncertainties of less than $3~\text{mas yr}^{-1}$ \RM{} still gives unbiased parameter results. Even if the proper motion errors are not perfectly known. 

The measurement uncertainties of Gaia in proper motions (already in the first data release $\delta \mu \sim 1~\text{mas yr}^{-1}$; \citealt{2016A&A...595A...4L}) and in distances (at least for the final data release within $3~\text{kpc}$ and for bright stars; \citealt{2014EAS....67...23D}) lie below these limits. 

In paper I we focused on recovering completely unbiased model parameters and found that \RM{} is robust against moderate deviations of the model assumptions. In this work we released the condition that the model parameters themselves had to be recovered accurately, but allowed \RM{} to simply find an overall best fit for the data strongly affected by spiral arms---which was surprisingly successful in recovering the local potential even if the model parameters were not recovered. 

We therefore presume that in reality we probably have an even larger margin of error than we found in Paper I, and before the measurement uncertainties noticeably muddle the constraints.

In addition, we found in this work that a volume of $r_\text{max} = 3~\text{kpc}$ should be already big enough to find an overall best fit axisymmetric model for the Galaxy. At larger distances dust starts affecting the measurements. And inside of $R=3-4~\text{kpc}$ the stellar motions become increasingly non-axisymmetric, possibly because of the Galactic bar (e.g., \citealt{2014ApJ...783..130R,2015ApJ...800...83B}, and others, see Introduction in Section \ref{sec:intro}).

Overall we should therefore be very well-off by applying \RM{} to the final Gaia data set within $r_\text{max}=3~\text{kpc}$ only.

How well we can do with the first few Gaia data releases remains to be seen. The Gaia DR1 from September 2016 has parallax measurements of $\sim16\%$ for red clump giants at a distance of $\sim 500~\text{pc}$ from the Sun \citep{2014EAS....67...23D,2015A&A...574A.115M}, which is not precise enough for \RM{}. We could, however, use photometric distances, which should be precise enough for red clump stars and even extend over a larger volume. The Gaia DR2 in April 2018 might however already cover $\sim 2~\text{kpc}$ with good parallaxes for fainter stars as well.

\subsection{Spiral arms in the solar neighbourhood} \label{sec:discussion_sun_location}

The Sun is located in one of the smaller spiral arms of the MW, the local Orion spur/arm \citep{1953ApJ...118..318M}. Two of the MW's major spiral arms pass by the Sun within a few kpc: The Perseus arm is $\sim2~\text{kpc}$ from the Sun (towards the outer MW) \citep{2006Sci...311...54X}, and the Sagittarius arm at $\sim1~\text{kpc}$ \citep{2010PASJ...62..287S} (towards the Galactic center). It is, however, still under dispute which arms are actually major arms of the MW \citep{1985IAUS..106..335B,2013ApJ...769...15X,2013ApJ...775...79Z}.

How reliable the \RM{} results from the Gaia DR1 will be depends on the strength of the local Orion arm, which will dominate the survey volume. \RM{} had, for example, some difficulties recovering the circular velocity curve for small volumes (see Figure \ref{fig:vcirc_surfdens_suite_small}).

However, recent measurements of the MW's rotation curve \citep{2012ApJ...759..131B,2014ApJ...783..130R} confirm again that it is flat. We could impose this condition as a prior constraint in \RM{} (or fix the rotation curve slope as \citet{2013ApJ...779..115B} did in their \RM{} analysis). Based on independent information on the location and strength of the MW spiral arms, we could also use the present approach to estimate systematic uncertainties.

For later Gaia data releases, where the tracers extend further into the Galaxy ($\sim2~\text{kpc}$ from the Sun), the Sagittarius and Perseus arms could also play a role. In general, \RM{} should do much better with larger volumes, and if several spiral arms and inter-arm regions within the survey volume average each other out. This averaging should work especially well if the MW is---like the simulation in this study---a four-armed spiral. The exact number of spiral arms is however still disputed (see Section \ref{sec:comparison_with_MW}). If Gaia should show that the MW is a two-armed grand design spiral instead, having a very large survey volume would become even more important to achieve a good axisymmetric average \RM{} model for the MW. Local measurements of the potential as in Figure \ref{fig:vcirc_surfdens_suite_small} should however still be possible.

\subsection{Absence of a central bar in the simulation}

The simulation we analyzed in this work does not have a prominent bar, and so we have not explicitly explored the impact of such a feature. Bars can play an important role in the dynamics when very small volumes near a resonance are considered (see, e.g., \citealt{2000AJ....119..800D}, and references in Section \ref{sec:comparison_with_MW}). When considering volumes of $\gtrsim 1~\text{kpc}$, we have no reason to believe that this should severely affect the robustness of such an analysis.

\subsection{Interpreting \RM{} results}

The two main findings of this work give a clear directive regarding how we should deal with any future result about the MW's potential derived with \RM{}. (i) If the data spans a large volume, a significant proportion of the disk (at least $R\sim 5-11~\text{kpc}$), and averages over several spiral arms and inter-arm regions, we can trust and use the resulting model as an overall axisymmetric potential for the MW. For $R\sim 3-13~\text{kpc}$ we should even be able to make definite statements about the DM distribution. (ii) If the data do not span such a large volume, we can still believe the local constraints. In particular, the surface density within 1-2 disk scale heights, the circular velocity within the survey volume, and the average gravitational forces where the majority of stars is located.

Fortunately, this is consistent with the procedure by \citet{2013ApJ...779..115B}, who used \RM{} to constrain the vertical force $F_z$ for each MAP at only one radius that corresponds to a typical radius. It will be interesting to see whether \RM{} potential constraints from Gaia data will agree with their findings.

In addition, one should always compare the distribution of the data and the recovered model in configuration space $(\vect{x},\vect{v})$ and in action space, as we did in Section \ref{sec:results_part1}. This is not only a sanity check to confirm the goodness of the fit, but it might also reveal some substructure in the data that only becomes visible when comparing it to an axisymmetric smooth model.

\subsection{Recovering non-axisymmetric structures in the potential from modeling small volumes}

As we saw in Sections \ref{sec:circvel_surfdens} and \ref{sec:local_grav_forces}, \RM{} also makes a very good attempt at constraining the local potential for volumes as small as $r_\text{max}=500~\text{pc}$ or $1~\text{kpc}$. It recovers, for example, the higher surface density in volumes dominated by spiral arms, or correctly estimates the circular velocity at the median radial position of the stars (see Figure \ref{fig:vcirc_surfdens_suite_small}). 

At a time after the final Gaia data release, when we have data of high accuracy covering a large proportion of the disk, we could make use of this interesting property of \RM{} modeling. We could split the data set not only into different MAPs analogous to \citet{2013ApJ...779..115B}, but also into different spatial bins in the $(x,y)$ plane of the MW and model each of the smaller volumes separately. This approach would only probe the local potential, even when using an axisymmetric potential model, and should be sensitive to the overdensities induced by the spiral arms. In this way it should be possible to build up a non-axisymmetric map of the MW potential---albeit with very large spatial pixels---with constraints from dynamical modeling only.

\section{Conclusion} \label{sec:conclusion}

\RM{} is a well-tested axisymmetric dynamical modeling machinery that simultaneously fits an action-based DF and gravitational potential to the individual 6D phase-space coordinates of stellar populations in the MW disk. \RM{} builds on previous work by \citet{2011MNRAS.413.1889B}, \citet{2012MNRAS.426.1324B}, \citet{2015ApJS..216...29B}, and was first applied by \citet{2013ApJ...779..115B}. \RM{} was improved and tested in detail against the breakdowns of its modeling assumptions by \citet{2016ApJ...830...97T}. 

In this paper we investigated the robustness of \RM{} when modeling a non-axisymmetric system. We explore this for the first time explicitly, by modeling a simulated spiral galaxy from \citet{2013ApJ...766...34D}, and by comparing the results to the true potential. This simulation has stronger spiral arms than we expect in the MW, and---except of the absence of a bar and thick and gas disk components---it has matter components similar to the MW, as discussed in Section \ref{sec:comparison_with_MW}. We find that \RM{}-like action-based dynamical modeling is very robust against perturbations of spiral arms in this simulation, especially if the survey volume is large enough to encompass both spiral and inter-arm regions. In Section \ref{sec:results_part1} we demonstrated this in detail for a single \RM{} analysis of a data set with a spatial coverage of radius $r_\text{max}=4~\text{kpc}$ around a position equivalent to that of the Sun.

In Section \ref{sec:results_part2} we have investigated the role of survey volumes differing in size and position with respect to the spiral arms and inter-arm regions within the simulated galaxy. We find that the gravitational forces are mostly well-recovered at the locations of the stars that entered the analysis. For survey volume sizes $r_\text{max} \geq 3~\text{kpc}$ the recovered potential model already becomes a good average potential model for a large portion of the galaxy. For some positions of the survey volume center, e.g., in a smooth and not-too-depleted inter-arm region, smaller volumes can also give good overall constraints. If a small volume is dominated by a very strong spiral arm the constraints become less reliable, as expected. The correct DM halo scale length was, however, only recovered for a survey volume as large as $r_\text{max}=5~\text{kpc}$. 

This overall robustness of \RM{} is particularly notable, as the breakdown of the assumption of axisymmetry implies a breakdown of several model assumptions simultaneously: (i) orbital actions are not fully conserved anymore, (ii) the true potential is not spanned by the family of model potentials, (iii) the quasi-isothermal DF need not, or will not, describe the orbit distribution within spiral arms. However, the qDF seems to be informative enough to guide the fit to potential shapes that correctly measure the average surface density (within $\sim2 \times$ the disk scale height) and the circular velocity where most of the stars that entered the analysis are located---even for small volumes with $r_\text{max}=500~\text{pc}$ dominated by spiral arms. 

The results of this paper imply that \RM{} should be well-suited to making new measurements of the MW's gravitational potential with the upcoming Gaia data releases. It might even potentially work with the Gaia DR1 with its smaller coverage of the disk ($r_\text{max}\sim 1~\text{kpc}$) because the local Orion arm, in which the Sun is located, is thought to be only a minor spiral arm in the MW and should not significantly disturb the modeling. 


\section{Acknowledgments}

W.H.T. thanks Stephen Pardy for helpful advice on handling $N$-body simulations. W.H.T.\ and H.-W.R.~acknowledge funding from the European Research Council under the European Union’s Seventh Framework Programme (FP 7) ERC Grant Agreement n.~${\rm [321035]}$. J.B. received support from the Natural Sciences and Engineering Research Council of Canada and the Alfred P. Sloan Foundation. E.D. gratefully acknowledges the support of the Alfred P. Sloan Foundation and ATP NASA Grant No NNX144AP53G, and support by Sonderforschungsbereich SFB 881 "The Milky Way System" (A3) of the German Research Foundation (DFG). 

\software{\texttt{GADGET-3} \citep{2005MNRAS.364.1105S}, \texttt{galpy} \citep{2015ApJS..216...29B}, \texttt{emcee} \citep{2013PASP..125..306F}, \texttt{Matplotlib} \citep{Hunter:2007}}

\bibliography{references}{}
\bibliographystyle{aasjournal}

\end{document}